\documentclass[showpacs,twocolumn,aps,preprintnumbers,letterpaper]{revtex4}
\usepackage{amsmath,amssymb}
\usepackage{epsfig}
\usepackage{graphicx}
\usepackage{amsmath}
\usepackage{slashed}
\usepackage{amsfonts}
\usepackage{epstopdf}
\usepackage{color}
\usepackage[caption=false]{subfig}
\usepackage[pdftex, pdfborder= 0 0 0, citecolor=blue, urlcolor=blue, linkcolor=blue, colorlinks=true, bookmarksopen=true]{hyperref}

\newcommand{\be}{\begin{equation}}
\newcommand{\ee}{\end{equation}}
\newcommand{\bear}{\begin{eqnarray}}
\newcommand{\eear}{\end{eqnarray}}
\newcommand{\ba}{\begin{array}}
\newcommand{\ea}{\end{array}}

\def\be{\begin{eqnarray}}
\def\ee{\end{eqnarray}}
\def\bea{\be}
\def\eea{\ee}

\def\roughly#1{\mathrel{\raise.3ex\hbox{$#1$\kern-.75em%
\lower1ex\hbox{$\sim$}}}}

\def\abs#1{{\left| #1 \right|}}

  \long\def\comment#1{ }

  \newcommand{\Tr}{{\rm Tr}}

  \newcommand{\beq}{\begin{eqnarray}}
  \newcommand{\eeq}{\end{eqnarray}}
  
 \def\simge{\mathrel{%
   \rlap{\raise 0.511ex \hbox{$>$}}{\lower 0.511ex \hbox{$\sim$}}}}
\def\simle{\mathrel{
   \rlap{\raise 0.511ex \hbox{$<$}}{\lower 0.511ex \hbox{$\sim$}}}}

\begin{document}

\title{Diffractive photoproduction of $J/\psi$ and $\Upsilon$ using holographic QCD: \\
gravitational form factors and GPD of gluons in the  proton}

\author{Kiminad A. Mamo}
\email{kiminad.mamo@stonybrook.edu}
\affiliation{Department of Physics and Astronomy, Stony Brook University, Stony Brook, New York 11794-3800, USA}
\author{Ismail Zahed}
\email{ismail.zahed@stonybrook.edu}
\affiliation{Department of Physics and Astronomy, Stony Brook University, Stony Brook, New York 11794-3800, USA}



\date{\today}
\begin{abstract}
We present a holographic analysis of diffractive photoproducton of charmonium $J/\psi$ and upsilonium  $\Upsilon$ on a proton, considered as a bulk Dirac fermion, for all ranges of $\sqrt{s}$, i.e., from near threshold to very high energy. Using the bulk wave functions of the proton and vector mesons, within holographic QCD, and employing Witten diagrams in the bulk, we compute the diffractive photoproduction amplitude of $J/\psi$ and $\Upsilon$. The holographic amplitude shows elements of the strictures of vector meson dominance (VMD). It is dominated by the exchange of a massive graviton or $2^{++}$ glueball resonances near threshold, and its higher spin-j counterparts that reggeize at higher energies. Both the differential and total cross sections are controlled by the gravitational form factor $A(t)$, and compare well to the recent results reported by the GlueX collaboration near threshold and the world data at large $\sqrt{s}$.  The holographic gravitational form factors, including the D-term, which is due to the exchange of massive spin-0 glueballs, are in good agreement with lattice simulations. We use it to extract the holographic pressure and shear forces inside the proton. Finally, using a pertinent integral representation of the holographic gravitational form factor $A(t)$ near threshold, and its Pomeron counterpart way above threshold,  we extract the generalized parton distribution (GPD) of gluons inside the proton at different resolutions.
\end{abstract}


\maketitle

\setcounter{footnote}{0}


\section{Introduction}
Exclusive production of heavy mesons such as charmonia and bottomonia through the use of
photo- or electroproduction processes  provides the optimal framework for diffractive physics.
In the limit when the coherence length of the virtual photon is large in comparison to the proton
size, the scattering virtual photon on a hadron is equivalent to the scattering of a hadron onto
a hadron. The process is mostly dominated by the exchange of gluons with vacuum quantum
numbers, leading to a slowly rising cross section at high energy. The rise is
due to the exchange of a Pomeron, an effective object  lying on the highest Regge trajectory.
First principle perturbative QCD calculations ~\cite{Kuraev:1977fs,Balitsky:1978ic} provide
insights to the nature of this exchange, although the softness of the exchange suggests an
altogether non-perturbative approach.

Soft electroproduction on a nucleon is analogous to a hadron of varying size scattering off a
nucleon, with a virtual photon wavefunction of squared transverse size $1/Q^2$. In the
photoproduction limit with $Q^2\rightarrow 0$,
the size is hadronic and non-perturbative physics applies.
The diffractive and non-perturbative production process
whereby the soft virtual photon turns to a heavy meson is analogous to the scattering of two
dipoles with light-cone wavefunctions for the in-out virtual photon states. It is inherently
non-perturbative at small $Q^2$.  Throughout, we will focus on electroproduction close to the
photon point or photoproduction for heavy mesons such as charmonium and bottomium.

Holographic QCD provides a non-perturbative framework for discussing structure and scattering
of hadrons. It  stems from a conjecture that observables in strongly coupled gauge theories
in the limit of a large number of colors, can be determined from classical fields interacting through
gravity in generally an anti-de-Sitter space in higher dimensions~\cite{HOLOXX}. The original conjecture was put
forth and demonstrated for conformal ${\cal N}=4$ Yang-Mills theory, and argued by many to hold
under some assumptions for non-conformal gauge theories such as QCD. Exclusive production
of heavy mesons has been analyzed in the context of holographic QCD at high energy~\cite{DJURIC,LEE}, where the
exchange reggeizes~\cite{Rho:1999jm,Janik:2000aj,Brower:2006ea,Stoffers:2012zw,Stoffers:2012ai,Basar:2012jb,Hatta:2007he,Hatta:2009ra}.
Diffractive production of vector mesons  in the non-holographic context  can be found in~\cite{MANY}.

Recently, the GlueX collaboration has put forth measurements of threshold charmonium production
using virtual photons close to the photon point~\cite{GLUEX}. Additional measurements  at JLab in this
channel  with higher accuracy using the SoLID detector should improve further the statistics~\cite{MEZIANI}.
One purpose of these experiments is the extraction of  the
gluonic component  entering the composition of the nucleon mass. In this spirit,
a new analysis of these threshold data was carried in~\cite{Hatta:2018ina,Hatta:2019lxo} using a hybrid holographic construction
combining general QCD arguments and lattice results. One of the  purposes of this paper is to carry an analysis
of the new GlueX data near threshold~\cite{GLUEX}  and the existing world data well above threshold, all within a holographic QCD model using the bottom-up approach. This analysis complements  the earlier investigations in~\cite{DJURIC, LEE} at high energy, all the way
 to threshold. For completeness, we note the earlier  suggestion to use the photoproduction process near
 threshold to probe the gluon content of the nucleon~\cite{BRODSKY}.

The holographic photoproduction  amplitude is dominated by the exchange of a massive
$2^{++}$ graviton at threshold, and higher spin-j exchanges away from threshold that rapidly reggeize.
The $0^{++}$ glueballs are found to
decouple owing to their vanishing coupling to the virtual photons, while the dilatons are shown to
decouple from the bulk Dirac fermion. At threshold, the holographic photoproduction amplitude directly probes
a pertinent gravitational form factor which maps on the gluonic contribution to the energy momentum tensor of the nucleon as a Dirac fermion in the bulk.

This paper consists of several new results:
 1/ The derivation of all three holographic gravitational form factors and
their comparison to recent lattice data;
2/ The derivation of the gluonic pressure and shear forces inside the proton;
3/ The derivation that the holographic processes $\gamma p\rightarrow V p$ and $\gamma p\rightarrow \gamma^* p$ are related
in bulk by vector meson dominance (VMD);
4/ The derivation of the holographic photoproduction
differential and total cross sections for $J/\Psi$ and their comparison to current data for all energies;
5/ The derivation  that the threshold cross section
is dominated by only one invariant gravitational form factor $A(t)$, due to the exchange of a $2^{++}$ glueball in bulk;
6/ The extraction of the value of $A(0)$ from the data for different  brane embeddings;
7/ The derivation of the holographic gluonic GPD   of  the nucleon as a bulk Dirac fermion;
8/ The prediction for the diffractive photoproduction of $\Upsilon$.

The organization of the paper is as follows:  In section II we review the kinematics for a general $2\rightarrow 2$ process. In
section III,  we detail the general structures of the Witten diagrams for exclusive process, like the diffractive photoproduction of $J/\psi$, by using the bulk wave functions of hadrons in holographic QCD. In section IV, we introduce in detail the bottom holographic holographic model we use, and derive the bulk vertices for the Witten diagrams from the bulk action of the model. In section V, we derive the holographic gravitational form factors using Witten diagrams, and campare them
to the recent lattice results. In section VI, we use our holographic D-term to calculate the pressure distribution and shear forces inside the proton.
In section VII, we show how vector meson dominance (VMD) holds in the present holographic construction, and derive the scattering amplitude for the diffractive photoproduction by approximating the bulk-to-bulk glueball propagator near the boundary which will enable us to write down the scattering amplitude explicitly in terms of the gravitational form factor $A(t)$ of spin-2 glueball exchanges. In section VIII, the photoproduction differential and total
cross sections close to the photon point are detailed at threshold in the single graviton exchange limit.
In section IX, we generalize the result beyond threshold through reggeization by including the higher spin-j exchanges and their re-summation. In section X, we derive the gluonic GPD from a pertinent integral representation of the form factor $A(t)$. Our conclusions are in section XI, and details of the calculations are given in several appendices.

\section{\label{kinematics} kinematics of  the $\gamma^*p\rightarrow Vp$ process}

Throughout, we will refer to real and virtual photoproduction by $\gamma^*$ in the general presentation, but we will specialize to photoproduction in most
of the specific analyses and results. All our arguments extend readily to diffractive electroproduction of heavy mesons $V=J/\Psi, \Upsilon$  with minor changes.

We start by briefly reviewing  the kinematics for the
process $\gamma^*p\rightarrow Vp$.
We first define the Lorentz scalars as $s=W^2=(p_1+q
_1)^2$, and $t=(p_1-p_2)^2=(q_1-q_2)^2$ where $q_{1,2}$ are the four-vectors of the virtual photon and vector meson, respectively (note that we occasionally use the notation $q\equiv q_1$ and $q'\equiv q_2$), and $p_{1,2}$ are the four vector of the proton. Throughout
we will work with mostly negative signature, i.e., $\eta_{\mu\nu}=(+1,-1,-1,-1)$. Note that our convention is different from the mostly positive signature used in most
holographic analyses.

\begin{figure}[!htb]
\includegraphics[height=5cm]{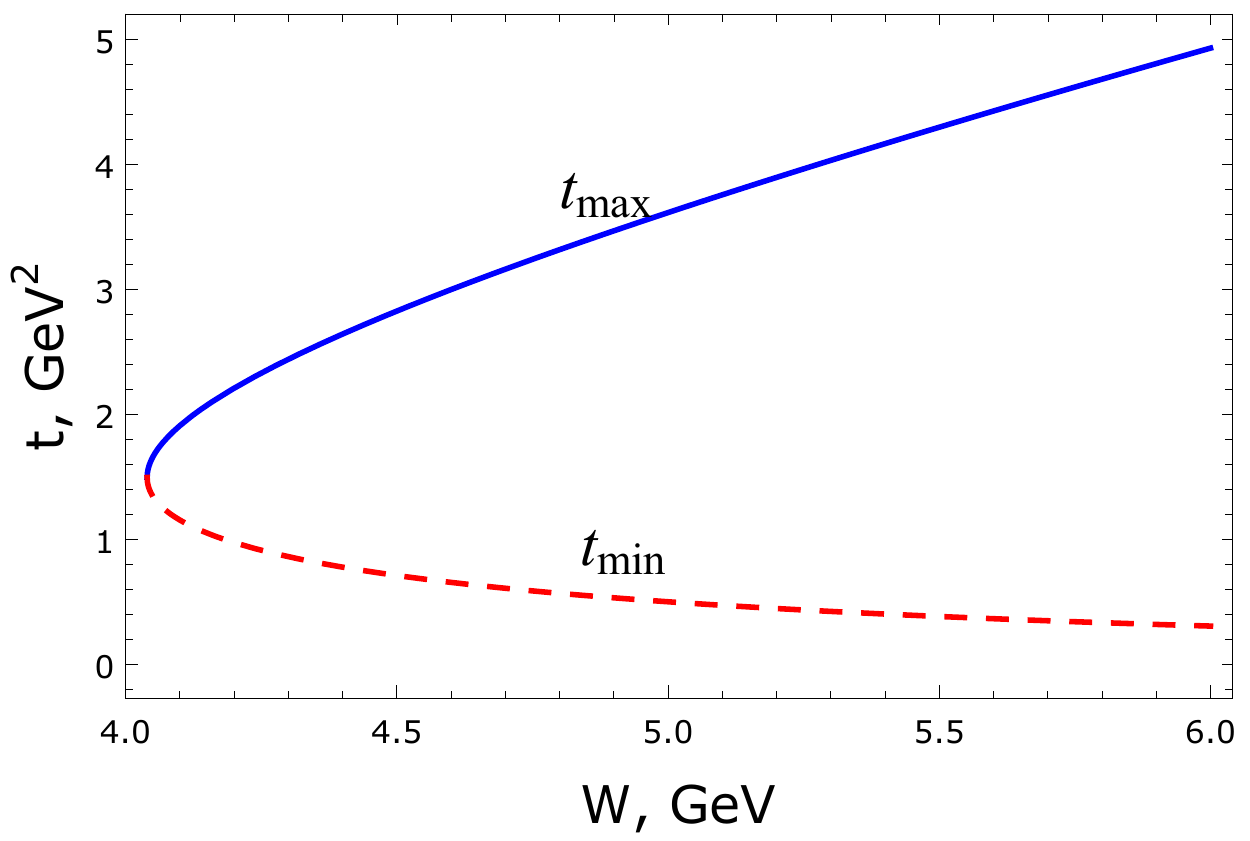}
  \caption{$t_{min}$ and $t_{max}$ vs $W=\sqrt{s}$ for $M_{V}=M_{J/\psi}=3.10~GeV$, $m_{N}=0.94~GeV$, and $Q=0$. Note that at the threshold energy $W_{tr}=\sqrt{s_{tr}}=m_{N}+M_{V}=4.04~GeV$, we have $t_{min}=t_{max}$.}
\label{tmx}
\end{figure}

We will work in the center-of-mass (CM) frame of the pair composed of the virtual photon $\gamma^*$ and the proton. In this frame, one can derive the mathematical relationships between the three-momenta of the virtual photon and vector meson ($\mathbf{q}_{\gamma}$, $\mathbf{q}_V$) and Lorentz scalars ($s$, $t$, $q_1^2=-Q^2$, $q_2^2=M_{V}^2$, $p_1^2=p_2^2=m_{N}^2$) as (see, for example, Eqs.11.2-4 in~\cite{Srednicki:2007qs})

\be
\vert\mathbf{q}_\gamma\vert=\frac{1}{2\sqrt{s}}\sqrt{s^2-2(-Q^2+m_N^2)s+(-Q^2-m_N^2)^2}\,,\nonumber\\
\ee
\be
\vert\mathbf{q}_V\vert=\frac{1}{2\sqrt{s}}\sqrt{s^2-2(M_V^2+m_N^2)s+(M_V^2-m_N^2)^2}\,,\nonumber\\
\ee
and

\be
\label{tminmax}
t=-Q^2+M_V^2-2E_{\gamma}E_V+2\vert\mathbf{q}_\gamma\vert\vert\mathbf{q}_V\vert\cos\theta\,,\nonumber\\
\ee
Here  $E_{\gamma}=(-Q^2+\mathbf{q}_{\gamma}^2)^{\frac 12}$ is the energy of the virtual photon, and
$E_{V}=(M_V^2+\mathbf{q}_{V}^2)^{\frac 12}$ is the energy of the vector meson. The t-transfer at low $\sqrt{s}$
is bounded by $t_{min}\equiv |t\vert_{\cos\theta=+1}|$ and $t_{max}\equiv|t\vert_{\cos\theta=-1}|$
as illustrated in Fig.~\ref{tmx}.

We now note that at threshold and for example $V=J/\Psi$ with $s_{\rm tr}=(m_N+M_V)^2=4.04\,{\rm GeV}^2$

 \bea
 -t_{\rm min}(s=s_{\rm tr})=&&\frac{m_NM^2_V}{m_N+M_V} \nonumber\\
 =&&1.5^2\,{\rm GeV}^2\ll 4.04^2\,{\rm GeV}^2=s_{\rm tr}\nonumber\\
 \label{6}
 \eea
 and away from threshold

 \be
- t_{\rm min}(s\gg s_{\rm tr})\sim \left(\frac {m_NM_V}s\right)^2\ll s
 \label{7}
 \ee
 The photoproduction kinematics for charmonium and also  bottomium, is dominated
 by the diffractive process all the way to threshold.

The differential cross section for the photoproduction process $\gamma^*p\rightarrow Vp$
is given by (see for example, Eq.11.34 in \cite{Srednicki:2007qs})

\be
\frac{d\sigma}{dt}=\frac{e^2}{64\pi s\vert\mathbf{q}_\gamma\vert^2}\vert\mathcal{A}_{\gamma*p\rightarrow Vp}(s,t)\vert^2\,.
\ee
and the total cross section for small $\sqrt{s}$ close to threshold is

\be
\label{5}
\sigma(s)=\int_{t_{min}}^{t_{max}}\,dt\left (\frac{d\sigma}{dt}\right)\,.
\ee
We  now show how to use Witten diagrams in AdS with bulk wavefunctions for the vector mesons,
bulk-to-boundary and bulk-to-bulk propagators within pertinent holographic models in the bottom-up approach.

 \section{\label{hphoto}holographic photoproduction of vector mesons}

\begin{figure}[!htb]
\includegraphics[height=5cm]{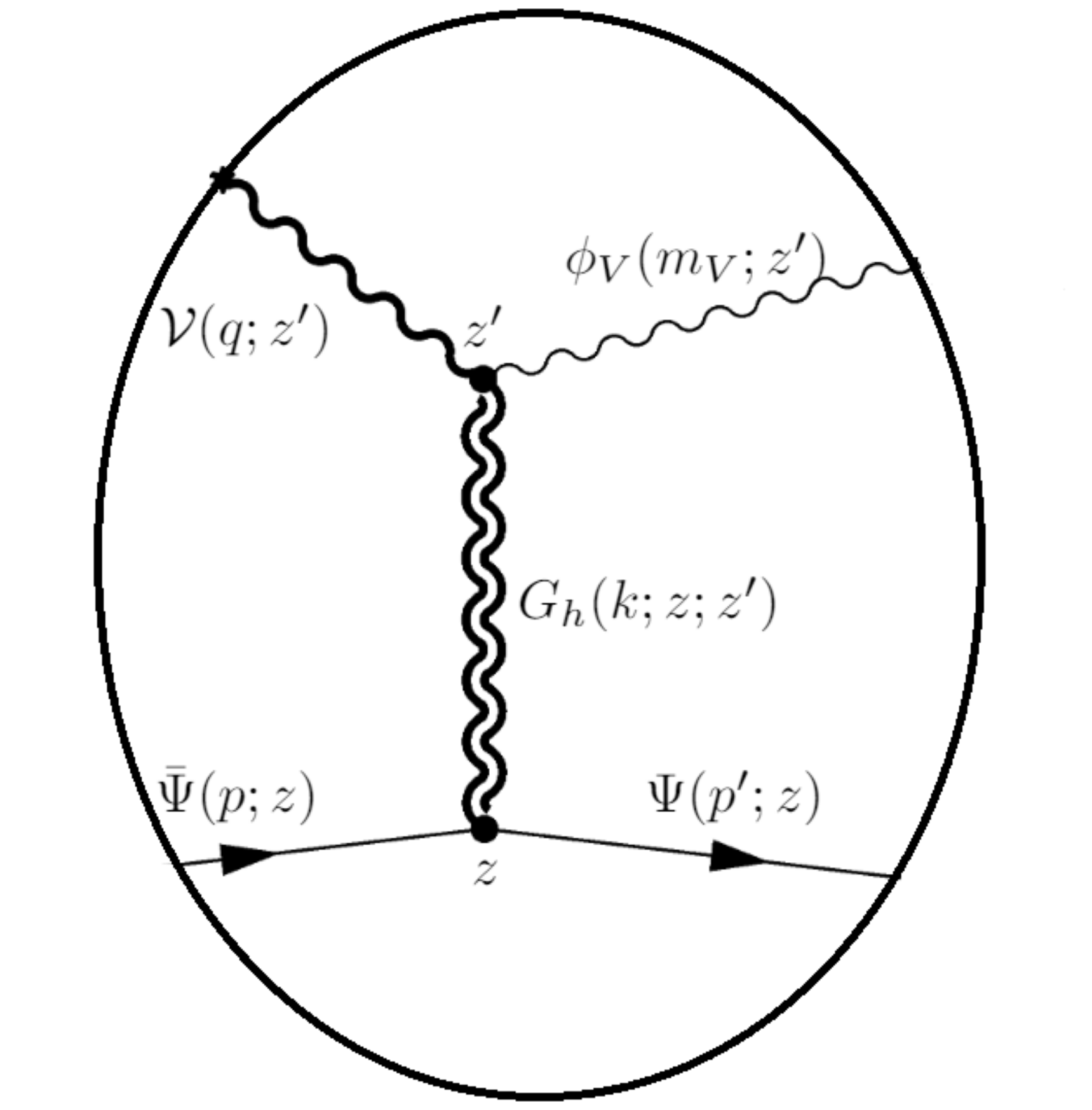}
  \caption{Witten diagram for the diffractive photoproduction of vector mesons with a bulk wave function $\phi_V$. The thick lines or thick wiggles represent the propagators of summed over vector meson or glueball resonances. The thin lines or thin wiggles correspond to a single vector meson and proton. For scalar glueball resonances, due to the dilaton and the trace-full part of the metric fluctuation, we simply replace the bulk-to-bulk propagator $G_h(k,z,z')$ of spin-2 glueballs by $G_{\varphi,f}(k,z,z')$.}
  \label{wdiagram2}
\end{figure}

The diffractive  amplitude for the photoproduction of a vector meson, in a given holographic model of QCD, can be computed by using the Witten diagram shown in Fig.~\ref{wdiagram2}, where bulk VMD is manifest as we will detail below.
The structure of the Witten diagram is pretty general, and can be applied to any holographic model to QCD with  a mass-gap, and a discrete mass spectrum of hadrons.


The main elements of the Witten diagrams shown in Fig.~\ref{wdiagram2} (also in Figs.~\ref{wdiagram2},~\ref{wdiagram3} for the  gravitational form factor) are composed of:
\\
\\
{\bf 1/}  the bulk-to-boundary propagator of the vector mesons (or virtual photons for space-like momenta $q^2=-Q^2$)   as

\be
V(q,z)=\mathcal{V}(q=iQ,z)=C_V\times g_5\sum_n \frac{F_n\phi_n(z)}{Q^2+m_n^2}\,,\nonumber\\ \label{vps01}
\ee
where $\phi_n(m_n,z)$, $m_n$, $f_n\equiv-{F_{n}}/{m_n}$, and $g_5$ are the bulk wave function, mass, decay constant, and hadronic coupling constant of each meson resonances, respectively. $C_V$ is a normalization constant for the  mesons which can be identified with the value of the electromagnetic form factor of the proton at zero-momentum transfer (and $C_V=F_1^{(P)}(Q=0)=1$ since the electric charge of the proton is normalized to one in units of $e$);
\\
\\
{\bf 2/} the bulk-to-boundary propagator of the spin-2 glueballs (for space-like momenta $k^2=-K^2$)

\be
h(k,z)=\mathcal{H}(K,z)=C_h\times\sqrt{2}\kappa\sum_n \frac{F_n\psi_n(z)}{K^2+m_n^2}\,, \label{hps1}
\ee
where $\psi_n(m_n,z)$, $m_n$, $f_n\equiv-{F_{n}}/{m_n}$, and $\kappa$ are the bulk wave function, mass, decay constant, and hadronic coupling constant of each glueball resonances. $C_h$ is the normalization constant for glueballs (which will be identified with the gravitational form factor of the proton at zero momentum transfer, i.e., $C_h\equiv A(t=0)$);
\\
\\
{\bf 3/} the bulk-to-bulk propagators of the vector meson and glueball resonances

\be
G_V(q',z,z')=C_{V}\times\sum_n \frac{\phi_n(z)\phi_n(z')}{q'^2-m_n^2} , \label{vbbt01}
\ee
and
 \be
G_h(k,z,z')=C_{h}\times\sum_n \frac{\psi_n(z)\psi_n(z')}{k^2-m_n^2}\,; \label{vbbt01}
\ee
and the bulk wave function of the proton (a Dirac fermion in the bulk) is denoted as $\Psi(p,z)$.
 \\
 \\
More specifically, for the hard-wall and soft-wall holographic models of QCD, that we focus on in this paper, all the ingredients of the Witten diagram  Fig.~\ref{wdiagram2} are determined in terms of their bulk wave functions, the normalization constants $C_{V,h}$, the mass scale parameters $z_0$ for the hard-wall ($\tilde{\kappa}_{V,N}$ and $c_V$ for the soft-wall), and the hadronic coupling constants $g_5$ (for mesons) and $\kappa$ (for glueballs).

The mass scale parameters $z_0$ or $\tilde{\kappa}_{\rho,N}$ are simultaneously fixed to the proton's and the $\rho$ meson's
mass, $\tilde{\kappa}_{V}$ and $c_V$ for $V=(J/\psi,\Upsilon)$ are fixed by their mass $m_V=(m_{J/\psi},m_{\Upsilon})$ and decay constants $f_V=(f_{J/\psi},f_{\Upsilon})$. The hadronic coupling constant of glueballs $\kappa$ is fixed by using Type II supergravity action on $AdS_5\times S^5$, and the hadronic coupling of vector mesons is fixed by using the DBI action for D7 or D9 flavor branes. Finally, we will extract the gravitational form factor $A(0)={C_h}/{g_5^2}$ by comparing the holographic scattering amplitude to experimental data in the low energy regime.

Note that, in general, the normalized bulk wave function of one of the vector meson resonances $\phi_{n=0}\equiv\phi_V$ takes the form
\be
\phi_V=c_V zJ(M_Vz)=\frac{f_V}{M_V}\times M_VzJ(M_Vz)
\ee
where $J(M_Vz)$ is a special function that depends on the details of the holographic model. And, the decay constant $f_V$, for a meson at rest, defined as
\be
<0\vert J_{V,i}\vert V_j>=f_{V}M_V \delta_{ij}
\ee
is calculable in a given holographic model to QCD, and can be extracted experimentally from the leptonic width as
\be
\Gamma(V\rightarrow \ell^+\ell^-)=\frac{4\pi}{3}\alpha^2_{QED}e_V^2\dfrac{f_{V}^2}{M_V}
\ee
where $e_V$ is the electric charge of the constituent quarks of the vector meson.
For $V=(J/\Psi, \Upsilon)$:   $e_V=(2/3, 1/3)$, $M_V=(3.097, 9.460)$ GeV  and $e_Vf_V=(270, 238)$ MeV.


 \section{holographic model}

 We consider AdS$_5$  with a background metric  $g_{MN}=(\eta_{\mu\nu},-1)/z^2$ and $\eta_{\mu\nu}=(1,-1,-1,-1)$.
 Confinement will be described by  a background dilaton $\phi=\tilde{\kappa}_{V}^2z^2$ for mesons,
 $\phi=\tilde{\kappa}_{N}^2z^2$ for protons and  $\phi=2\tilde{\kappa}_{N}^2z^2$ for glueballs in the soft wall model.
 In the hard wall model, $\phi=0$ and confinement  is set at $z=z_0$. The bulk graviton and dilaton fields will be
 described by $\varphi$ and $h_{\mu\nu}$ respectively, while the bulk  U(1) vector gauge field and a spin-$\frac 12$ Dirac
 fermion by $V^M$ and $\Psi$ respectively.

 \subsection{Bulk Dirac fermion and vector meson}

The bulk Dirac fermion action in curved AdS$_5$ with minimal coupling to the U(1) vector meson is ~\cite{CARLSON}

\be
S=\int d^{5} x \sqrt{g}\,\big(\mathcal{L}_F+\mathcal{L}_V \big)+\int d^4 x \sqrt{-g^{(4)}}\mathcal{L}_{UV}\,,\nonumber\\
\label{Action}
\ee
with the fermionic, gauge field and boundary actions

\bea
\label{fermionAction}
\mathcal{L}_F=&& \frac{1}{2g_{5}^2}e^{-\phi(z)} \nonumber\\
&&\times \bigg( \frac{i}{2} \bar{\Psi} e^N_A \Gamma^A\big(\overrightarrow{D}_N-\overleftarrow{D}_N\big)\Psi-(M+V(z))\bar{\Psi}\Psi\bigg)\,,\nonumber\\
\mathcal{L}_V= &&-\frac{1}{4g^2_5}\,e^{-\phi(z)}\,g^{\mu\alpha}g^{\beta\nu}\,F_{\mu\nu}^V\,F_{\alpha\beta}^V\,,\nonumber\\
\mathcal{L}_{UV}=&&
 \frac{1}{2g_5^2}\left(\bar{\Psi}_L \Psi_R +\bar{\Psi}_R \Psi_L\right)_{z=\varepsilon}\,,
\ee
We have fixed the potential $V(z)=\tilde{\kappa}_{N}^2z^2$ for both the hard and soft wall model.
We have denoted by  $e^N_A=z \delta^N_A$ the inverse vielbein, and defined the covariant derivatives

\bea
\overrightarrow{D}_N=&&\overrightarrow{\partial}_N +\frac{1}{8}\omega_{NAB}[\Gamma^A,\Gamma^B]-iV_N\nonumber\\
\overleftarrow{D}_N=&&\overleftarrow{\partial}_N +\frac{1}{8}\omega_{NAB}[\Gamma^A,\Gamma^B]+iV_N
\eea
The components of the spin connection are $\omega_{\mu z\nu}=-\omega_{\mu\nu z}=\frac{1}{z}\eta_{\mu\nu}$, the Dirac gamma matrices  satisfy anti-commutation relation $\{\Gamma^A,\Gamma^B\}=2\eta^{AB}$, that is, $\Gamma^A=(\gamma^\mu,-i\gamma^5)$, and $F^V_{MN}=\partial_M V_N-\partial_N V_M$.
The  equation of motions for the bulk Dirac fermion and the U(1) gauge field follow by variation

\bea
&&\big[i e^N_A \Gamma^A D_N -\frac{i}{2}(\partial_N\phi)\, e^N_A \Gamma^A- (M+\phi(z))\big]\Psi =0\,,\nonumber\\
&&\frac{1}{\sqrt{g}}\partial_M\big(\sqrt{g}e^{-\phi}F^{MN}\big)=0\,.\nonumber\\
\eea

The coupling $g_5$ is inherited from the nature of the brane embeddings in bulk:
${1}/{g_{5}^2}\equiv{3N_cN_f}/(12\pi^2)$ (D7-branes),
and ${1}/{g_{5}^2}\equiv ({3\sqrt{\lambda}}/{2^{5/2}\pi}){N_cN_f}/({12\pi^2})$ (D9-branes).
The brane embeddings
with $N_f=1$ are more appropriate for describing heavy mesons in bulk, as the U(1) field mode decompose in
an infinite tower of massive vector  mesons on these branes as we discussed above.
When ignoring these embedding, the standard assignement is:
${1}/{g_{5}^2}\equiv{N_c}/(12\pi^2)$.

We note that in (\ref{fermionAction}), we  have excluded a Yukawa-type coupling between the dilaton and  the bulk Dirac fermion,  since neither the fermionic part of the Type IIB supergravity action (see, for example, Eq.~A.20 in \cite{DHoker:2016ncv}) nor the fermionic part of the  DBI action in string theory (see, for example, Eq.~56 in \cite{Kirsch:2006he}) support such a coupling.

\subsection{Spectra}

The spectrum for the hard wall model is fixed by the zeros of the Bessel function $J_1(m_nz_0)=0$
and does not Reggeize. It does in the soft wall model by solving the equation of motion for $V^N$ following
from (\ref{Action}). The results for the heavy meson masses and decay constants are~ \cite{Grigoryan:2010pj}

\bea
\label{SOFTMF}
m_n^2=&&4\tilde\kappa^2_V(n^*+1)\nonumber\\
g_5f_n=&&\sqrt{2}\tilde\kappa_V\left(\frac{n+1}{n^*+1}\right)^{\frac 12}
\eea
with $n^*=n+c_V^2/4\tilde\kappa_V^2$. The additional constant $c_V$ is
fixed as ${c_{V}^2}/{4\tilde{\kappa}_{V}^2}={M_{V}^2}/{4\tilde{\kappa}_{V}^2}-1$ for $n=0$ for the heavy mesons  $V=(J/\psi , \Upsilon)$,
and $c_\rho=0$ for the light mesons. The mass spectrum of the bulk Dirac fermions is given by~\cite{CARLSON}

\be
\label{SOFTMFD}
m_n^2=&&4\tilde\kappa^2_N(n+\tau-1)\,,
\ee
with the  twist factor $\tau$.  For the specific soft wall applications
to follow we will set $\tilde\kappa_N=\tilde\kappa_V=\tilde\kappa_\rho$ for simplicity,
unless specified otherwise.

\subsection{Bulk graviton and dilaton}

The graviton  in bulk is dual to a glueball on the boundary. It is a rank-2 tensor with reducible parts in general.
To decompose the graviton tensor $h_{\mu\nu}$ to its transverse and traceless part $h$, and trace-full part $f$
we follow~\cite{Kanitscheider:2008kd} and define

\be
\label{exp}
h_{\mu\nu}=\epsilon_{\mu\nu}^{TT}\,h+\tilde{k}^2\epsilon_{\mu\nu}^{T}\,f-\tilde k_\mu \tilde k_\nu H+\tilde k_\mu A_{\nu}^{\perp}+\tilde k_\nu A_{\mu}^{\perp}\nonumber\\
\ee
where

\be
k^\mu\epsilon_{\mu\nu}^{TT}=&&\eta^{\mu\nu}\epsilon_{\mu\nu}^{TT}=0\nonumber\\
 \epsilon_{\mu\nu}^{T}=&&\frac{1}{4}\eta_{\mu\nu}
\ee
with $\alpha\equiv {\tilde{k}}/{kz_0}$  a dimensionless normalization constant which can be fixed empirically.
Here $z_0$ is the hard-wall scale, and $k^\mu A_{\mu}^{\perp}=0$. A similar rescaling follows in the soft-wall
model  with $z_0\rightarrow 1/\tilde\kappa_V$.

In a gauge where $A_{\mu}^{\perp}=0$, the equation of motion for $h$ decouples.
In contrast, the equations for  $f$, $H$, and $\varphi$ (denoted as $k$ in \cite{Kanitscheider:2008kd}) are coupled
(see Eqs.7.16-20 in \cite{Kanitscheider:2008kd}). Diagonalizing the equations, one can show that $f$ satisfies the same equation of motion as $h$ \cite{Kanitscheider:2008kd}. Also note that $f_{0}=f(z=0)$ couples to $T^{\mu}_{\mu}$ of the gauge theory, while $H_{0}=H(z=0)$ couples to $k^{\mu}k^{\nu}T_{\mu\nu}\equiv 0$ (see Eq.7.6 of \cite{Kanitscheider:2008kd}).

\subsubsection{Action}

The effective action for the gravitaton  ($\eta_{\mu\nu}\rightarrow\eta_{\mu\nu}+h_{\mu\nu}$) and dilaton fluctuations ($\phi\rightarrow\phi+\varphi$) follows from the Einstein-Hilbert action plus dilaton by expanding to quadratic order, and after adding the  background de-Donder gauge fixing term. The result is

\be
S=\int d^{5} x \sqrt{g}\,e^{-2\phi}\big(\mathcal{L}_{h+f}+\mathcal{L}_\varphi \big)\,,\nonumber\\
\label{Action2}
\ee
with

\bea
\label{kinetic}
\mathcal{L}_{h+f} =&& -\frac{1}{4\tilde{g}_5^2}\,g^{\mu\nu}\,\eta^{\lambda\rho}\eta^{\sigma\tau}\partial_{\mu}h_{\lambda\sigma}\partial_{\nu}h_{\rho\tau}\nonumber\\
&&+\frac{1}{8\tilde{g}_5^2}\,g^{\mu\nu}\eta^{\alpha\beta}\eta^{\gamma\sigma}\,\partial_{\mu}h_{\alpha\beta}\,\partial_{\nu}h_{\gamma\sigma}\,,\nonumber\\
\mathcal{L}_\varphi=&&+\frac{1}{2\tilde{g}_5^2}\,g^{\mu\nu}\,\partial_{\mu}\varphi\,\partial_{\nu}\varphi\,,
\ee
and $\tilde{g}_5^2=2\kappa^2=16\pi G_N={8\pi^2}/{N_c^2}$.

\subsubsection{Spectrum}

In the soft wall model, the glueball spectrum is determined by solving the equation of motion for $h_{\mu\nu}$ following from (\ref{Action2}). The results for the spin-2 glueball masses and decay constants are

\bea
\label{SOFTMFG2}
m_n^2=8\tilde\kappa^2_N(n+1)\qquad
\tilde{g}_5f_n=2\tilde\kappa_N
\eea
They differ from their vector meson counterparts in (\ref{SOFTMF}) by the replacements
 $\tilde\kappa_V\rightarrow\sqrt{2}\tilde\kappa_N$ and $g_5\rightarrow\tilde g_5$ due to the difference in the bulk actions.
For spin-0 glueballs, we have for the trace-full part of the metric fluctuation

\bea
\label{SOFTMFG3}
m_n^2=8\tilde\kappa^2_N(n+1)\qquad
\sqrt{2}\tilde{g}_5f_n=2\tilde\kappa_N\
\eea
 after replacing $\tilde g_5\rightarrow\sqrt{2}\tilde g_5$ in the results for spin-2 glueballs.
For the dilaton fluctuations we have

\bea
\label{SOFTMFG4}
m_n^2=8\tilde\kappa^2_N(n+1)\qquad
\tilde{g}_5f_n=2\tilde\kappa_N
\eea

\subsubsection{Couplings}

For the graviton in the axial gauge $h_{\mu z}=h_{zz}=0$.
The pertinent couplings in Fig.~\ref{wdiagram2}, which follow from linearizing the action (\ref{Action}) by replacing $\eta_{\mu\nu}\rightarrow\eta_{\mu\nu}+h_{\mu\nu}$, are

  \be
 h\overline\Psi\Psi:\quad &&-\frac{\sqrt{2\kappa^2}}{2}\int d^5x\,\sqrt{g}\,h_{\mu\nu}T_F^{\mu\nu}\nonumber\\
  h AA:\quad && -\frac{\sqrt{2\kappa^2}}{2}\int d^5x\,\sqrt{g}\,h_{\mu\nu}T_V^{\mu\nu}\nonumber\\
  \label{vertices1}
 \ee
with the energy-momentum tensors

 \bea
T_F^{\mu\nu}&=&e^{-\phi}\frac{i}{2}\,z\,\overline\Psi\gamma^\mu\overset{\leftrightarrow}{\partial^\nu}\Psi-\eta^{\mu\nu}\mathcal{L}_F\,,\nonumber\\
T_V^{\mu\nu} &=&-e^{-\phi}\Big(z^4\eta^{\rho\sigma}\eta^{\mu\beta}\eta^{\nu\gamma}\,F^V_{\beta\rho}F^V_{\gamma\sigma}\nonumber\\
&&-z^4\,\eta^{\mu\beta}\eta^{\nu\gamma}\,F^V_{\beta z}F^V_{\gamma z}\Big)-\eta^{\mu\nu}\mathcal{L}_V\,.
  \label{EMT}
 \eea
Note that the UV-boundary term in the (\ref{Action}) vanishes for the normalizable modes of the fermion.
For the dilaton the couplings are

  \begin{widetext}
  \bea
  \label{vertices2}
 \varphi \bar\Psi\Psi:\quad&& \sqrt{2\kappa^2}\int d^5x\,\sqrt{g}\,\frac{e^{-\phi}}2 \,\left(\frac z2 \,\partial_z\varphi\right)\,
 \overline\Psi\gamma^5\Psi+\sqrt{2\kappa^2}\int d^5x\,\sqrt{g}\,\frac{e^{-\phi}}{2}\,\left(\frac {iz}2 \,\partial_\mu\varphi\right)\,
 \overline\Psi\gamma^\mu\Psi\nonumber\\
\varphi AA:\quad && \sqrt{2\kappa^2}\int d^5x\,\sqrt{g}\,e^{-\phi}\,(-\varphi)\,
 \left(-\frac 1{4}\,g^{\mu\alpha}g^{\beta\nu}\,F_{\mu\nu}^V\,F_{\alpha\beta}^V\right)
\eea
We have canonically normalized the bulk fields through the substitutions

\bea
\label{SUBX}
\Psi\rightarrow g_5\Psi\qquad V_{N}\rightarrow g_5V_{N}\qquad \varphi\rightarrow\sqrt{2\kappa^2}\,\varphi\qquad h_{\mu\nu}\rightarrow\sqrt{2\kappa^2}\,h_{\mu\nu}
\eea
\end{widetext}
which makes the couplings  and power counting manifest  in Witten diagrams. Note that after this rescaling, the meson decay constants
in (\ref{SOFTMF}) and the glueball decay constants in (\ref{SOFTMFG2}-\ref{SOFTMFG4}) redefine through  $g_5f_n\rightarrow f_n$. 
This will be understood in most of our analysis.


Evaluating the couplings or the vertices (\ref{vertices1})-(\ref{vertices2}) on the solutions, Fourier transforming the fields to momentum space, and integrating by part the trace-full part for the fermions, we find for the couplings to the fermions ($h\overline \Psi \Psi$) and gauge fields ($hAA$)

\bea
 h\overline\Psi\Psi:\quad &&\int \frac{d^4 p_2  d^4 p_1d^4k}{(2\pi)^{12}}(2\pi)^4 \delta^4(p_2-k-p_1)\nonumber\\
 &&\times\big(S^k_{h\bar\Psi\Psi}+S^k_{f\bar\Psi\Psi}\big)\nonumber\\
h AA:\quad && \int \frac{d^4q'  d^4qd^4k}{(2\pi)^{12}}(2\pi)^4 \delta^4(q'-k-q)\nonumber\\
&&\times\big(S^k_{hAA}+S^k_{fAA}\big)\nonumber\\
\label{vertices3}
\eea
The corresponding couplings to the dilatons are

\bea
 \varphi\overline\Psi\Psi:\quad &&\int \frac{d^4 p_2  d^4 p_1d^4k}{(2\pi)^{12}}(2\pi)^4 \delta^4(p_2-k-p_1)\,S^k_{\varphi\bar\Psi\Psi}\nonumber\\
\varphi AA:\quad && \int \frac{d^4q'  d^4qd^4k}{(2\pi)^{12}}(2\pi)^4 \delta^4(q'-k-q)\,S^k_{\varphi AA}\nonumber\\
\label{vertices4}
\eea
with

\begin{widetext}
\bea
S^{k}_{h\bar\Psi\Psi}&=&-\frac{\sqrt{2\kappa^2}}2\int dz\sqrt{g}\,e^{-\phi}z\,\epsilon^{TT}_{\mu\nu}h(k,z)\bar\Psi(p_2,z)\gamma^\mu p^\nu\Psi(p_1,z)\,,\nonumber\\
S^{k}_{f\bar\Psi\Psi}&=&- \frac{\sqrt{2\kappa^2}}2\int dz\sqrt{g}\,e^{-\phi}z\,\bar\Psi(p_2,z)\Big(\epsilon^{T}_{\mu\nu}f(k,z)\tilde{k}^2\gamma^\mu p^\nu+\partial_z\big(\epsilon^{T}_{\mu\nu}f(k,z)\big)k^2\eta^{\mu\nu}\gamma^5
+\epsilon^{T}_{\mu\nu}f(k,z)\tilde{k}^2\eta^{\mu\nu}k_{\alpha}\gamma^\alpha\Big)\Psi(p_1,z)\,,\nonumber\\
S^{k}_{hAA}&=&\sqrt{2\kappa^2}\int dz\sqrt{g}\,e^{-\phi}z^4\,\epsilon^{TT}_{\mu\nu}h(k,z)K^{\mu\nu}(q,q',n,n',z)\,,\nonumber\\
S^{k}_{fAA}&=&\frac{\sqrt{2\kappa^2}}{2}\int dz\sqrt{g}\,e^{-\phi}z^4\,\epsilon^{T}_{\mu\nu}f(k,z)\tilde{k}^2\Big(K^{\mu\nu}(q,q',n,n',z)-\frac{1}{4}\eta^{\mu\nu}K(q,q',n,n',z)\Big)\,,\nonumber\\
\label{vertices5}
\eea
and

\bea
S^{k}_{\varphi\bar\Psi\Psi}&=&\frac{\sqrt{2\kappa^2}}{2}\int dz\sqrt{g}\,e^{-\phi}z\,\bar\Psi(p_2,z)\Big(\partial_z\varphi(k,z)\gamma^5+\varphi(k,z)k_{\alpha}\gamma^\alpha\Big)\Psi(p_1,z)\,,\nonumber\\
S^{k}_{\varphi AA}&=&\frac{\sqrt{2\kappa^2}}{4}\int dz\sqrt{g}\,e^{-\phi}z^4\,\varphi(k,z)K(q,q',n,n',z)\,,\nonumber\\
\eea
We  have set  $q^2=-Q^2$, $q^{\prime 2}=-Q^{\prime 2}$  for space-like momenta, and defined

\bea
&&K^{\mu\nu}(q,q',n,n',z)\equiv  B_1^{\mu\nu}\mathcal{V}(Q,z)\mathcal{V}(Q',z)
-B_0^{\mu\nu}\partial_z\mathcal{V}(Q,z)\partial_z\mathcal{V}(Q',z)\,,\nonumber\\
&&B_{0}^{\mu\nu}(n,n')\equiv n^\mu n'^\nu\,,\nonumber\\
&&B_{1}^{\mu\nu}(q,q',n,n')\equiv
n\cdot n' \,q^\mu q'^\nu-q\cdot n'\, n^\mu q'^\nu
-q^\prime \cdot n\, q^\mu n^{\prime \nu}+q\cdot q^\prime\, n^\mu n^{\prime \nu}\,.
\label{BK}
\eea
\end{widetext}
with $B_{1,0}=\eta_{\mu\nu}B_{1,0}^{\mu\nu}$, and $K=\eta_{\mu\nu}K^{\mu\nu}$. The non-normalizable wave function for the virtual photon $\mathcal{V}(Q,z)$ is given in Appendix XII.

\begin{figure}[!htb]
\includegraphics[height=4cm]{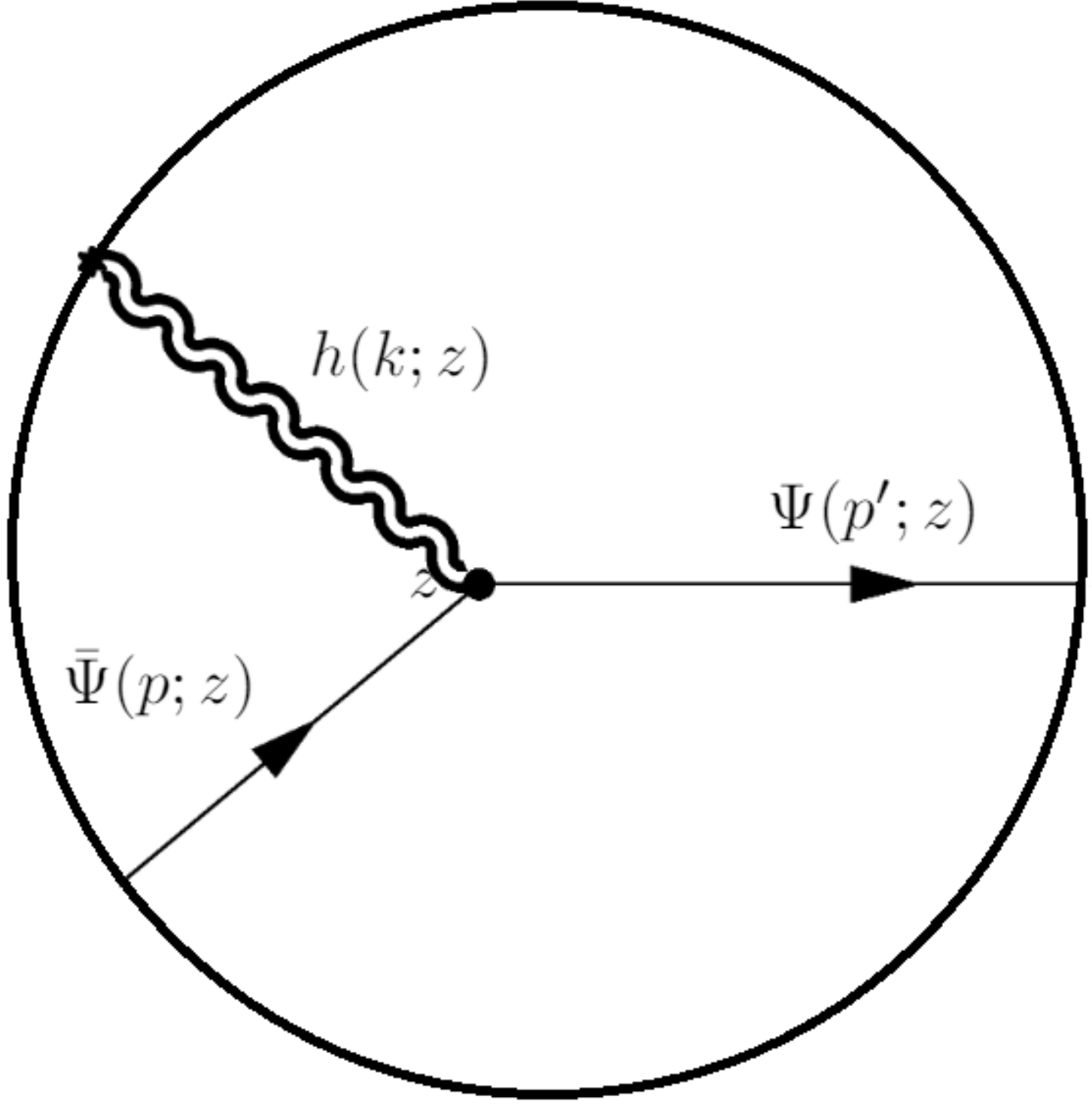}
  \caption{Witten diagram for the gravitational form factor $A(t)$ due to the exchange of spin-2 glueball resonances.}
  \label{wdiagram3}
\end{figure}

\begin{figure}[!htb]
\includegraphics[height=4cm]{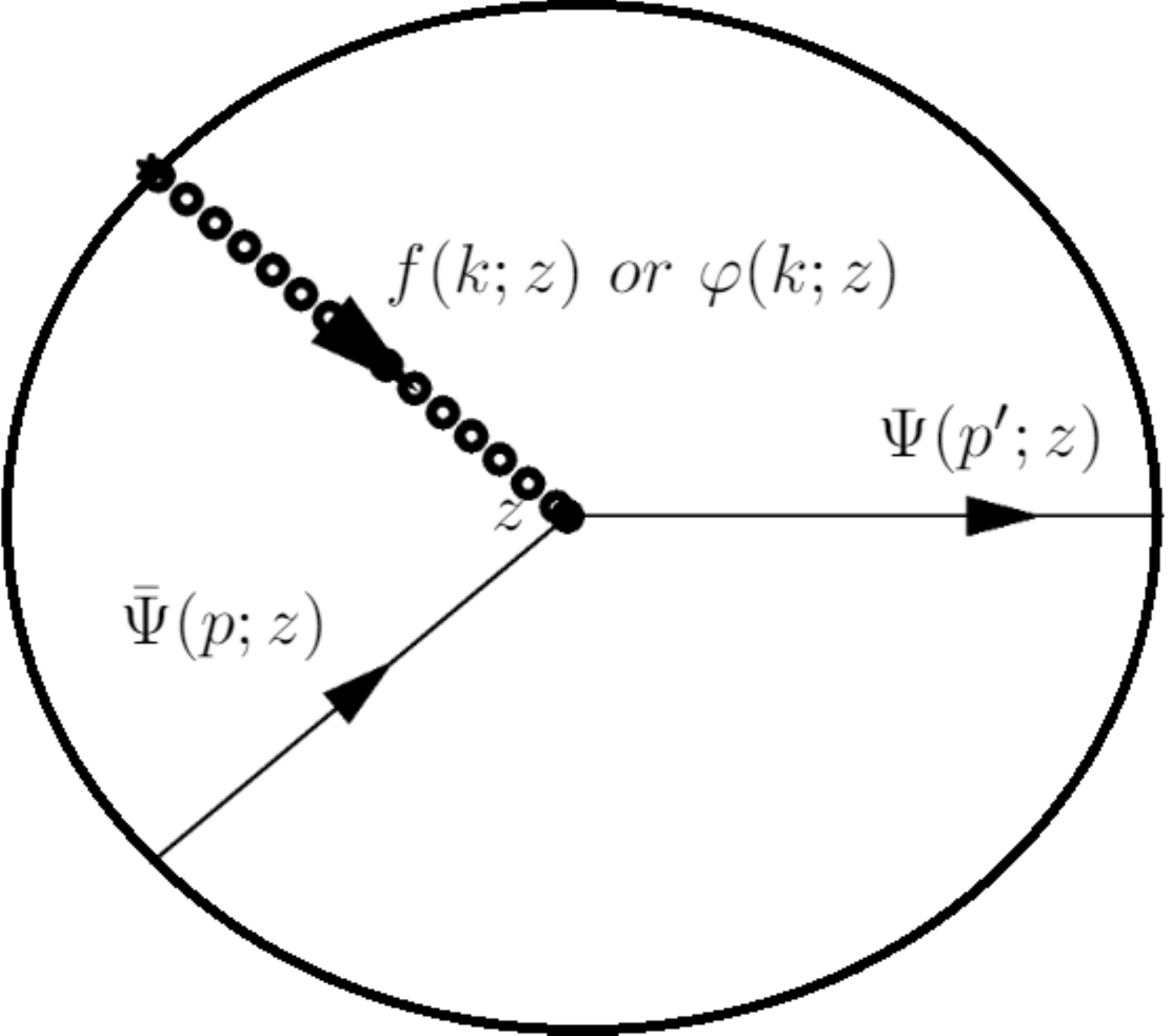}
  \caption{Witten diagram for the gravitational form factor $C(t)$ due to the exchange of scalar glueball resonances from the trace-full part of the metric fluctuation $f(k,z)$. Also shown is a form factor due to the exchange of the dilatonic scalar glueball resonances $\varphi(k,z)$.}
  \label{wdiagram4}
\end{figure}

\section{Gravitational form factors}

The graviton coupling to the Dirac fermion in bulk is through its energy momentum tensor. In the conformally broken geometry (hard or soft wall), the corresponding energy momentum tensor traces to the normalization of the bulk Dirac fermion as a nucleon state, modulo the source field normalization at the boundary (see below). More importantly, since the holographic
construction operates in the limit of a large number of colors, it follows that the energy momentum tensor of the bulk Dirac fermion is dual to the quenched energy momentum tensor of the nucleon. In other words, only the gluonic contribution to the energy momentum tensor is picked by the photoproduction amplitude close to threshold in the present holographic analysis.

More specifically, the energy momentum tensor to the bulk Dirac fermion involves both the $2^{++}$ tensor glueball field $h$ and the $0^{++}$ scalar glueball field $f$, see  Fig.~\ref{wdiagram3} and \ref{wdiagram4},

\begin{widetext}
\be
\label{EMT1}
i\left<p_2|T^{\mu\nu}(0)|p_1\right>=(-i)V_{h\bar\Psi\Psi}^{\mu\nu(TT)}(p_1,p_2,K)+(-i)V_{f\bar\Psi\Psi}^{\mu\nu(T)}(p_1,p_2,K)\,,\nonumber\\
\ee
with the explicit vertices

\be
\label{EMT2}
V_{h\bar\Psi\Psi}^{\mu\nu(TT)}(p_1,p_2,K)=&&-\frac{1}{2g_5^2}\int dz\sqrt{g}\,e^{-\phi}z\bar\Psi(p_2,z)\gamma^\mu p^\nu\Psi(p_1,z)\mathcal{H}(K,z)\nonumber\\
=&&-\frac{1}{2g_5^2}\int dz\sqrt{g}\,e^{-\phi}z\,\big(\psi_R^2(z)+\psi_L^2(z)\big)\mathcal{H}(K,z)\times\bar u(p_2)\gamma^\mu p^\nu u(p_1)\,,\nonumber\\
V_{f\bar\Psi\Psi}^{\mu\nu(T)}(p_1,p_2,K)=&&-\frac{1}{2g_5^2}\int dz\sqrt{g}\,e^{-\phi}z\,\big(\psi_L(z)\psi_R(z)-\psi_R(z)\psi_L(z)\big)\partial_z \mathcal{F}(K,z)\times \tilde{k}^2\eta^{\mu\nu}\times\bar u(p_2)u(p_1)\nonumber\\
&&-\frac{1}{16g_5^2}\int dz\sqrt{g}\,e^{-\phi}z\,\big(\psi_R^2(z)+\psi_L^2(z)\big)\mathcal{F}(K,z)\times \tilde{k}^2\eta^{\mu\nu}\times\bar u(p_2)\big(\gamma_\alpha p^\alpha +4k_{\alpha}\gamma^\alpha \big)u(p_1)\,.\nonumber\\
\ee
\end{widetext}
They follow
 by substituting the normalizable mode $J_h(m_n, z)$ and $J_f(m_n,z)$ by the non-normalizable mode $\mathcal{H}(K,z)$ (given in \ref{hps1}) and $\mathcal{F}(K,z)$ (given in \ref{fps1}) in the second vertices of (\ref{vh}) and (\ref{vf}) for space like momenta $k^2=-K^2$, with the boundary value for the source set generically to ${\cal H}(K,0)=1$.  Below, we show that this boundary condition is tied to the normalization of the (gluonic) trace of the energy momentum tensor in the bulk Dirac fermion state as a nucleon and will relax it, since it is arbitrary in  holography.

With this in mind, a comparison of  (\ref{EMT1}-\ref{EMT2}) to the standard decomposition of the energy-momentum form factor

\begin{widetext}
\be
\label{EMT2}
\left<p_2|T^{\mu\nu}(0)|p_1\right>=\overline{u}(p_2)\left(
A(k)\gamma^{(\mu}p^{\nu)}+B(k)\frac{ip^{(\mu}\sigma^{\nu)\alpha}k_\alpha}{2m_N}+C(k)\frac{k^\mu k^\nu-\eta^{\mu\nu}k^2}{m_N}\right)u(p_1)\,,
\ee
yields

\be\label{Aff}
A(K)=-\frac{C(K)}{(\alpha z_0m_N/2)^2}=\frac{1}{2g_{5}^2}\int dz\sqrt{g}\,e^{-\phi}z\,\big(\psi_R^2(z)+\psi_L^2(z)\big)\,\mathcal{H}(K,z)\,.
\ee
For  the soft wall model,

\be \label{FFj2}
A(K)=A(0)\,(a_K+1)\bigg(-\left(1+a_K+2a_K^2\right)+2\left({a_K}+2{a^3_K}\right)\Phi(-1,1,a_K)\bigg) \,,
\ee
or equivalently

\be
A(K)=A(0) \bigg((1-2a_K)(1+a_K^2)+a_K(1+a_K)(1+2a_K^2)\bigg(\psi\bigg(\frac{1+a_K}{2} \bigg)-\psi\bigg(\frac{a_K}{2}\bigg)\bigg)\bigg)
\ee
with $a_K={K^2}/{8\tilde\kappa_N^2}$. Here $\Phi(-1,1,a')$ refers to the LerchPhi function, and $\psi(x)$ refers to the digamma function 
or harmonic number $H_x=\psi(x)+\gamma$.  Modulo $A(0)$,
(\ref{FFj2}) is in agreement with the result in~\cite{CARLSON}.
The gravitational form factor $C(K)$ is proportional to $A(K)$ modulo a negative  overall constant
$-(\alpha z_0 m_N/2)^2<0$ which is left undetermined since $\alpha$ is arbitrary in the tensor
decomposition  (\ref{exp}). We note that   (\ref{EMT1}) gives $\left<p|T^{\mu}_\mu|p\right>=2A(0)m_N^2$.
Since the boundary value ${\cal H}(K,0)={\cal H}(0,z)$ is arbitrary as we just noted above, it follows that
$A(0)$ is not fixed in holography. This will be understood from here on.


\begin{figure}[!htb]
\includegraphics[height=5.5cm]{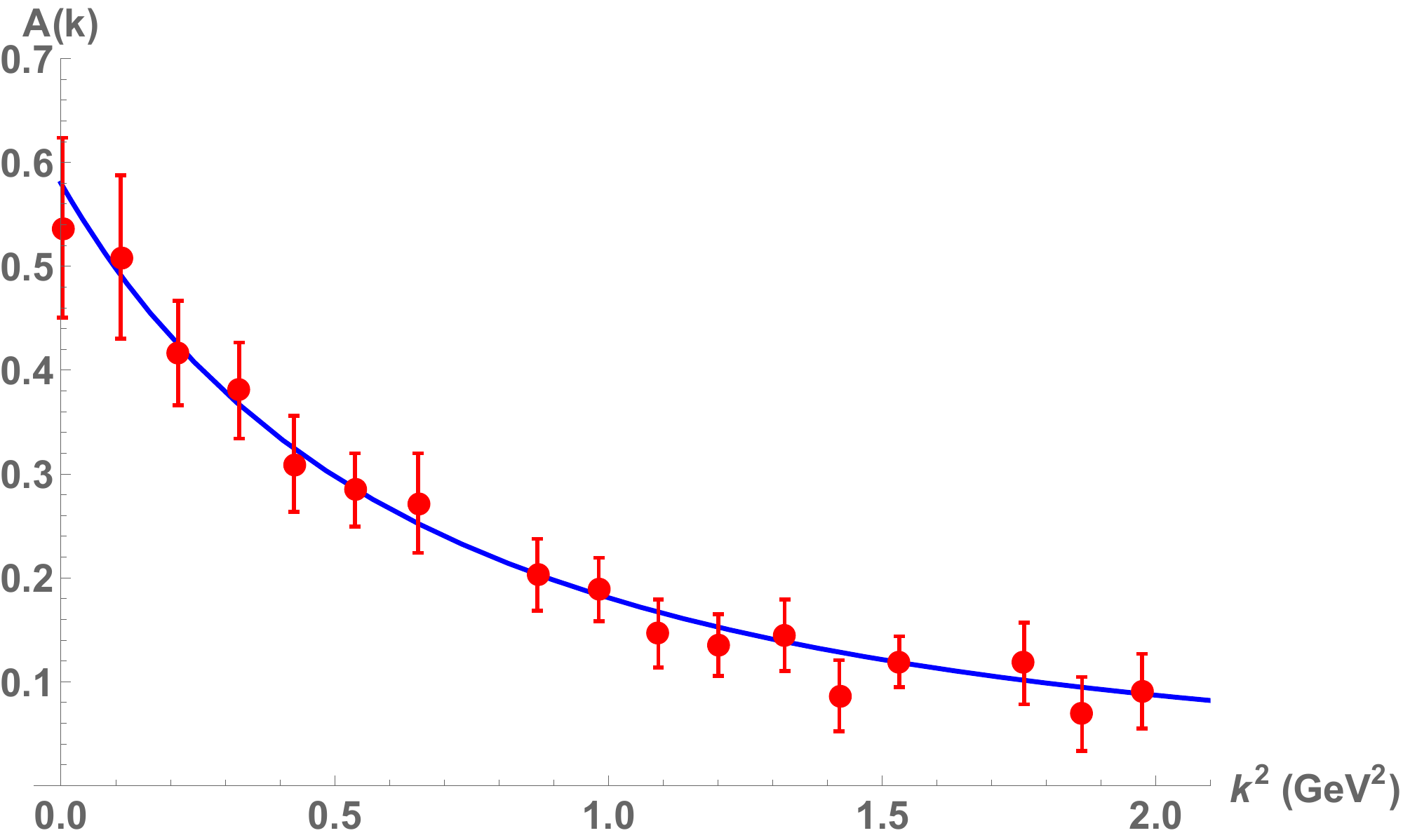}
  \caption{Holographic gravitational form factor $A(k)$ (for $k^2\geq 0$) shown in solid-blue curve versus the lattice data in red-squares~\cite{MIT}.}
  \label{fig_AK}
\end{figure}

\begin{figure}[!htb]
\includegraphics[height=5.5cm]{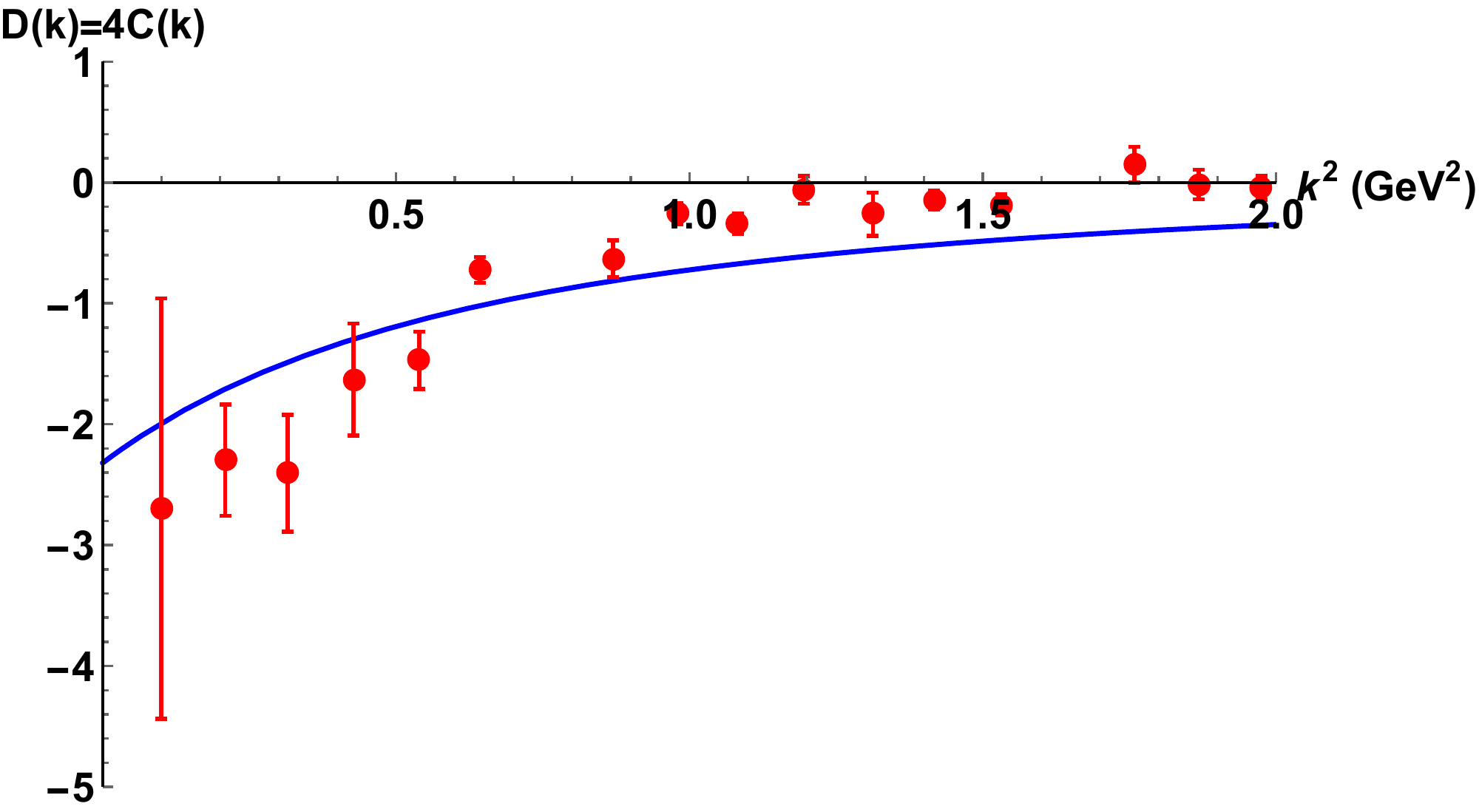}
  \caption{Holographic gravitational form factor $D(k)=4C(k)$ (for $k^2\geq 0$) shown in solid-blue curve versus the lattice data in red-squares~\cite{MIT}.}
  \label{fig_DK}
\end{figure}

\begin{figure}[!htb]
\includegraphics[height=5.5cm]{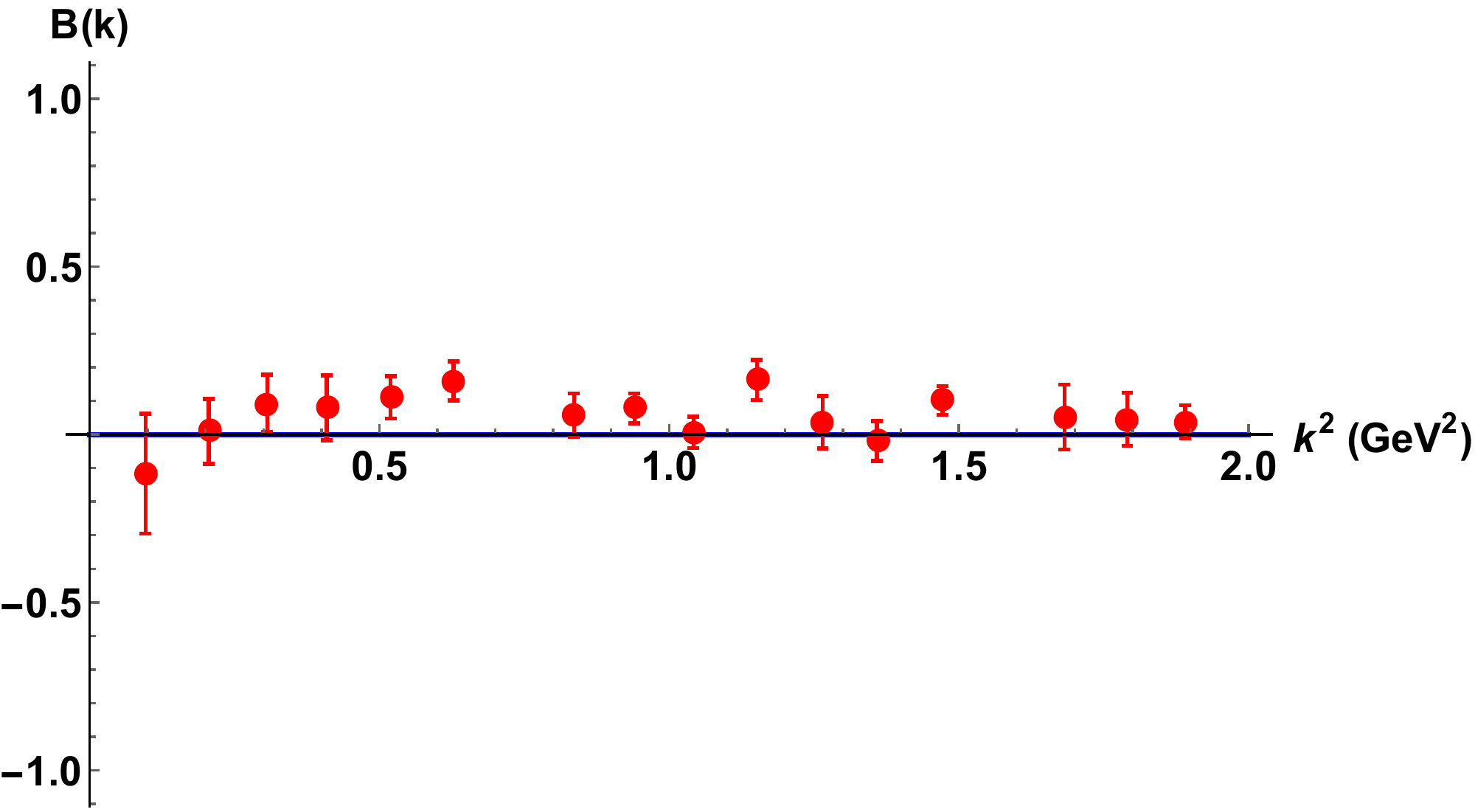}
  \caption{Holographic gravitational form factor $B(k)=0$ (for $k^2\geq 0$) shown in solid-blue curve versus the lattice data in red-squares~\cite{MIT}.}
  \label{fig_BK}
\end{figure}

\end{widetext}



The invariant form factors  $A(k), B(k), C (k)$ (for $k^2\geq 0$) measure the gluonic content of the energy momentum tensor in the
nucleon state, as the holographic dual of the energy momentum tensor of the dilation in bulk in the double  limit of large $N_c, \lambda$. This limit maps the bulk fields in a soft or hard wall metric, to a pure Yang-Mills theory at the boundary
in the confining regime. More specifically,  the form factor
$A(k)$ through ${\cal H}(K,z)$ in bulk resums the $2^{++}$ or tensor glueball Regge trajectory as given  in (\ref{hps1}).
For the soft wall model, the result is in agreement with the one reported in~\cite{CARLSON}. The form factor
$C(k)$ through ${\cal F}(K,z)$ in bulk resums the $0^{++}$ or scalar glueball Regge trajectory as shown in  (\ref {fps1}). In holography,
the scalar and tensor glueball spectra  are degenerate as we noted earlier (same bulk equations for $h,f$), so ${\cal H}(K,z)$ and ${\cal F}(K,z)$ are
tied, i.e ${\cal F}=-2{\cal H}$. The factor of 2 reflects on the $\frac 12$ difference in the normalization of the kinetic energies in (\ref{kinetic}).
Finally, the Pauli-like  form factor $B(k)=0$ as  the coupling of the graviton to the bulk Dirac fermion through the spin-connection in (\ref{fermionAction}) vanishes,

\be
\frac 18 \omega_{NAB}\,\overline{\Psi}e^N_C\Gamma^C\frac i2[\Gamma^A, \Gamma^B]\Psi
\rightarrow \frac i8 h^\mu_\alpha\,\overline{\Psi}\Gamma^\alpha [\Gamma_\mu, \Gamma^z]\Psi =0\nonumber\\
\ee

The soft walll results for the gravitational form factor $A(k)$ compares well with the recently reported lattice results,
as shown in Fig.~\ref{fig_AK}. The solid-blue curve is our result for the soft wall model,
and the red-squares are the recent lattice data~\cite{MIT}.
The re-summed $A(k)$ (for $k^2\geq 0$) in the soft wall model is well reproduced by the dipole form factor

\be
\label{AKKX}
A(k)=\frac {A(0)}{\bigg(1+\frac {k^2}{m_A^2}\bigg)^2}
\ee
with  $m_{A}=1.124$ GeV in comparison to the reported lattice value $m_{A,\,\rm lattice}=1.13$ GeV. The arbitrary normalization
$A(0)=0.58$ was adjusted to the lattice data~\cite{MIT}. Recall that the gravitational form factor $A(k)$ is saturated by the $2^{++}$
 glueball trajectory without any quark mixing, essentially a quenched result.  In Fig.~\ref{fig_DK}  we show in the solid-blue curve the holographic
 gravitational form factor $D(k)\equiv 4C(k)=-4A(k)$ with $\alpha =2/(z_0 m_N)$ in the soft wall model, versus the reported lattice results
 in red-squares~\cite{MIT}. In holography $C(k)$ is saturated by the $0^{++}$ massive glueballs which are degenerate
 with the $2^{++}$ ones, hence $m_A=1.124$ GeV in comparison to $m_A=0.48$ GeV from  the lattice.The difference is likely due to the strong scalar-isoscalar quark mixing to the $0^{++}$ gueball channell in the unquenched lattice simulations,  in particular to the light sigma meson with a  mass of about $0.5$ GeV. In Fig.~\ref{fig_BK} we show the lattice results in red-squares for $B(k)$ which are consistent  with  $B(k)=0$ in  holography shown   as a solid-blue curve.


\begin{figure*}
\subfloat[The pressure distribution inside the proton (\ref{sp}) for soft-wall holographic QCD with $m_A=1.124$\,GeV.\label{figxgp5}]{%
  \includegraphics[height=6cm,width=.49\linewidth]{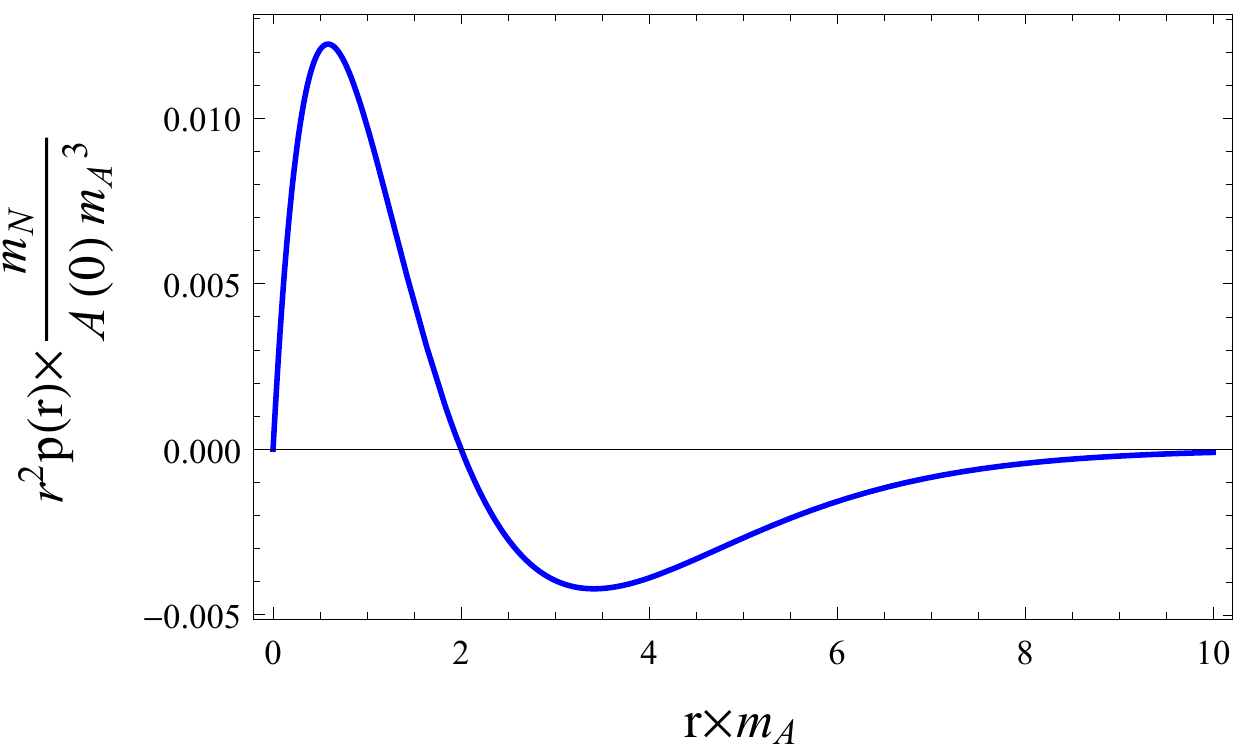}%
}\hfill
\subfloat[The shear force inside the proton (\ref{sp}) for soft-wall holographic QCD with $m_A=1.124$\,GeV.\label{figxgp6}]{%
  \includegraphics[height=6cm,width=.49\linewidth]{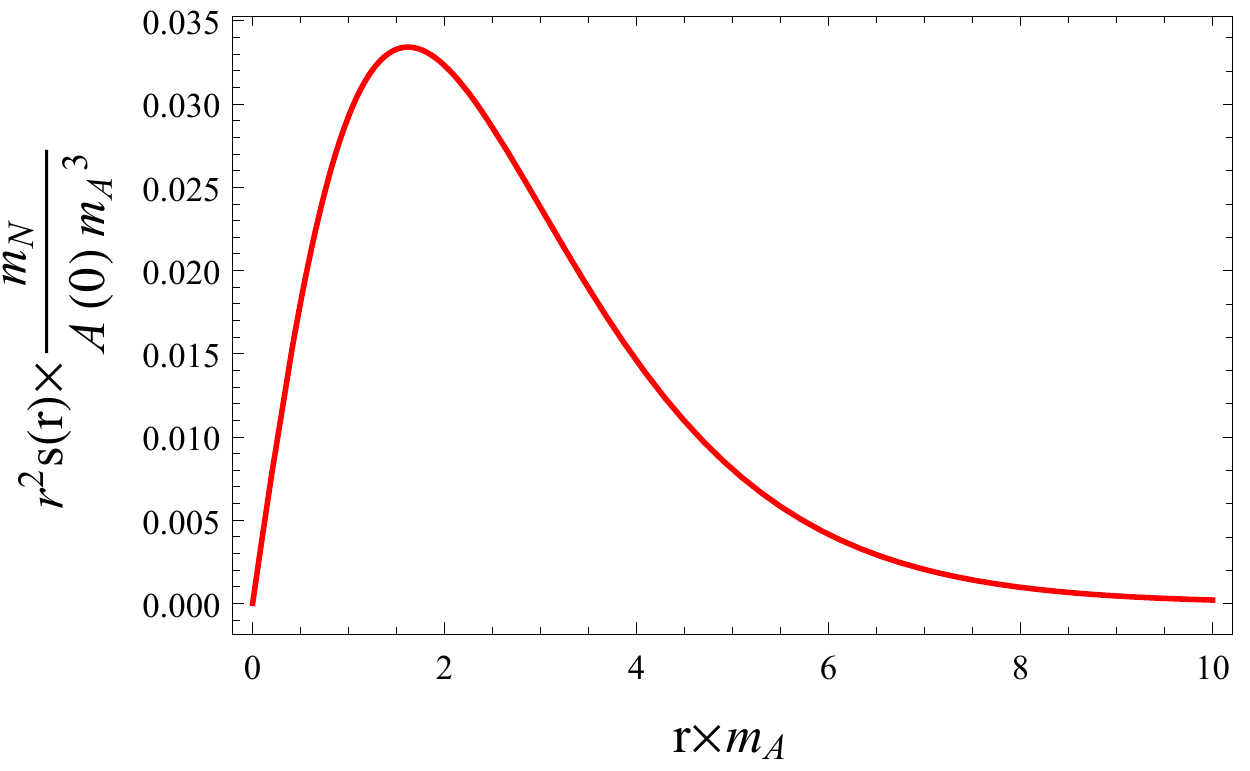}%
}
\caption{Holographic pressure and shear inside the proton.}
\label{fig_pxgp}
\end{figure*}

\section{Holographic pressure and shear  inside the proton}

Using the dipole representation for $A(K)$ (\ref{AKKX}) which is a good parametrization
of our holographic results, the D-term as $D(K)=-4A(K)$ can be written as

\be
\label{DKK}
D(K)=\frac {-4A(0)}{\bigg(1+\frac {K^2}{m_A^2}\bigg)^2}
\ee
with $m_A=1.124$\,GeV.  The Fourier transform (\ref{DKK}) of the three-dimensional coordinate space gives
($E=m_N$)

\begin{widetext}
\be
\tilde{D}(r)=-4A(0)\int\frac{d^3{K}}{2E(2\pi)^3}\frac{e^{-i{K}\cdot{r}}}{\bigg(1+\frac {{K}^2}{m_A^2}\bigg)^2}
=-A(0)\frac{m_A^3}{4\pi m_N}\,e^{-m_Ar}\,,
\ee
\end{widetext}

The holographic shear $s(r)$ and pressure $p(r)$ distributions in the proton can be expressed in terms of $\tilde{D}(r)$ as~\cite{POLYAKOV}
\bea
\label{sp}
s(r)= &&-\frac{r}{2} \frac{d}{dr} \frac{1}{r} \frac{d}{dr}
{\tilde{D}(r)} \nonumber\\
p(r)=&&\frac{1}{3} \frac{1}{r^2}\frac{d}{dr} r^2\frac{d}{dr}
{\tilde{D}(r)}
\eea
In Fig.~\ref{fig_pxgp}a,b  we show the holographic gluon contribution to the  pressure $p(r)$ distribution and shear force $s(r)$, respectively.
The  results  are in agreement with the lattice QCD result~\cite{Shanahan:2018nnv} for the gluon contribution.
They are also comparable to the experimentally extracted quark contributions in~\cite{Burkert:2018bqq}. Below, we will argue that the holographic
relationship $D(K)=-4A(K)$ will allow the extraction of the pressure and shear of the proton from the threshold photoproduction data of heavy
vector mesons $V=J/\Psi, \Upsilon$.

\section{Holographic Vector Meson Dominance}

The diffractive scattering amplitude with a single graviton and dilaton exchanges is detailed in Appendix XIII.
For photoproduction or electroproduction close to the photon point $Q^2= 0$,  and we may set $\mathcal{V}(Q= 0,z)=1$
in ${\cal V}_{hAA}$ in  (\ref{nAmph}). This will be indicated by the re-labeling of the entry photon $A\rightarrow \gamma$.
This will be understood in the remainder of our analysis. With this in mind,
The combined amplitudes (\ref{nAmpD})  read

\begin{widetext}
\be
\label{vvx}
-{\cal A}_{\gamma p\rightarrow A p} (s,t)=&&{\cal A}^{\varphi}_{\gamma p \rightarrow A p} (s,t)+{\cal A}^{h}_{ \gamma p\rightarrow A  p} (s,t)+{\cal A}^{f}_{\gamma p\rightarrow A p} (s,t)\nonumber\\
=&&\frac{1}{2g_{5}^4}\mathcal{V}_{hAA}B^{1}_{\alpha\beta}\mathcal{V}^{\alpha\beta(TT)}_{h\bar\Psi\Psi}+\frac{1}{g_{5}^4}\mathcal{V}_{fAA}B^{1}\mathcal{V}^{(T)}_{f\bar\Psi\Psi}+\frac{1}{g_{5}^4}\tilde{\mathcal{V}}_{\varphi AA}B^{1}\mathcal{V}_{\varphi\bar\Psi\Psi}\,,\nonumber\\
\ee
\end{widetext}

\newpage
The effective vertices for the hard-wall model are
\be
\label{vv}
&&\mathcal{V}_{hAA}=\frac{\sqrt{2\kappa^2}}{2}\int_{0}^{z_0} dz\sqrt{g}\,z^4\mathcal{V}(Q',z)\frac{z^4}{4}\,,\nonumber\\
&&\mathcal{V}_{fAA}=\eta_{\mu\nu}\mathcal{V}^{\mu\nu(T)}_{fAA}=0\,,
\ee
\be\label{vv2}
&&\tilde{\mathcal{V}}_{\varphi AA}=\frac{\sqrt{2\kappa^2}}{4}\int_{0}^{z_0} dz\sqrt{g}\,z^4\mathcal{V}(Q',z)\frac{z^4}{4}\,,\nonumber\\
&&\mathcal{V}_{\varphi\bar\Psi\Psi}=0\,.
\ee
The corresponding vertices for the soft wall model follows through the substitution
$\sqrt{g}\rightarrow \sqrt{g}e^{-{\tilde\kappa_V}^2z^2}$ with $\tilde\kappa_V$ the soft wall scale.
$B^1_{\alpha\beta}$ and $B^1=\eta^{\mu\nu}B^1_{\mu\nu}$  are defined in (\ref{BK}).

The TT-part of the transverse and traceles $2^{++}$ glueball  contribution  corresponds to  $\alpha, \beta=x,y$.
The T-coupling of the transverse and traceful $0^{++}$ glueball  to the virtual photons involves the full photon energy
momentum tensor and vanishes after contraction with $\epsilon_{\mu\nu}^T$. The TT-coupling
involves only the non-trace part of  the photon energy momentum tensor and does not vanish after contraction with $\epsilon_{\mu\nu}^{TT}$.
The Yukawa coupling of the dilaton to the bulk Dirac fermion is null as we noted earlier.
As a result, the scattering amplitude (\ref{vvx}) is solely due to the exchange of the  $2^{++}$ glueball.

The result (\ref{vvx}-\ref{vv2}) is for a general bulk-to-boundary current ${\cal V}(Q^\prime, z)$ which sums over a tower of vector meson resonances. The production of a specific meson at the boundary, say charmonium or upsilonium, amounts to the substitution 
\begin{widetext}
\bea
\mathcal{V}(Q',z)\rightarrow \phi_{n}(z)=c_{n}zJ_{1}(m_{n}z)
=\frac{f_{n}}{m_{n}}(m_{n}z)J_1(m_{n}z)
\eea
\end{widetext}
 in (\ref{vv}-\ref{vv}) with $c_{n}=\frac{\sqrt{2}}{z_{0}J_1(m_{n}z_0)}$ and $f_{n}$ the decay constant of a heavy meson of mass $m_{n}$, with the identification $n=0$ for
 $J/\Psi$.
 As a result, the total amplitude for the photoproduction of $J/\Psi$ can be written in the block form
\bea
&&{\cal A}_{\gamma p\rightarrow J/\Psi p} (s,t)=-\frac{1}{2g_{5}^4}\mathcal{V}_{hAA}B^{1}_{\alpha\beta}\mathcal{V}^{\alpha\beta(TT)}_{h\bar\Psi\Psi}\,,\nonumber\\
\eea
with the vertices for a hard wall
\begin{widetext}
\be
\label{vvJPSI1}
V_{h\bar\Psi\Psi}^{\mu\nu(TT)}(p_1,p_2,K)=&&\ -\frac{\sqrt{2\kappa^2}}{2}
\int dz\sqrt{g}\,e^{-\phi}z\,\big(\psi_R^2(z)+\psi_L^2(z)\big)\mathcal{H}(K,z)\times\bar u(p_2)\gamma^\mu p^\nu u(p_1)\,,\nonumber\\
\mathcal{V}_{hAA}=&&\Big(\frac{f_{n}}{m_{n}}\Big)\times\frac{\sqrt{2\kappa^2}}{2}\int_{0}^{z_0} dz\sqrt{g}\,z^4\times(m_{n}z)J_1(m_{n}z)\times\frac{z^4}{4}\nonumber\\
\approx&&\Big(\frac{f_{n}}{m_{n}}\Big)\times\bigg(\frac{\sqrt{2\kappa^2}}{16m_{n}^4}\int_{0}^{w_0} dw\,w^5\bigg)
\equiv \bigg(\frac {f_V}{M_V}\bigg)\,{\mathbb V}_{hAA}\,,
\ee
\end{widetext}
with $w=m_{n}z$, $w_0=m_{n}z_0$.
The  wave function for the emitted meson near the boundary is
$J_{1}(w)\approx{w}/{2}$. In comparison, the same arguments for the soft wall model give

\begin{widetext}
\be \label{vvJPSI2soft}
\mathcal{V}_{hAA}=&&\Big(\frac{f_{n}}{m_{n}}\Big)\times\frac{\sqrt{2\kappa^2}}{2}\int_{0}^{\infty} dz\sqrt{g}e^{-z^2\tilde{\kappa}_{V}^2}\,z^4\times (2\tilde{\kappa}_{V}^2z^2)L_{n}^1(z^2\tilde{\kappa}_{V}^2)\times\frac{z^4}{4}\nonumber\\
\approx&&\Big(\frac{f_{n}}{m_{n}}\Big)\times\bigg(\frac{\sqrt{2\kappa^2}}{2}\frac{L_{n}^1(0)}{4\tilde{\kappa}_{V}^4}\int_{0}^{\infty} d\xi \,e^{-\xi^2}\,\xi^2\bigg)
\equiv \bigg(\frac {f_V}{M_V}\bigg)\,{\mathbb V}_{hAA}\,,
\ee
with $\xi=\tilde{\kappa}_{V}^2z^2$ and $\tilde{\kappa}_{V}$ the soft wall parameter, and $n=0$.

(\ref{vvJPSI1}) and (\ref{vvJPSI2soft}) embody the general strictures of
VMD with the emergence of $f_{n}/m_{n}\equiv f_V/M_V$, the ratio of the leptonic decay constant to the mass of the heavy meson emitted,
as  illustrated in Fig.~\ref{wdiagram2}.
This result shows that in holographic QCD, the  photoproduction amplitude  $\gamma p\rightarrow Vp$
follows from the {\bf inverse of the diffractive part} of the deeply virtual Compton scattering amplitude
 $\gamma^* p\rightarrow \gamma p$ through VMD with $\gamma^*\approx  (ef_V/M_V)V$.

 The triple coupling ${\mathbb V}_{hAA}$  is the coupling of the bulk graviton with wavefunction
near the boundary $h\approx z^2 J_2(m_n z)\approx {z^4}$ (heavy $2^{++}$ glueball),  to a virtual photon near mass shell
with  ${\cal V}(Q\approx 0, z)\rightarrow 1$,  and a virtual photon off mass shell with
${\cal V}(Q^\prime , z)\rightarrow ({f_{n}}/{m_{n}})\times(m_{n}z)J_1(m_nz)$ (hard wall) or
${\cal V}(Q^\prime , z)\rightarrow  ({f_{n}}/{m_{n}})\times(2\tilde{\kappa}_{V}^2z^2)L_{n}^1(z^2\tilde{\kappa}^2)$ (soft wall).
The masses and decay constant for the soft wall are given in~(\ref{SOFTMF}) with the proviso that $g_5 f_n\rightarrow f_n$
following the canonical rescaling (\ref{SUBX}).


\section{Differential cross section for photoproduction}

Although our analysis for vector meson production applies equally well to both photoproduction and electroproduction, we now
specialize to the photoproduction of heavy mesons  given the recent experimental
interest in extracting the gluon contribution to the proton state from threshold data
at current electron machine facilities. With this in mind, the differential cross section for photoproduction of $V=J/\Psi$ can now be constructed from leading spin $j=0,2$ glueball exchanges near threshold.  The contribution of higher spin-j exchanges and their reggeization
will follow.  The pertinent differential cross section is of the form

\be\label{DCC}
\left(\frac{d\sigma}{dt}\right)
=\frac{e^2}{16\pi(s-m_N^2)^2}\,\frac 12\sum_{{\rm pol}}
\frac 12\sum_{{\rm spin}}
\Bigg|{\cal A}^{h}_{\gamma p\rightarrow  J/\Psi p} (s,t)\Bigg|^2\,,\nonumber\\
\ee
\end{widetext}
which is dominated by the TT-part of the graviton or $2^{++}$ glueball exchange as we noted earlier.
The first sum over the photon and $J/\Psi$ polarizations  is carried out using the identities

\be
&&\sum_{{\rm s=1,2}}n_s^{\mu}n_s^{*\nu}=-\eta^{\mu\nu}\,,\nonumber\\
&&\sum_{{\rm s'=1,2,3}}n_{s'}'^{\mu}n_{s'}'^{*\nu}=-\eta^{\mu\nu}+\frac{q'^{\mu}q'^\nu}{M_V^2}\,,
\ee
The second sum is over the initial and final bulk Dirac fermion as a proton spin

\be
\frac 14 \Tr\Big(\big(\gamma_\mu p_2^\mu+m_N\big)\big(\gamma_\mu p_1^\mu+m_N\big)\Big)=2K^2+8m_N^2\nonumber\\
\ee
Carrying explicitly these summations yield the differential cross section for photoproduction of heavy meson in the spin $j=2$
exchange approximation as

\begin{widetext}
\be
\label{diff1}
\left(\frac{d\sigma}{dt}\right)
&&=\frac{e^2}{64\pi (s-m_N^2)^2}\times\bigg(\frac {f_V}{M_V}\bigg)^2\mathbb V_{hAA}^2\times \frac{\kappa^2}{2g_5^8}\times \frac{g_5^4A^2(K)}{m_N^2}\times F(s,t=-K^2,M_V,m_N)\times(2K^2+8m_N^2)\,,\nonumber\\
&&=\mathcal{N}^2\times\frac{e^2}{64\pi (s-m_N^2)^2}\times\frac{A^2(K)}{4m_N^2A^2(0)}\times F(s,t=-K^2,M_V,m_N)\times(2K^2+8m_N^2)
\ee
\end{widetext}
with all vertex insertions following the rescaling (\ref{SUBX}) are shown explicitly and, in the last line, we have defined the normalization factor $\mathcal{N}$ as
\be
\label{diff1N}
\mathcal{N}^2=\bigg(\frac {f_V}{M_V}\bigg)^2\mathbb V_{hAA}^2\times \frac{2\kappa^2}{g_5^8}\times g_5^4A^2(0)\,,\nonumber\\
\ee
where $A(K)$ is the gravitational form factor (\ref{Aff}), which reduces to (\ref{FFj2}) for the soft wall model.
The kinematical function $F(s,t,M_V,m_N)$ follows from the contractions of the
various spins emanating from the photon and graviton vertices, and reads

\begin{widetext}
\bea
&&F(s,t,M,m)=\nonumber\\
&&\frac{1}{4096 M^2}\Big[-9 M^{10}+M^8\Big(-32+68m^2+28s+37t\Big)+2M^6\Big(256m^4+8m^2(32s-3t)+t(56-40s-29t)\Big)+\nonumber\\
&&2M^4\Big(-136m^6+64s^2-56s^3+8m^4(8+27s-64t)+3t^2(-24+7t)+4st(-4+9t)-4m^2(6s^2+32s(1+4t)+\nonumber\\
&&t(-4+25t))\Big)+M^2\Big(144m^8+144s^4-192s^2t+96s^3t -16s(-4+t)t^2+(80-13t)t^3+96 m^6(-6s+7t)+\nonumber\\
&&32m^4(27s^2-6t-39st+8t^2)+16m^2(-36s^3+30s^2t+24st(1+2t)+t^2(-4+17t))\Big)-\nonumber\\
&&t(2m^2-2s-t)\Big(64m^4+8m^6-8s^3+76m^4t-16t^2-90m^2t^2+t^3
+4s^2(16+6m^2+3t)-\nonumber\\
&&2s(12m^4+3t^2+m^2(64+44t))\Big)\Big]\,,\nonumber\\
\eea
\end{widetext}
with  $M_V=M$, $m\equiv m_{N}$, and $V=J/\Psi, \Upsilon$.
In the double limit of large $N_c,\lambda$,  the differential cross section (\ref{diff1}) scales as

\begin{widetext}
\be
\frac{d\sigma}{dt}\sim f_V^2\left(\frac{\kappa^4}{g_5^4}\right) \sim \frac 1{N_c^0}
\left(\lambda^0\,\,{\rm :\, soft\,\, wall}; \,\,\lambda^0\,\,{\rm :\, D7\,\, brane}; \,\,\lambda^2\,\,{\rm :\, D9\,\, brane}; \right)
\ee
since $f_V\sim N_c^0$ after the rescaling (\ref{SUBX}).
It  differs from the scaling of the surface exchange in~\cite{LEE},
 where their bulk Dirac fermion action is not normalized with $1/g_5^2$.
For large $s$,  we note that  $F(s,t)\sim s^4$ and
the differential cross section is seen to grow like $s^2$ as expected from a $2^{++}$ glueball exchange as a graviton. The corresponding  amplitude is purely real in this limit. These features reflect on the shortcomings of the $j=2$ exchange and its lack of reggeization at large $\sqrt{s}$.  They will be addressed below.

This notwithstanding, the differential cross section for photoproduction of a heavy meson is proportional to the gravitational form factor
$A(K)$ with $A(0)$ the sought after gluonic contribution to the trace of the energy momentum tensor. However, it
is folded with various couplings and kinematical factors that makes its extraction at threshold challenging.  For  the numerical analysis to follow, we will use the soft wall model with a fixed scale $\tilde\kappa_N=0.350$ GeV,  $\kappa^2=4\pi^2/N_c^2$ as fixed by the normalization of the kinetic part of the gravitational action in (\ref{kinetic}),  and  set $1/g_5^2$ through the D7 or D9 brane embeddings.
The coupling $\mathbb V_{hAA}$ is fixed by setting $V=J/\Psi$ in bulk.


\begin{figure}[!htb]
\includegraphics[height=5.5cm]{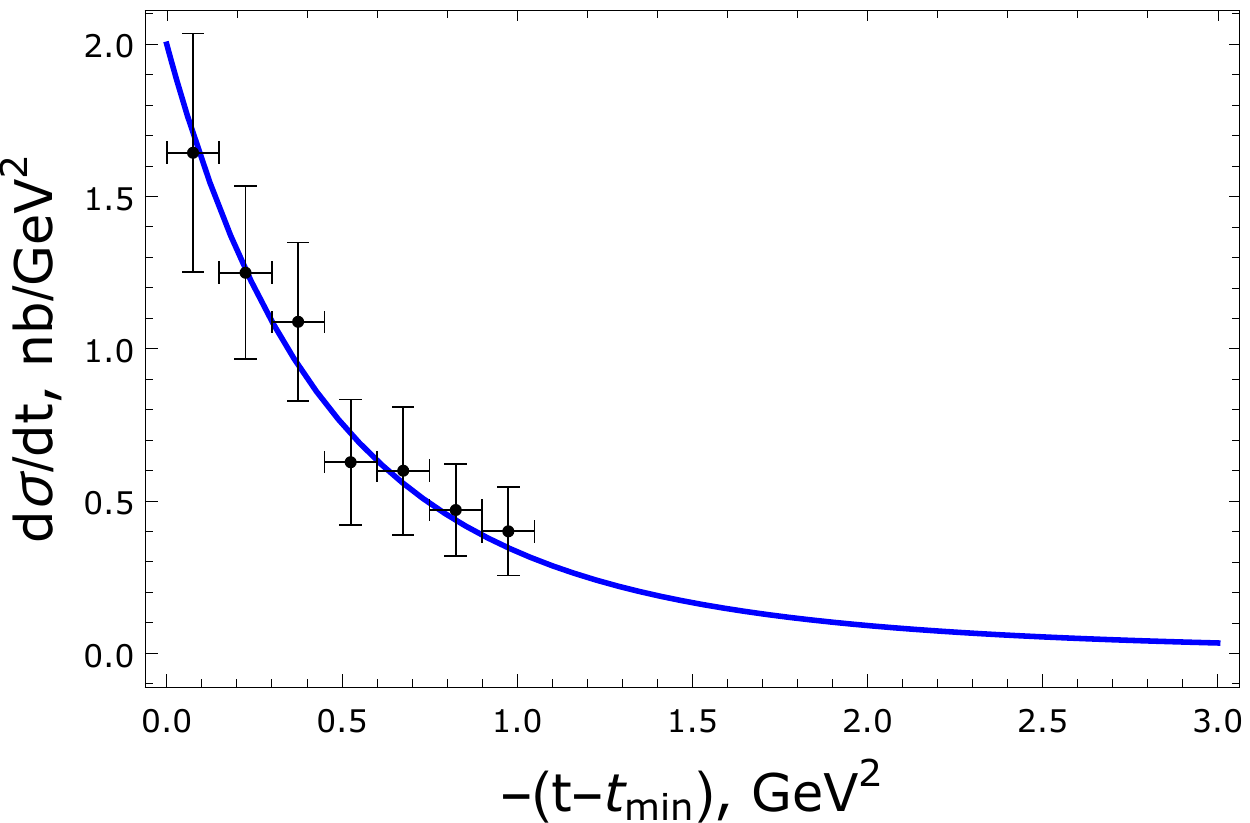}
  \caption{Differential cross section  for $V=J/\Psi$ photoproduction for $E_\gamma=10.72$ GeV}. The solid-blue  curve is our result for the soft-wall model. The data near threshold are from are from   GlueX~\cite{GLUEX}.
  \label{fig_diff}
\end{figure}

\begin{figure}[!htb]
\includegraphics[height=5.5cm]{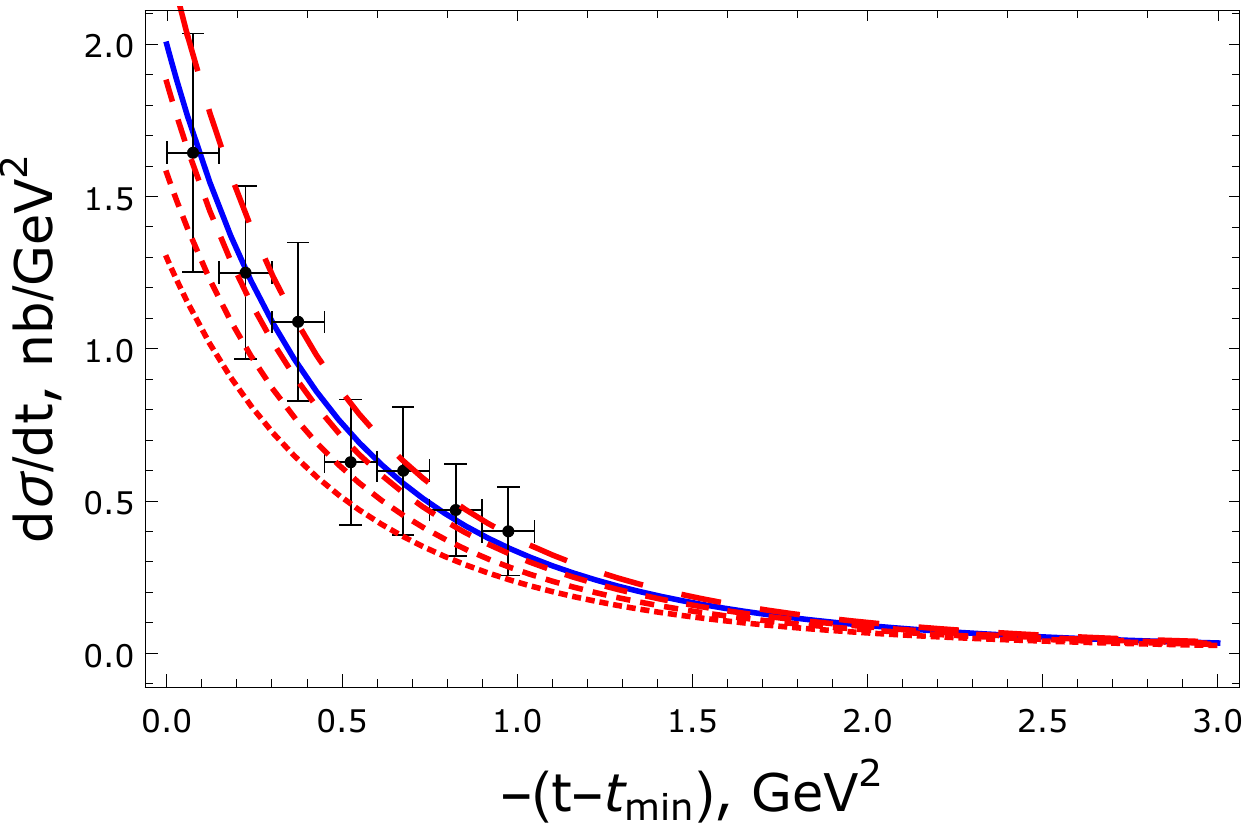}
  \caption{Same differential cross section for $V=J/\Psi$ photoproduction but different photon energies: $E_\gamma=11$ GeV large-red-dashing,
   $E_\gamma=10.72$ GeV solid-blue-curve, $E_\gamma=10.6$ GeV medium-red-dashing, $E_\gamma=10.3$ GeV small-red-dashing,
   and $E_\gamma=10$ GeV dotted-red curve. The data are from GlueX~\cite{GLUEX}.}
  \label{fig_diffE}
\end{figure}

\begin{figure}[!htb]
\includegraphics[height=5.5cm]{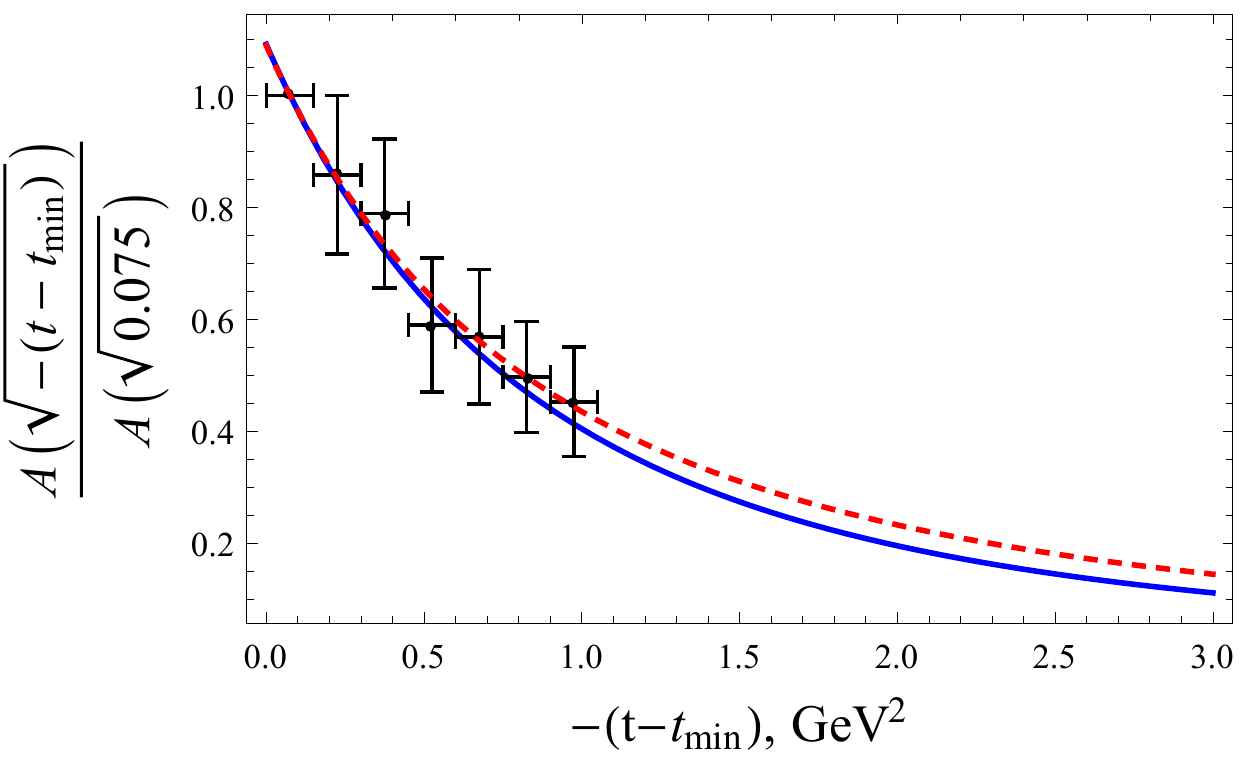}
  \caption{The gravitational form factor $A(\sqrt{-(t-t_{min})})$ (normalized by $A(\sqrt{0.075})$) for $\tilde{\kappa}_{\rho}=0.350~GeV$, $m_N=0.94~GeV$, and $m_{J/\psi}=3.10~GeV$. Blue solid line is our result, and dashed red line is from lattice QCD. We used the data from GlueX~\cite{GLUEX} with the errors added in quadrature.}
  \label{A_ratio}
\end{figure}

\begin{figure}[!htb]
\includegraphics[height=5.5cm]{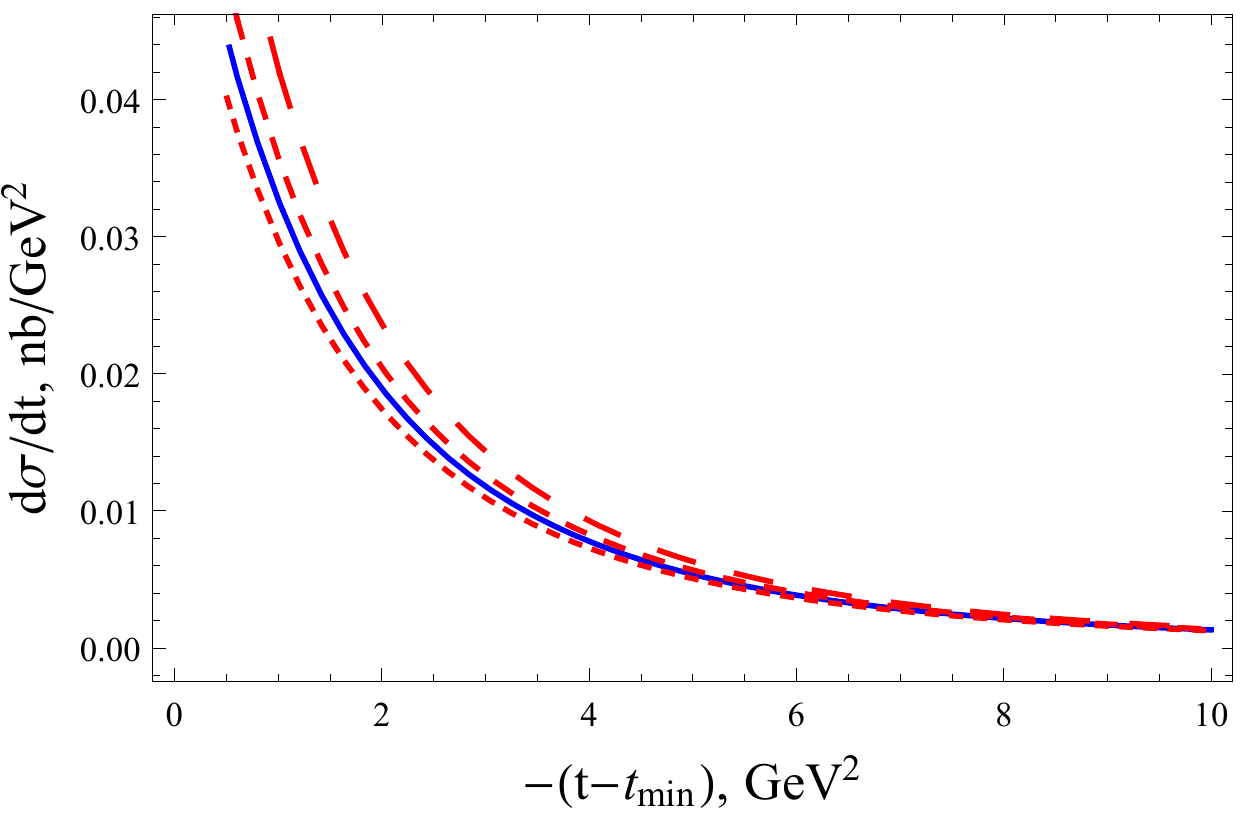}
  \caption{Differential cross section for $V=\Upsilon$
and different photon energies: $E_\gamma=58.9$ GeV large-red-dashing, $E_\gamma=58.6$ GeV medium-red-dashing, $E_\gamma=58.45$ GeV solid-blue-curve,
$E_\gamma=58.3$ GeV small-red-dashing, and $E_\gamma=58$ GeV dotted-red curve. }
  \label{fig_diffEupsilon}
\end{figure}




\begin{figure}[!htb]
\includegraphics[height=5.5cm]{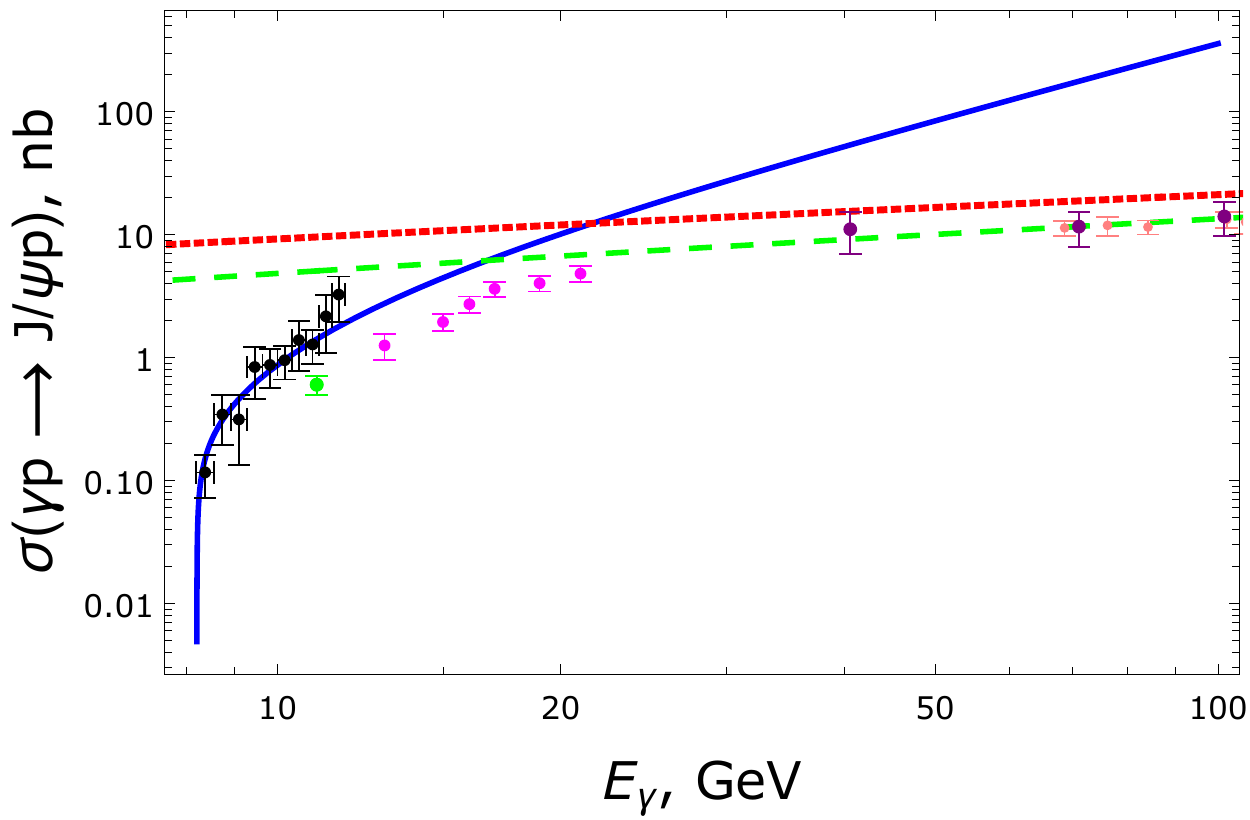}
  \caption{The total cross section for $J/\Psi$ photoproduction with the same parameters as in Fig.~\ref{fig_tot22}, but zoomed in near threshold. The data points are from: GlueX~\cite{GLUEX} (black), \cite{Camerini:1975cy} (magneta), \cite{Gittelman:1975ix} (green), \cite{Barate:1986fq} (purple), \cite{Binkley:1981kv} (pink).}
  \label{fig_tot11}
\end{figure}

\end{widetext}

In Fig.~\ref{fig_diff}, we show the behavior of the differential cross section (\ref{diff1}) for $V=J/\Psi$ photoproduction
for a photon energy $E_\gamma=10.72$ GeV in comparison to the GlueX recent data near threshold~\cite{GLUEX}. The solid-blue curve is our result for the soft-wall model. The data near threshold are from GlueX~\cite{GLUEX}. The mesonic parameters were fixed using (\ref{SOFTMF})
with $m_{0}=M_{J/\psi}=3.10$ GeV and $f_0=f_{J/\psi}=0.405$ GeV for the soft-wall model, and

\bea
\label{KAPPAV}
\tilde{\kappa}_V=\frac{2^{3/8}\pi^{3/4}}{3^{1/4}}\frac{\sqrt{f_V\,m_V}}{\lambda^{1/8}}\qquad &&{\rm  (D9\, model)}\nonumber\\
\tilde{\kappa}_V=\frac{2^{3/8}\pi^{3/4}}{3^{1/4}}\frac{\sqrt{f_V\,m_V}}{(2^{5/2}\pi)^{1/4}} \qquad &&{\rm  (D7\, model)}\nonumber\\
\tilde{\kappa}_V=\frac{2^{3/8}\pi^{3/4}}{3^{1/4}}\frac{\sqrt{f_V\,m_V}}{(2^{5/2}\pi/3)^{1/4}} \qquad &&{\rm (Original)}\nonumber\\
  \eea
Here the label $^\prime$original$^\prime$ refers the original soft-wall model.
The value of the form factor $A(0)$ is model dependent and follows from fitting the normalization factor $\mathcal{N}$, defined in (\ref{diff1N}), to data as
\begin{widetext}
\bea
\label{RATIOA0}
\mathcal{N}=\frac{20\,\sqrt{\lambda} N_fA(0)}{(10\tilde{\kappa}_{V})^4}=&&\sqrt{\frac{f_{J/\psi}m_{J/\psi}}{f_Vm_V}}\times7.768 \,{\rm GeV^{-4} \qquad (D9\, model)} \nonumber\\
\mathcal{N}=\frac{20\,2^{5/2}\pi\times N_fA(0)}{(10 \tilde{\kappa}_{V})^4}=&&\sqrt{\frac{f_{J/\psi}m_{J/\psi}}{f_Vm_V}}\times 7.768\,{\rm  GeV^{-4}\qquad  (D7\, model)}\nonumber\\
\mathcal{N}=\frac{ 20\,{2^{5/2}\pi}\times N_fA(0)}{3(10 \tilde{\kappa}_{V})^4}=&&\sqrt{\frac{f_{J/\psi}m_{J/\psi}}{f_Vm_V}}\times 7.768\, {\rm GeV^{-4} \qquad (Original)}.\nonumber\\
 \eea
\end{widetext}


In Fig.~\ref{fig_diffE} we show the same differential cross section for other  photon energies in dashed-red in comparison
to $E_\gamma=10.72$ GeV in solid-blue and the GlueX recent data~\cite{GLUEX}. The large-red-dashing curve is for $E_\gamma=11$ GeV,
the medium-red-dashing curve is for $E_\gamma=10.6$ GeV, the small-red-dashing curve is for $E_\gamma=10.3$ GeV
and the dotted-red curve is for $E_\gamma=10$ GeV.

In Fig.~\ref{A_ratio} we show the empirical ratio of the differential cross sections as a proposal for the ratio
of the gravitational form factors $A(\Delta t)/A(\Delta t_{\rm min})$ with $\Delta t=(-(t-t_{\rm min}))^{\frac 12}$ and $\Delta t_{\rm min}=\sqrt{0.0075}$  versus $\Delta t^2$ in GeV$^2$,

\begin{widetext}
\bea
\label{ATA0}
\frac{A(\Delta t)}{A(\Delta t_{\rm min})}=
\bigg(\frac{ F(s,t=t_{\rm min},M_V,m_N)(-2t_{\rm min}+8m_N^2)}{ F(s,t=-K^2,M_V,m_N)(2K^2+8m_N^2)}\bigg)
\frac{\bigg(\frac{d\sigma}{dt}\bigg)}{\,\,\,\,\,\,\,\bigg(\frac{d\sigma}{dt}\bigg)_{{\rm min}}}
\eea
\end{widetext}
The blue-solid line is our holographic result (\ref{diff1}), the red-dashed line is the fitted lattice gluonic contribution
from the recent simulations in~\cite{MIT}, and the data are the ratio of the data from from GlueX~\cite{GLUEX}. The empirical
errors for the ratio have been added in quadrature. (\ref{ATA0}) provides for a
model independent extraction of the gravitational form factor, under two generic assumptions:
1/ The Yukawa coupling of the dilaton to the bulk Dirac fermion vanishes in holography;
2/ The tensor $2^{++}$ glueball couplings map on the graviton couplings in bulk.

In Fig.~\ref{fig_diffEupsilon} we show the  differential cross section  for $V=\Upsilon$ production close to threshold
for different photon energies which is a prediction, for the same parameter set as the one used for $J/\Psi$ production.
The photon energies are: $E_\gamma=58.9$ GeV large-red-dashing,
$E_\gamma=58.6$ GeV medium-red-dashing, $E_\gamma=58.45$ GeV solid-blue-curve,
$E_\gamma=58.3$ GeV small-red-dashing,  and $E_\gamma=58$ GeV dotted-red curve.
We have used $m_{0}=M_{\Upsilon}=9.460$ GeV,   $f_0=f_{\Upsilon}=0.714$ GeV and $\kappa_V$ as in (\ref{KAPPAV}) for the models
with a soft wall. Note that in this case, $A(0)$ is fixed by the same ratios as in (\ref{RATIOA0}) with the numbers rescaled
by the factor $({{f_{J/\psi}m_{J/\psi}}/{f_{V}m_{V}}})$ to correct for the $V=\Upsilon$ parameters.

In Fig.~\ref{fig_tot11} (solid blue line), we show the total cross section for photoproduction of $V=J/\Psi$  versus the photon
energy close to threshold.  The total cross section follows by integrating the differential cross section in (\ref{diff1}) using
the dipole parametrization (\ref{AKK}) with $k^2\rightarrow -t$. The comparison is to GlueX data~\cite{GLUEX} (black ones). All other holographic
parameters are kept unchanged.

\section{Reggeized photoproduction}

The differential cross section (\ref{diff1}) grows rapidly as $s^2$ at large $s$ as expected from the exchange of a graviton
as a tensor glueball exchange with spin-2.  The physical cross section grows much slower due to the exchange of a
Pomeron instead. The transmutation from a graviton to a Pomeron was originally discussed in~\cite{Brower:2006ea}.
With increasing $\sqrt{s}$, higher spin-j exchanges contribute  leading to a reggeized amplitude with the emergence of a
Pomeron.  In this section and in supportive material given in the Appendices, we detail the spin-j contribution to (\ref{diff1})
and then re-sum these exchanges to extend the photoproduction results to all $\sqrt{s}$.

\subsection{Spin-j amplitude}

The spin-j exchange amplitude follows from the same considerations as the spin-2 exchange given earlier.
Here we summarise the results for the soft wall model with more details given in Appendix XIII together with
the results for the hard wall model. With this in mind,
 the spin-j glueball contribution to the TT-part of the photoproduction amplitude $\gamma p\rightarrow Ap$
with an  arbitrary virtual photon $A$, reads

\begin{widetext}
\be
&&i{\cal A}^{h}_{\gamma p\rightarrow A p} (j,s,t)\approx(-i)\mathcal{V}^{\mu\nu(TT)}_{hAA}(j,q_1,q_2,k_z)\times \bigg(\frac{i}{2}\eta_{\mu\alpha}\eta_{\nu\beta}\bigg)\times(-i)\mathcal{V}^{\alpha\beta(TT)}_{h\bar\Psi\Psi}(j,p_1,p_2,k_z)\,,	\nonumber\\\label{nAmphj}
\ee
\end{widetext}
The explicit form of the tensor TT-vertices ${\cal V}^{TT}$ depend on the model used.
For the {soft-wall} model, the normalized wave functions and bulk-to-bulk propagator
are detailed in Appendix XII. The result for the spin-j contribution to the vertices is

\begin{widetext}
\bea
\label{VSWjSW}
\mathcal{V}^{\mu\nu(TT)}_{hAA}(j,q_1,q_2,k_z)=&&\frac{\sqrt{2\kappa^2}}{2}\int dz\sqrt{g}\,e^{-\phi}\,z^{4+2(j-2)}K^{\mu\nu}(q,q^\prime,n,n^\prime, z)\times C(j)\times z^{\Delta(j)-{j-2}}\nonumber\\
\mathcal{V}^{\alpha\beta(TT)}_{h\bar\Psi\Psi}(j,p_1,p_2,k_z)=&&-\frac{\sqrt{2\kappa^2}}{2}\int dz\,\sqrt{g}\,e^{-\phi}\,z^{1+2(j-2)}\bar\Psi(p_2,z)\gamma^\alpha p^\beta\,\Psi(p_1,z)z^{-(j-2)}\mathcal{H}(j,K,z)\nonumber\\
=&&-\frac{\sqrt{2\kappa^2}}{2}\int dz\,\sqrt{g}\,e^{-\phi}\,z^{1+2(j-2)}\,\big(\psi_R^2(z)+\psi_L^2(z)\big)\,z^{-(j-2)}\mathcal{H}(j,K,z)\times\bar u(p_2)\gamma^\alpha p^\beta  u(p_1)\nonumber\\
=&&-\sqrt{2\kappa^2}\times g_5^2A(j,K)\times\bar u(p_2)\gamma^\alpha p^\beta u(p_1)\,,
\eea
with the parameters

\bea
\label{PARA}
&&C(j)= \tilde{\kappa}_V^{2\Delta(j)-4} \times\frac 1{\Delta(j)}
\frac{2^{\Delta(j)-2}\Gamma(a_K+\frac{\Delta(j)}{2})}{\Gamma(\Delta(j)-2)}\qquad\nonumber\\
&&\Delta(j)=2+\sqrt{2\sqrt{\lambda}(j-j_0)}\qquad{\rm and}\qquad a_K=\frac a2=\frac{K^2}{8{\tilde\kappa}^2}\qquad{\rm and}\qquad j_0=2-\frac{2}{\sqrt{\lambda}}
\eea
 For completeness, the analogue vertex $\mathcal{V}^{\mu\nu(TT)}_{hAA}$ for the hard wall model is

\be \label{VSWj}
\mathcal{V}^{\mu\nu(TT)}_{hAA}(j,q_1,q_2,k_z)=&&\frac{\sqrt{2\kappa^2}}{2}\int dz\sqrt{g}\,e^{-\phi}\,z^{4+2(j-2)}K^{\mu\nu}(q,q',n,n',z)\,\frac{\frac{2^{1-\tilde{\Delta}(j)}}{\pi}\times K^{\tilde{\Delta}(j)}\times z^{\tilde{\Delta}(j)+2-(j-2)}}{\tilde{\Delta}(j)+2}\,\,,\nonumber\\
\ee

Using (\ref{BBSWj}) in (\ref{VSWjSW}), we can write the spin-j form factor $A(j,K)$ of the proton as a bulk Dirac fermion in the soft-wall model as

\be
\label{Aj1}
A(j,K)=\frac{1}{2g_5^2}\frac{2^{2-\Delta(j)}\tilde{\kappa}_N^{j-2-\Delta(j)}}{\Gamma(\tilde{a}(j))}
\int_{0}^{1}dx\,{x^{\tilde{a}(j)-1}}{(1-x)^{-\tilde{b}(j)}}\big(I_z^R(x)+I_z^L(x)\big)\,,
\ee
with

\be
\label{PARAX}
\tilde a(j)=a_K+2-\frac 12 \Delta(j)\qquad {\rm and}\qquad \tilde b(j)=3-\Delta(j)
\ee
The integrals ($\xi=\tilde\kappa_N^2 z^2$)

\be
I_z^{R/L}(x)=\int dz\,\sqrt{g}\,e^{-\phi}\,z^{1+2(j-2)}\,
\psi_{R/L}^2(z)\,\xi^{\frac{-(j-2)}{2}}\xi^{2-\frac{\Delta(j)}{2}}
{\rm exp}\Big(-\frac{2x\xi}{1-x}\Big)\,,
\label{Iz1}
\ee
\end{widetext}
are over the wavefunctions of the proton as a Dirac fermion in bulk in the soft-wall model.
Specifically, we have

\be
&&\psi_R(z)=\frac{\tilde{n}_R}{\tilde{\kappa}_N^{\tau-2}} z^{\frac{5}{2}}\xi^{\frac{\tau-2}{2}}L_0^{(\tau-2)}(\xi)\,,\nonumber\\
&&\psi_L(z)=\frac{\tilde{n}_L}{\tilde{\kappa}_N^{\tau-1}} z^{\frac{5}{2}}\xi^{\frac{\tau-1}{2}}L_0^{(\tau-1)}(\xi)\,,\nonumber\\
\ee
with the twist parameter $\tau=7/2-1/2=3$.
Here  $L_n^{(\alpha)}(\xi)$ are the generalized Laguerre polynomials, and

\be
\tilde{n}_R=\tilde{n}_L \tilde{\kappa}_N^{-1}\sqrt{\tau-1}\qquad
\tilde{n}_L=\tilde{\kappa}_N^{\tau}\sqrt{{2}/{\Gamma(\tau)}}\nonumber\\
\ee
 Using the wave functions,  the integrals in (\ref{Iz1}) can be carried out explicitly, with the results

\begin{widetext}
\bea
\label{Iz2}
I_z^{R}(x)&=&\frac{1}{2}\times\tilde{\kappa}_N^{-2(j-2)}\times
\bigg(\frac{\tilde{n}_R}{\tilde{\kappa}_N^{\tau-1}}\bigg)^2\times\int d\xi\,\xi^{\frac{j-2}{2}+\tau-\frac{\Delta(j)}{2}}\,\bigg(L_{0}^{(\tau-2)}(\xi)\bigg)^2\,{\rm exp}\bigg(-\bigg(\frac{1+x}{1-x}\bigg)\xi\bigg)
\,,\nonumber\\
I_z^{L}(x)&=&\frac{1}{2}\times\tilde{\kappa}_N^{-2(j-2)}\times
\bigg(\frac{\tilde{n}_L}{\tilde{\kappa}_N^{\tau}}\bigg)^2\times\int d\xi\,\xi^{\frac{j-2}{2}+\tau-\frac{\Delta(j)}{2}+1}\,\bigg(L_{0}^{(\tau-1)}(\xi)\bigg)^2\,{\rm exp}\bigg(-\bigg(\frac{1+x}{1-x}\bigg)\xi\bigg)\,,
\eea
where we used $\phi=e^{-\xi}$. Evaluating the integrals in (\ref{Iz2}) we obtain

\bea
\label{Iz3}
&&I_z^{R}(x)=\frac{1}{2}\times\tilde{\kappa}_N^{-2(j-2)}\times\bigg(\frac{\tilde{n}_R}{\tilde{\kappa}_N^{\tau-1}}\bigg)^2\times\Gamma\bigg(\frac{j-2}{2}+\tau -\frac{\Delta(j)}{2}+1\bigg)\times\bigg(\frac{1+x}{1-x}\bigg)^{-\frac{j-2}{2}-\tau +\frac{\Delta(j)}{2}-1}\,,\nonumber\\
&&I_z^{L}(x)=\frac{1}{2}\times\tilde{\kappa}_N^{-2(j-2)}\times\bigg(\frac{\tilde{n}_L}{\tilde{\kappa}_N^{\tau}}\bigg)^2\times\Gamma\bigg(\frac{j-2}{2}+\tau -\frac{\Delta(j)}{2}+2\bigg)\times\bigg(\frac{1+x}{1-x}\bigg)^{-\frac{j-2}{2}-\tau +\frac{\Delta(j)}{2}-2}\,.\nonumber\\
\eea

Using (\ref{Iz3}) in (\ref{Aj1}), the spin-j glueball form factor of the proton becomes

\bea
\label{Aj2}
A(j,K)=&&\frac{1}{4g_5^2}\frac{\tilde{\kappa}_N^{-(j-2)-\Delta(j)}}{\Gamma(\tilde{a}(j))}\int_{0}^{1}dx\,x^{\tilde{a}(j)-1}(1-x)^{-\tilde{b}(j)}\nonumber\\&&\times\bigg(\bigg(\frac{\tilde{n}_R}{\tilde{\kappa}_N^{\tau-1}}\bigg)^2\times\Gamma(c(j))\bigg(\frac{1+x}{1-x}\bigg)^{-c(j)}
+\bigg(\frac{\tilde{n}_L}{\tilde{\kappa}_N^{\tau}}\bigg)^2\times\Gamma(c(j)+1)\bigg(\frac{1+x}{1-x}\bigg)^{-(c(j)+1)}\bigg)\,,\nonumber\\
\eea
\end{widetext}
with $\Delta(j)$ given in (\ref{PARA}), $\tilde a(j), \tilde b(j)$ given in (\ref{PARAX}) and

\bea
c(j)=(\tau+1)+\frac{j-2}{2}-\frac{\Delta(j)}{2}
\eea
$A(j,K)$ generalizes the gravitational form factor for all $j\geq 2$.
 Evaluating the integral in (\ref{Aj2}), we obtain (\ref{Aj3}).
Inserting  (\ref{Aj3}) in (\ref{VSWjSW}), (\ref{nAmphj}) becomes

\begin{widetext}
\be
 \label{vvvJPSI1j}
&&{\cal A}_{\gamma p\rightarrow J/\Psi p} (j,s,t)=\mathcal{V}_{hAA}(j)\bigg(-\frac{1}{2}B_{1}^{\alpha\beta}\,
\bar u(p_2)\gamma_\alpha p_\beta u(p_1)\bigg)\mathcal{V}^{(TT)}_{h\bar\Psi\Psi}(j)\,,\nonumber\\
\ee
The spin-j vertices are

\bea
 \label{vvJPSI1j}
\mathcal{V}_{h\bar\Psi\Psi}^{(TT)}(j)=&&-g_5^2A(j,K)\,,\nonumber\\
\mathcal{V}_{hAA}(j)=&&\Big(\frac{f_{n}}{m_{n}}\Big)\times \Big(\frac{\sqrt{2\kappa^2}}{2}\int_{0}^{\infty} dz\sqrt{g}e^{-z^2\tilde{\kappa}_V^2}\,z^{4+2(j-2)}\times (2\tilde\kappa^2 z^2)L_{n}^1(z^2\tilde{\kappa}_V^2)\times C(j) \times z^{\Delta(j)-(j-2)}\Big)\nonumber\\
\approx&&\Big(\frac{f_{n}}{m_{n}}\Big)\times\bigg(\frac{\sqrt{2\kappa^2}}{2}
\frac{L_{n}^1(0)}{\Delta(j)\tilde{\kappa}_V^{\Delta(j)+j-2}}\times C(j)
\times\bigg(\int_{0}^{\infty} d\xi\,e^{-\xi^2}\,\xi^{\frac{\Delta(j)}{2}+\frac{j}{2}-1}\bigg)\bigg)
\equiv \bigg(\frac{f_V}{M_V}\bigg)\,{\mathbb V}_{hAA}(j)\nonumber\\
\eea
\end{widetext}
with  $B_{1}^{\alpha\beta}(q,q^\prime,n,n^\prime)$ defined in (\ref{BK}).
The heavy mesons  with $n=J/\Psi\,,\Upsilon$  are subsumed.
(\ref{vvvJPSI1j}) shows how VMD  extends to general spin-j exchange in holography,
with ${\mathbb V}_{hAA}(j)$ reflecting on  its coupling to the pair vector-meson-photon in bulk.

\subsection{Reggeized amplitude}

After summing over all contributions from the spin-j glueballs, the
photoproduction amplitude ${\cal A}^{tot}_{\gamma p\rightarrow J/\Psi p} (s,t)$ is

\begin{widetext}
\bea\label{pomeron}
{\cal A}^{tot}_{\gamma  p\rightarrow J/\Psi p} (s,t)=&&-\int_{\mathbb C}\frac{dj}{2\pi i}
\left(\frac{s^{j-2}+(-s)^{j-2}}{{\rm sin}\,\pi j}\right){\cal A}_{\gamma p\rightarrow J/\Psi p} (j,s,t)\nonumber\\
{\cal A}_{\gamma p\rightarrow J/\Psi p} (j,s,t)=&&\frac{1}{2}\mathcal{V}_{hAA}(j)\times B_{1}^{\alpha\beta}\times\frac{{2\kappa^2}}{g_5^4} \times g_5^2A(j,K)\times\bar u(p_2)\gamma_\alpha p_\beta u(p_1)\,,
\eea
\end{widetext}
The contour $\mathbb C$ is at the rightmost of the branch-point of $A(j,K)$.
The spin-j glueball form factor $A(j,K)$ of the proton  as a bulk Dirac fermion is given in  (\ref{Aj2}) for the soft wall model.
The integrals can be carried explicitly, with the result

\begin{widetext}
\be\label{Aj3}
&&A(j,K)=\frac{\tilde{\kappa}_N^{-(j-2)-\Delta(j)}}{4g_5^2}\,\frac{\Gamma(c)\Gamma(1-\tilde{b}+c)}{\Gamma(1-\tilde{b}+c+\tilde{a})}\nonumber\\&&\times\bigg(\bigg(\frac{\tilde{n}_R}{\tilde{\kappa}_N^{\tau-1}}\bigg)^2\,_2F_1(\tilde{a},c+1,1-\tilde{b}+c+\tilde{a},-1)
+\bigg(\frac{\tilde{n}_L}{\tilde{\kappa}_N^{\tau}}\bigg)^2\frac{c(1-\tilde{b}+c)}{1-\tilde{b}+c+\tilde{a}}\,_2F_1(\tilde{a}+1,c+1,2-\tilde{b}+c+\tilde{a},-1)\bigg)\,.\nonumber\\
\ee
\end{widetext}
The parameters are fixed in (\ref{PARA}) as

\bea
&&1-\tilde{b}+c=(\tau-1)+\frac{j-2}{2}+\frac{\Delta(j)}{2}\nonumber\\
&&1-\tilde{b}+c+\tilde{a}=(\tau+1)+\frac{j-2}{2}+a_K\nonumber\\
\eea
 Note that at $j=2$, (\ref{Aj3}) is exactly equal to the spin-2 gravitational form factor (\ref{FFj2}) (times $1/\tilde{\kappa}_V^{4}$ to compensate for the new normalization we used for the higher spin case).

From (\ref{pomeron}-\ref{Aj3}), we determine the single Pomeron amplitude (total amplitude) in momentum space,
after wrapping the j-plane contour ${\mathbb C}$ to the left,

\begin{widetext}
\bea\label{pomeron}
{\cal A}^{tot}_{\gamma  p\rightarrow J/\Psi p} (s,t)=-s^{j_{0}-2}\int_{-\infty}^{j_0}\frac{dj}{\pi}
\left(\frac{1 +e^{-i\pi}}{{\rm sin}\,\pi j}\right)s^{j-j_{0}}\,\text{Im}[{\cal A}_{\gamma p\rightarrow J/\Psi p} (j,s,t)]
\eea
\end{widetext}
The imaginary part follows from the discontinuity of the $\Gamma$-function

\begin{widetext}
\bea
\label{disc}
&&\text{Im}[{\cal A}_{\gamma p\rightarrow J/\Psi p} (j,s,t)]\approx \frac{\tilde{\kappa}_N^{-(j-2)-\Delta(j)}}{\tilde{\kappa}_N^{4-\Delta(j)+j-2}}\times\Big(\frac{\tilde{\kappa}_N}{\tilde{\kappa}_V}\Big)^{4-\Delta(j)+j-2}\times\frac{\sqrt{2\kappa^2}}{g_5^4}\times\nonumber\\
&&\bigg(\frac{1}{2}\tilde{\kappa}_V^{4-\Delta(j)+j-2}\Gamma(\Delta(j)-2)\mathcal{V}_{hAA}(j)\times B_{1}^{\alpha\beta}
\times\tilde{\kappa}_N^{j-2+\Delta(j)}g_5^2A(j,K)
\bar u(p_2)\gamma_\alpha p_\beta u(p_1)\bigg)\bigg\vert_{j\rightarrow j_0, \Delta(j)\rightarrow 2}\times\text{Im}\bigg[\frac{1}{\Gamma(\tilde{\Delta}(j))}\bigg]\nonumber\\
\eea
\end{widetext}
with the complex argument

\be
\tilde{\Delta}(j)=\Delta(j)-2=i\sqrt{2\sqrt{\lambda}(j_0-j)}\equiv iy\nonumber\\
\ee
and $j_0=2-{2}/{\sqrt{\lambda}}$. For $y\rightarrow 0$, we may approximate $1/\Gamma(iy)\approx  iy\,e^{i\gamma y}$,
 with the Euler-Mascheroni constant $\gamma=0.55772...$.
 The single Pomeron amplitude (total amplitude) in momentum space (\ref{pomeron}) can now be cast in block form

\be
\label{pomeron2}
{\cal A}^{tot}_{\gamma  p\rightarrow J/\Psi p} (s,t)= I_{j}(j_0,s)\times G_{5}(j_0,s,t)
\ee
with

\begin{widetext}
\bea
\label{IJ}
&&I_j(j_0,s)=-\tilde{s}^{j_{0}}\int_{-\infty}^{j_0}\frac{dj}{\pi}
\left(\frac{1 +e^{-i\pi}}{{\rm sin}\,\pi j}\right)\tilde{s}^{j-j_{0}}\,\sin\left[\tilde{\xi}\sqrt{2\sqrt{\lambda}(j_0-j)}\right]\nonumber\\
&&G_{5}(j_0,s,t)=\Big(\frac{\tilde{\kappa}_N}{\tilde{\kappa}_V}\Big)^{4-\Delta(j)+j-2}\times\frac{1}{s^2}\bigg(\frac{1}{2}\tilde{\kappa}_V^{4-\Delta(j)+j-2}\Gamma(\Delta(j)-2)\mathcal{V}_{hAA}(j)\times B_{1}^{\alpha\beta}\times\frac{\sqrt{2\kappa^2}}{g_5^4}\nonumber\\
&&\times\tilde{\kappa}_N^{j-2+\Delta(j)}g_5^2A(j,K)
 \bar u(p_2)\gamma_\alpha p_\beta u(p_1)\bigg)\bigg\vert_{j\rightarrow j_0,\,\Delta(j)\rightarrow 2}\nonumber\\
\eea
\end{widetext}
We have set $\tilde{s}\equiv {s}/{\tilde{\kappa}_N^2}$, and defined $\tilde{\xi}\equiv\gamma+{\pi}/{2}$. We note that the apparent pole
in the Gamma-function at the Pomeron intercept, cancels out in the combination  $\Gamma(\Delta(j_0)-2){\cal V}_{hAA}(j_0)$.

 In the block form
(\ref{pomeron2}), the spin-j integral $I_j(j_0,s)$ is similar to the spin-j integral in~\cite{Brower:2006ea}  (see Eq. 4.19),
with the identifications $\mathcal{K}(s,b^{\perp},z,z')\leftrightarrow {\cal A}^{tot}_{\gamma  p\rightarrow J/\Psi p} (s,t)$, $(zz'/R^4)G_3(j_0,v)  \leftrightarrow G_5(j_0,s,t)$, $\xi(v)\leftrightarrow\tilde{\xi}$, and $\widehat s\leftrightarrow \tilde{s}$. We then follow~\cite{Brower:2006ea}
to evaluate the spin-j integral by closing the j-contour appropriately. In the high energy limit $\sqrt{\lambda}/\tilde{\tau}\rightarrow 0$
($\tilde{\tau}\equiv\log\tilde{s}$),  the  single Pomeron contribution to the photoproduction amplitude is

\begin{widetext}
\bea\label{pomeron3}
{\cal A}^{tot}_{\gamma  p\rightarrow J/\Psi p} (s,t)\simeq e^{j_0\tilde{\tau}} \left[(\sqrt{\lambda}/\pi)+ i\right] ( \sqrt{\lambda}/ 2 \pi )^{1/2}\; \tilde{\xi}  \; \frac{e^ {-\sqrt\lambda  \tilde{\xi}^2 / 2\tilde{\tau}}}{\tilde{\tau}^{3/2}}\left(1 + {\cal O}\bigg(\frac{\sqrt{\lambda}}{\tilde{\tau}}\bigg) \right)
\times  G_{5}(j_0,s,t)
\eea
\end{widetext}
As expected, the amplitude develops both a real and imaginary part with  a  $\rho$-ratio about constant

\be
\rho=\frac{\text{Re}[{\cal A}^{tot}_{\gamma p\rightarrow J/\Psi p} (s,t=0)]}{\text{Im}[{\cal A}^{tot}_{\gamma  p\rightarrow J/\Psi p} (s,t=0)]}\simeq\frac{\sqrt{\lambda}}\pi
\ee
The single Pomeron contribution to the  total differential cross section is

\begin{widetext}
\bea
\label{DCCj}
\left(\frac{d\sigma}{dt}\right)_{\rm tot}
=&&\frac{e^2}{16\pi (s-m_N^2)^2}\,\frac 12\sum_{{\rm pol}}
\frac 12\sum_{{\rm spin}}
\Bigg|{\cal A}^{tot}_{\gamma p\rightarrow  J/\Psi p} (s,t)\Bigg|^2 \nonumber\\
&&\simeq\frac{e^2}{16\pi (s-m_N^2)^2}\times \bigg(e^{2j_0\tilde{\tau}} \left[(\lambda/\pi^2)+ 1\right] ( \sqrt{\lambda}/ 2 \pi )\; \tilde{\xi}^2  \; \frac{e^ {-\sqrt\lambda  \tilde{\xi}^2 / \tilde{\tau}}}{\tilde{\tau}^{3}}\bigg)\times \,\frac 12\sum_{{\rm pol}}
\frac 12\sum_{{\rm spin}}
\Bigg| G_5(j_0,s,t)\Bigg|^2 \nonumber\\
\eea
with the polarization-spin average

\bea
\label{DCCjX}
\sum_{{\rm pol, spin}}\Bigg| G_5(j_0,s,t)\Bigg|^2 =&&
\Big(\frac{\tilde{\kappa}_N}{\tilde{\kappa}_V}\Big)^{2(4-\Delta(j)+j-2)}\nonumber\\
&&\times\bigg(\frac {f_V}{M_V}\bigg)^2\,\bigg(\frac{2\kappa^2}{g_5^8}\,\tilde{\kappa}_V^{2(4-\Delta(j)+j-2)}\Gamma^2(\Delta(j)-2)\mathbb V_{hAA}^2(j)
\times {\tilde{\kappa}_N^{2(j-2+\Delta(j))}}\frac{g_5^4A^2(j,K)}{m_N^2}\bigg)\bigg\vert_{j\rightarrow j_0,\,\Delta(j)\rightarrow 2}\nonumber\\
&&\times \frac{F(s,t=-K^2,M_V,m_N)}{s^4}\times(2K^2+8m_N^2)\nonumber\\
\eea
\end{widetext}
and $j_0=2-2/\sqrt{\lambda}$.  Note that the resummed spin-j contribution to the gravitational form factor is now fixed by the Pomeron exchange with the form factor $A(K,j_0)$ at large $\sqrt{s}$. Remarkably, the emerging Pomeron exchange in the soft wall model in (\ref{DCCj}) which is a new result,
bears much in common with the original conformal Pomeron  kernel in~\cite{Brower:2006ea}.

The differential cross section rises with twice the conformal Pomeron intercept or $2\times(1-2/\sqrt{\lambda})$, and asymptotes

\be
\left(\frac{d\sigma}{dt}\right)_{\rm tot}\sim && s^{2-\frac{4}{\sqrt{\lambda}}}\times\bigg(1+\frac{\pi^2}{\lambda}\bigg)
\nonumber\\
&&\times\bigg(\bigg(\frac{\sqrt{\lambda}}{\log\tilde{s}}\bigg)^3+\mathcal{O}\bigg(\bigg(\frac{\sqrt{\lambda}}{\log\tilde{s}}\bigg)^4\bigg)\bigg)\nonumber\\
\ee
in the high energy limit with $\log\tilde{s}=\log({s}/{\tilde{\kappa}_V^2})\gg \sqrt{\lambda}$.
Using the optical theorem one can determine the total cross section $\sigma_V(s)$
for $\gamma p\rightarrow  V p$ with $V=J/\Psi, \Upsilon$ to be

\be
\sigma_V(s)=\bigg(\frac{16\pi}{1+\rho^2}\left(\frac{d\sigma}{dt}\right)_{{\rm tot}}\bigg)_{t=0}^{\frac 12}
\ee
with the Pomeron rise $\sigma_V(s)\sim s^{1-2/\sqrt\lambda}$
at large $\sqrt{s}$~\cite{Brower:2006ea}. Recall that close to threshold, the t-exchange is kinematically bounded as shown in Fig.~\ref{tmx}, and
the total cross section follows from the differential cross section (\ref{diff1}) by integration using (\ref{5}).

\begin{widetext}

\begin{figure}[!htb]
\includegraphics[height=5.5cm]{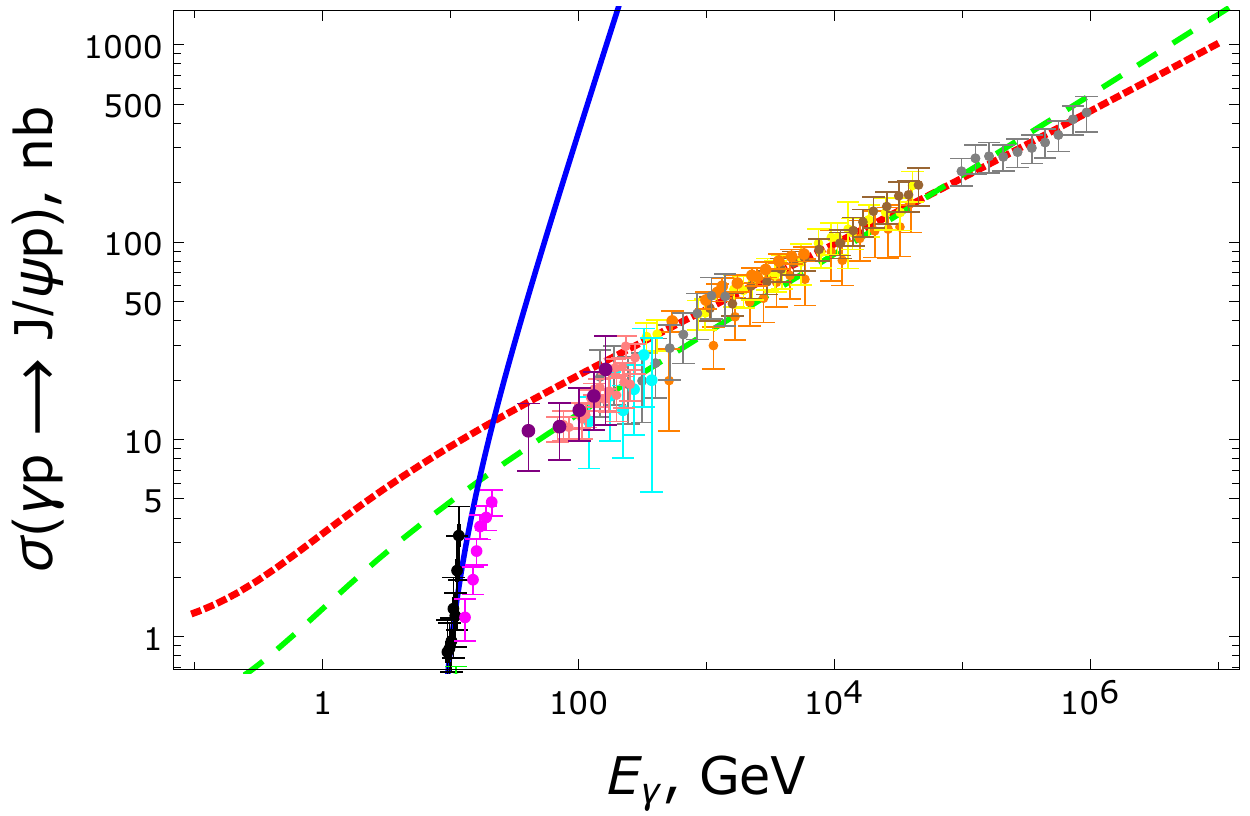}
  \caption{Total cross section for photoproduction of charmonium with $V=J/\Psi$, from close to threshold to very high energy.
The solid (blue) curve is the low-energy regime
 compared to the data from GlueX~\cite{GLUEX} (black). The red (tiny dashed) is the high energy regime. 
The green line (medium dashed) is found after fixing a normalization constant with one high energy data point but with the same high energy 't Hooft coupling constant $\lambda=11.243$ as the red (tiny dashed) one.
The data points are from: \cite{Camerini:1975cy} (magneta), \cite{Gittelman:1975ix} (green), \cite{Barate:1986fq} (purple), \cite{Adloff:1999kg} (orange),
\cite{Binkley:1981kv} (pink), \cite{Chekanov:2002xi} (yellow), \cite{Aktas:2005xu} (brown), \cite{Alexa:2013xxa} (orange), and \cite{Aaij:2013jxj} (grey).}
  \label{fig_tot22}
\end{figure}

\begin{figure}[!htb]
\includegraphics[height=5.5cm]{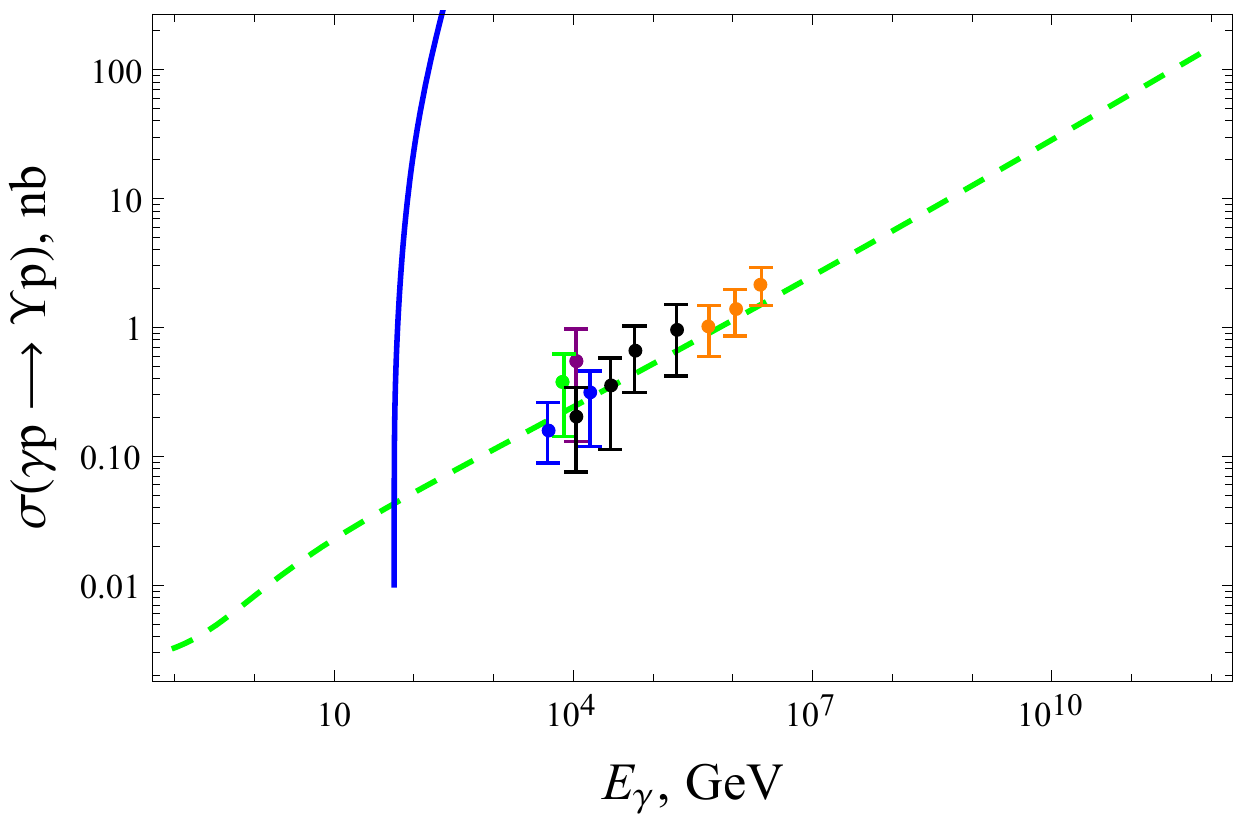}
  \caption{Total cross section for photoproduction of upsilonium with $V=\Upsilon$ from close to threshold
  to the very high energy regime.  The solid (blue) curve is the low-energy regime (near threshold). 
The green line (medium dashed) is found after fixing a normalization constant with one high energy data point but with the same 't Hooft coupling constant $\lambda=11.243$ as $J/\psi$. The data points are from: \cite{Breitweg:1998ki} (green), \cite{Adloff:2000vm} (purple), \cite{Chekanov:2009zz}  (blue), \cite{Aaij:2015kea}  (orange), and \cite{Sirunyan:2018sav} (black).}
  \label{fig_tot44}
\end{figure}

\begin{figure}[!htb]
\includegraphics[height=5.5cm]{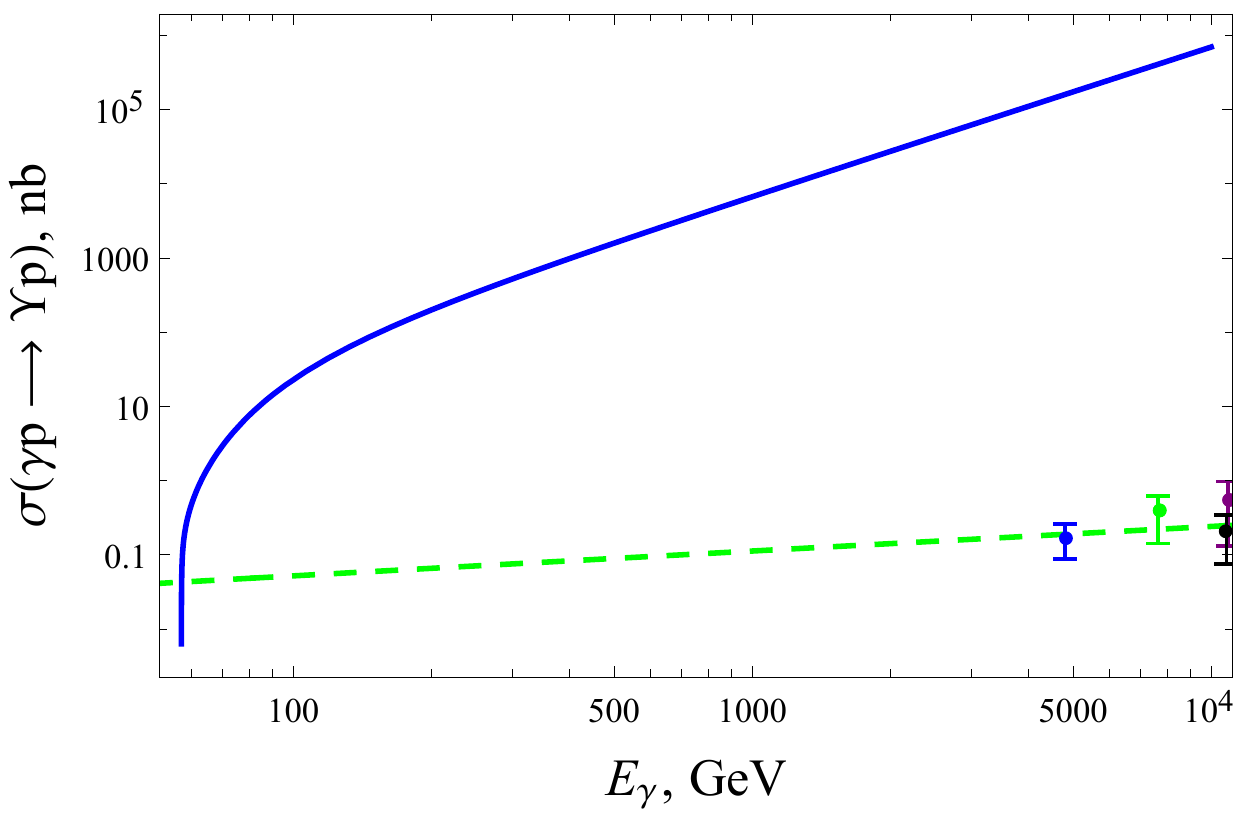}
  \caption{Total cross section for $\Upsilon$ photoproduction with the same parameters as in Fig.~\ref{fig_tot44} but zoomed in near the threshold.}
  \label{fig_tot33}
\end{figure}

\end{widetext}

In Fig.~\ref{fig_tot22}, we
show the total cross section for photoproduction of charmonium with $V=J/\Psi$ from threshold
to very high energy.  The same soft wall parameters~(\ref{KAPPAV}) and the same fitting condition on $A(0)$ as in (\ref{RATIOA0}) are used in the threshold region
for the solid-blue curve.  In this region, the parameter set is insensitive to the expansion of the vector meson wavefunction $L_n^1(z)$  near the holographic boundary.
At very high energy, we used the parameter set~(\ref{KAPPAV}) for the D9 model and adjusted $A(0,j_0)$ to

\be
\label{Aj0}
\Big(\frac{\tilde{\kappa}_N}{\tilde{\kappa}_V}\Big)^{j_0}\times\frac{10\,N_fA(0,j_0)}{\sqrt{\lambda}(10\tilde{\kappa}_{V})^4}=3.631\,{\rm GeV}^{-4}\qquad (\rm {D9\, model})\nonumber\\
\ee
with $\lambda=11.243$. The fit value (\ref{Aj0}) is sensitive to the expansion of $L^1_n(z)$ near the holographic boundary. The value of the coupling
$\lambda$ is not.  Similar fits are found for the other two holographic models.
The solid (blue) curve is the low-energy regime. The data  are from GlueX~\cite{GLUEX}(black). The red (tiny dashed) is the high energy regime.
The green line (medium dashed) is found after fixing a normalization constant with one high energy data point but with the same high energy 't Hooft coupling constant $\lambda=11.243$ as the red (tiny dashed) one. In Fig.~\ref{fig_tot11}, we zoomed in the total cross section for
$J/\Psi$ photoproduction near the threshold with the same parametrs as in Fig.~\ref{fig_tot22}, and compared to data from GlueX~\cite{GLUEX} in this regime.

In Fig.~\ref{fig_tot44}, we show the total cross section for $V=\Upsilon$ photoproduction from close to threshold
to very high energy regime, with the same parameter set.
The solid (blue) curve is the low-energy regime. 
The green line (medium dashed) is found after fixing a normalization constant with one high energy data point but with the same 't Hooft coupling constant $\lambda=11.243$ as $J/\psi$.
In Fig.~\ref{fig_tot33}, we show the total cross section for $\Upsilon$ photoproduction zoomed in close to the threshold with the same parameters as in Fig.~\ref{fig_tot44}.



\section{Generalized parton distribution  of gluons inside the proton}

The generalized parton distribution (GPD) can be viewed as the amplitude for removing a parton with momentum fraction-x and then re-inserting it,
while the nucleon is receiving a momentum kick $\vec K$ all the while travelling on the light cone.  It is related to the form factor of the energy-momentum
tensor by a sum rule as we now detail. The fact that the gluon GPD can be  picked in the diffractive
photoproduction of heavy mesons is not surprising. Indeed, as we noted earlier, the Witten diagram for the holographic photoproduction
amplitude is related to the amplitude for the inverse deeply virtual Compton scattering amplitude  through VMD.

\subsection{Gluon GPD: $j=2$}

The tensor coupling of the glueball  to the nucleon as a Dirac fermion is through its
 gravitational invariant form factors (\ref{EMT2}). For $j=2$ the exchange is dominated by the graviton at threshold,
with the contribution ($\tau=3$ and $a_K={K^2}/{8\tilde{\kappa}_N^2}$)

\begin{widetext}
\bea
\label{Aj222}
A(K)=&&A(0)\times \Gamma(a_K+2)\times g_5^2\tilde{\kappa}_N^{4}A(j=2,K)=A(0)\int_0^1dx \frac{a_K(a_K+1)}{x^{\alpha_G(t)}}\,\bigg(\frac{1-x}{1+x}\bigg)^\tau
\eea
\end{widetext}
with the graviton Regge trajectory $\alpha_G(t)=1+t/m^2_0$ and $-t=K^2\ll s$.
Here $m_0$ is fixed by the $2^{++}$ glueball mass in (\ref{SOFTMFG2}).
For a spin-2 and twist-2  exchange, the $A(K)$ form factor
obeys the sum rule ~\cite{POLYAKOV} (see also Eq.~3.154 in \cite{Belitsky:2005qn}, and reference therein)

\bea \label{AKK}
A(K)\equiv &&\int_{0}^{1}dx\,x^{j-1}{g}(x,K)
\eea
with $xg(x,K)$ the gluon GPD, at the renormalization scale set by the nucleon mass.

\begin{figure*}
\subfloat[The small-x gluon distribution $xg_<(x, b_\perp)$ inside the proton (\ref{xgb1}) as probed by the graviton.\label{figxgp1}]{%
  \includegraphics[height=6cm,width=.49\linewidth]{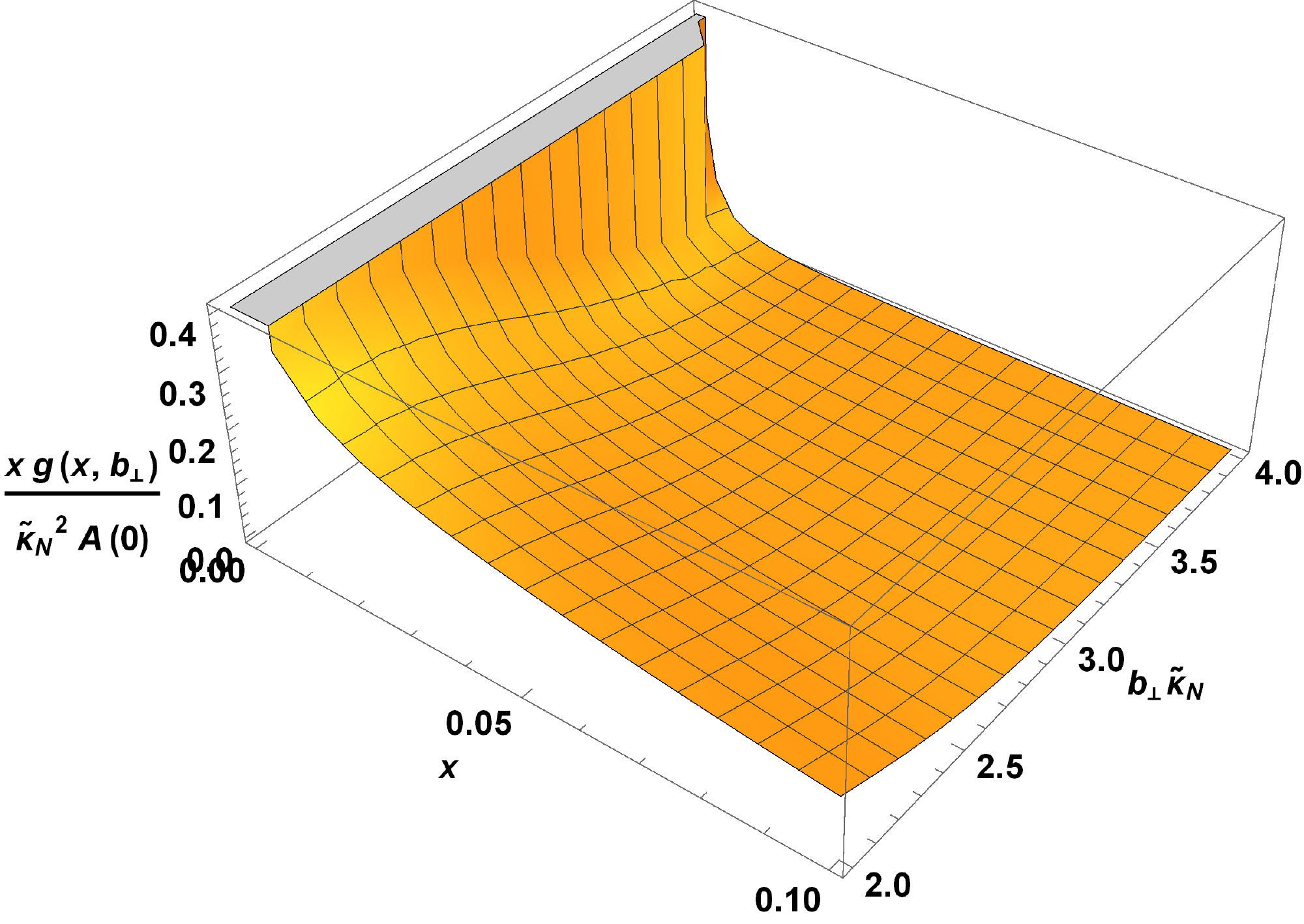}%
}\hfill
\subfloat[The small-x gluon $xg_<(x, b_\perp)$ distribution inside the proton (\ref{xgb1}) as probed by the graviton
with $\tau=3$, $x=0.1$ and $b_\perp=({b_x^2+b_y^2})^{\frac 12}$.\label{figxgp2}]{%
  \includegraphics[height=6cm,width=.49\linewidth]{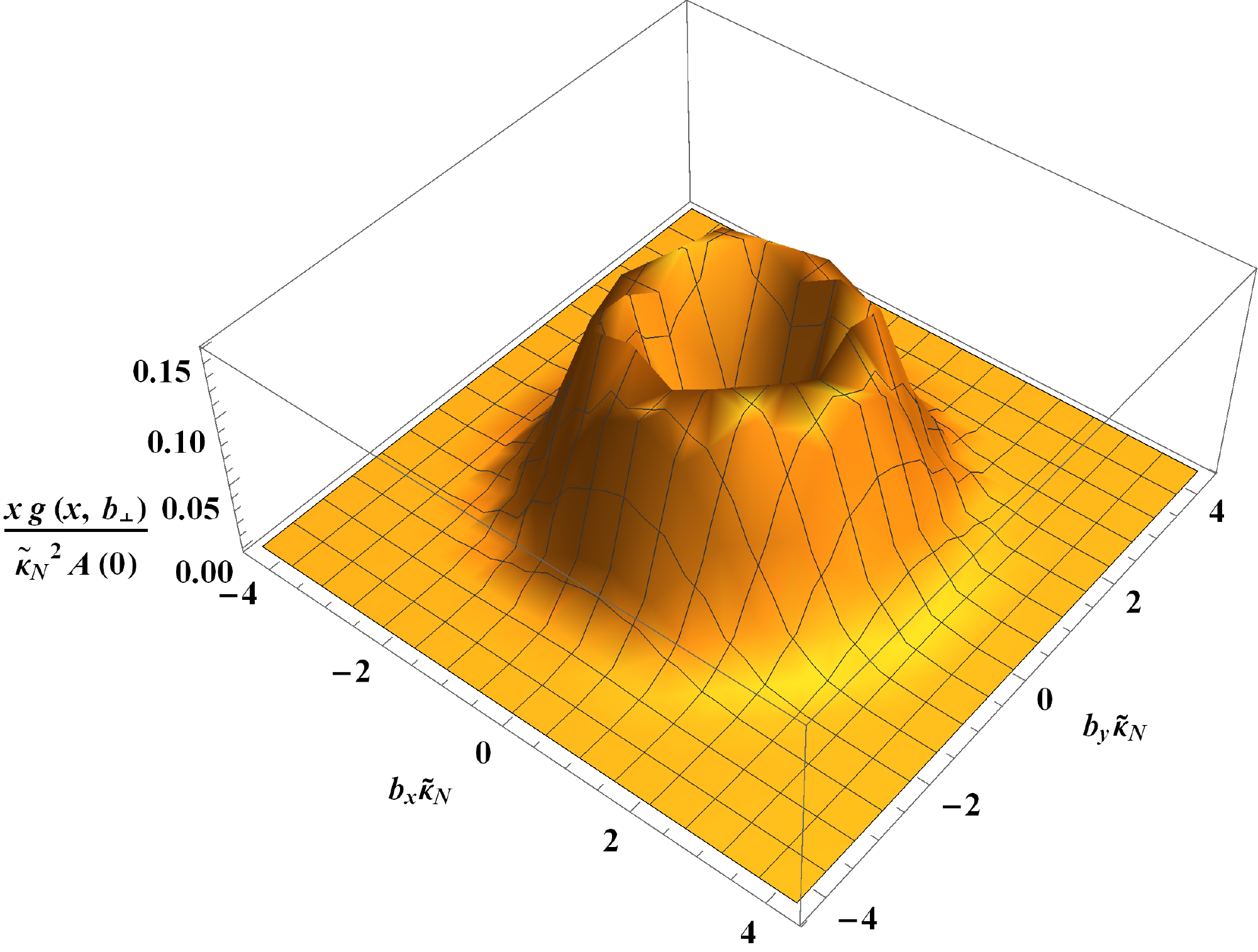}%
}
\caption{Small-x gluon GPD as probed by graviton exchange in photoproduction of a heavy meson close to threshold.}
\label{fig_smallx}
\end{figure*}

\begin{figure*}
\subfloat[The large-x gluon distribution $xg_>(x, b_\perp)$ inside the proton (\ref{XGX111}) as probed by the graviton with $\tau=3$.\label{figxgp3}]{%
  \includegraphics[height=6cm,width=.49\linewidth]{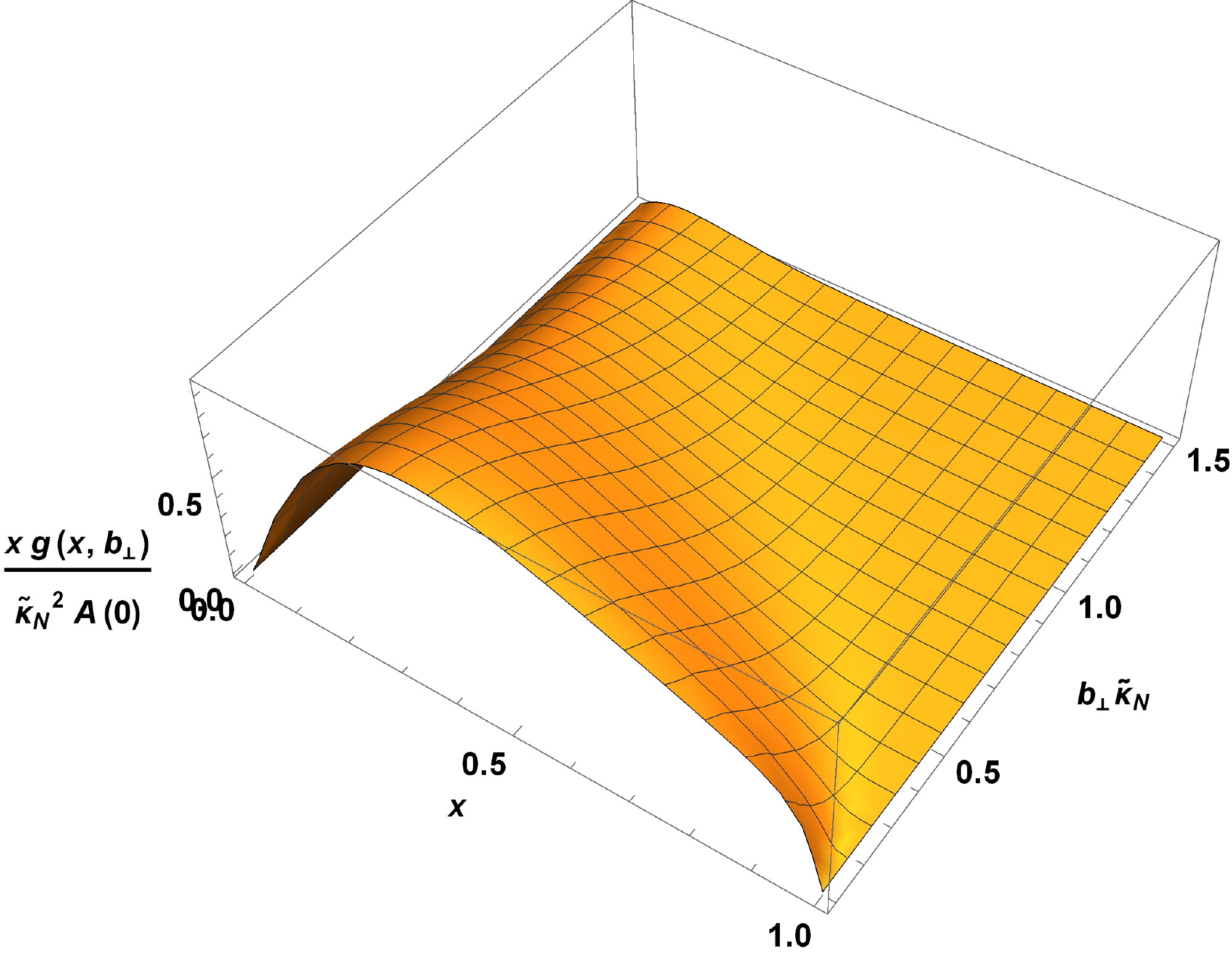}%
}\hfill
\subfloat[The large-x gluon distribution $xg_>(x, b_\perp)$ inside the proton (\ref{XGX111}) as probed by the graviton
with $\tau=3$, $x=0.5$ and $b_\perp=({b_x^2+b_y^2})^{\frac 12}$.\label{figxgp4}]{%
  \includegraphics[height=6cm,width=.49\linewidth]{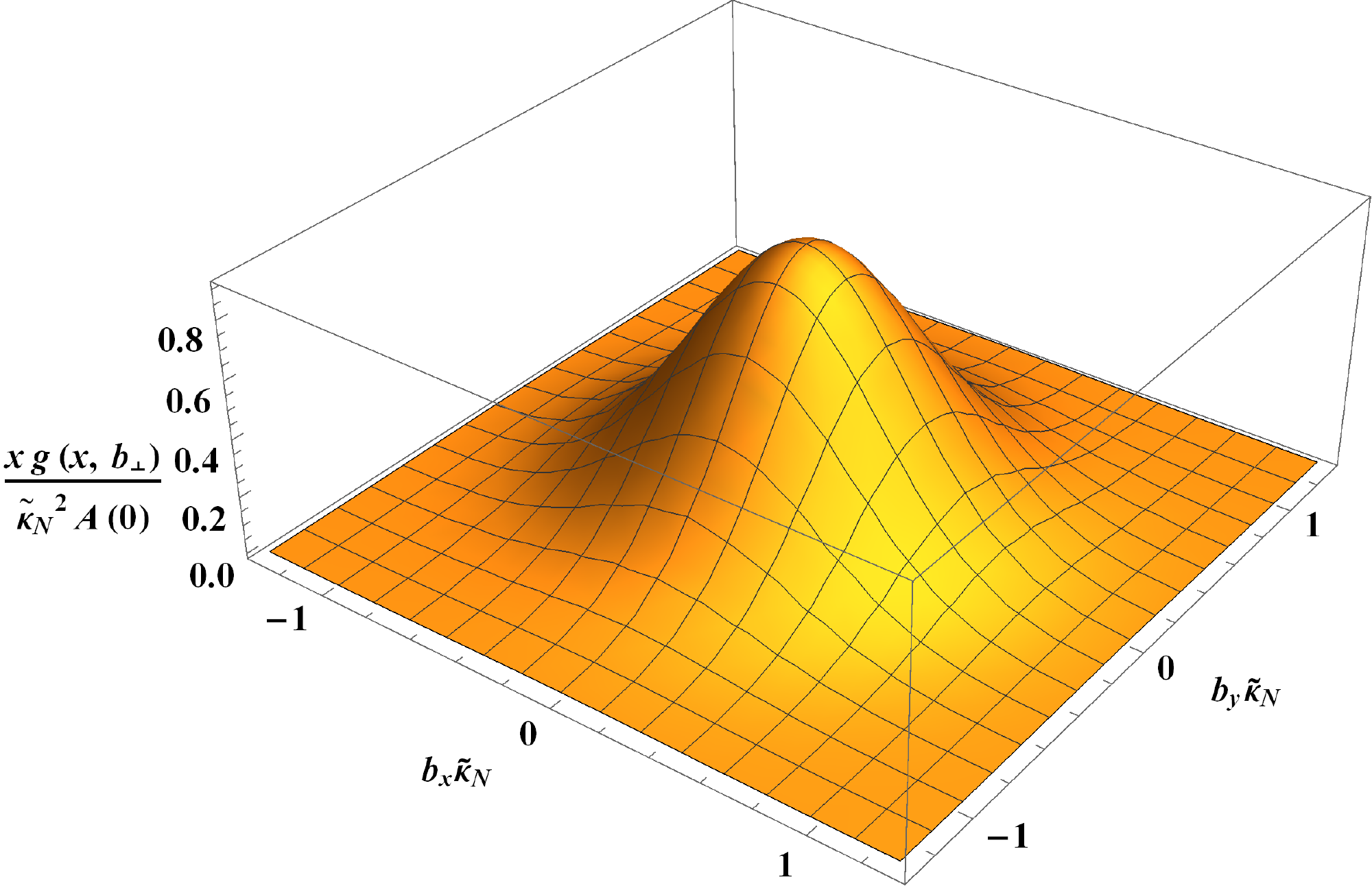}%
}
\caption{Large-x gluon GPD as probed by graviton exchange in photoproduction of a heavy meson close to threshold.}
\label{fig_largex}
\end{figure*}







The representation
(\ref{Aj222})  suggests that  $0\leq x\leq 1$  maybe interpreted as the x-momentum fraction of the gluons in the proton probed by the graviton.  At small-x, the exchange   is dominated by the graviton Regge trajectory which is manifest in the integral representation (\ref{Aj222})


\bea
\label{XGX1}
x{g}_<(x,K)\sim   A(0)\frac{a_K(a_K+1)}{x^{\alpha_G(t)}}\,\bigg(\frac{1-x}{1+x}\bigg)^\tau
\eea
For zero skewness ($\xi=0$), the
momentum transfer is purely transverse and the spatial and transverse Fourier transform of (\ref{XGX1}) samples the distribution
of an x-parton at a given transverse spatial distance in the light cone,
\be
xg_<(x, b_\perp)=\int \frac{dK_\perp}{(2\pi)^2}\,e^{-iK_\perp\cdot b_\perp}\,xg_<(x, K_\perp)
\ee
with


\begin{widetext}
\bea\label{xgb}
b_\perp^2xg_<(x, b_\perp)\sim  A(0)\frac{2\tilde{\kappa}_N^2b_\perp^2}{\pi x}\bigg(\frac{1-x}{1+x}\bigg)^\tau
\frac{e^{\frac{2\tilde{\kappa}_N^2b_\perp^2}{{\rm ln} x}}}{{\rm ln}^5x}
\bigg(-4\tilde{\kappa}_N^4b_\perp^4+{\rm ln}x\,(-8\tilde{\kappa}_N^2b_\perp^2+{\rm ln}x\,(-2+2\tilde{\kappa}_N^2b_\perp^2+{\rm ln}x))\bigg)
\eea
\end{widetext}
(\ref{xgb}) is seen to spread or diffuse (Gribov diffusion) in the transverse plane over a length scale fixed by
$l_\perp\sim ( 2{\rm ln}(1/x))^{\frac 12}/\tilde\kappa_N$, with

\bea\label{xgb1}
b_\perp^2xg_<(x, b_\perp)\sim  A(0)\frac{8(\tilde{\kappa}_Nb_\perp)^6}{\pi x}\bigg(\frac{1-x}{1+x}\bigg)^\tau
\frac{e^{-\frac{2\tilde{\kappa}_N^2b_\perp^2}{{\rm ln} \frac 1x}}}{{\rm ln}^5\frac 1x}\nonumber\\
\eea
which is enhanced at low-x as $1/(x{\rm ln}^5\frac 1x)$.
The diffusion ceases to be semi-positive for
$b_\perp<l_\perp$ or $K_\perp>1/l_\perp$.
In Fig.~\ref{fig_smallx} we show the behavior of the transverse gluon density (\ref{xgb1})  as probed by the graviton  at
small-x and small $K_\perp$ or large $b_\perp$. The central hole in Fig.~\ref{fig_smallx}b occurs at small $b_\perp<l_\perp$  and falls outside
the range of the  diffusive approximation in (\ref{xgb1}).

To probe large-x and small $b_\perp$ through (\ref{Aj222}),  it is best to remove the large K-factors in the integrand through two
integrations by parts {\bf without modifying}  the sum rule for $A(K)$. The result is

\bea
\label{XGX11}
x{g}_>(x,K)\sim   A(0)x^{a_K+1}\bigg(\bigg(\frac{1-x}{1+x}\bigg)^\tau\bigg)^{\prime\prime}
\eea
with the primes refering to x-derivatives. The corresponding transverse density  at large-x is semi-positive throughout, and reads

\bea
\label{XGX111}
b_\perp^2x{g}_>(x,b_\perp)\sim   A(0)x\bigg(\bigg(\frac{1-x}{1+x}\bigg)^\tau\bigg)^{\prime\prime}\,\frac{2(\tilde \kappa b_\perp)^2}{\pi}
\frac{e^{-\frac{2\tilde{\kappa}_N^2b_\perp^2}{{\rm ln} \frac 1x}}}{{\rm ln}\frac 1x}\nonumber\\
\eea

In ~Fig.~\ref{fig_largex}a we show the large-x behavior of the gluon GPD (\ref{XGX111}) as probed by the graviton,
as  a function of parton-x and the rescaled transverse size $\tilde\kappa_N b_\perp$ for $\tau=3$.
The GPD  distribution for large-x and  fixed $x=0.5$ in the transverse plane is shown in~Fig.~\ref{fig_largex}b. For comparison, one can look at the GPD
of valence quarks in the proton extracted from holographic QCD models in~\cite{Vega:2010ns}.

\begin{figure*}
\subfloat[The small-x gluon density inside the proton (\ref{POM3}) as probed  by the Pomeron
with $\lambda=11.243$, and $\tau=3$.\label{figxgp5}]{%
  \includegraphics[height=6cm,width=.49\linewidth]{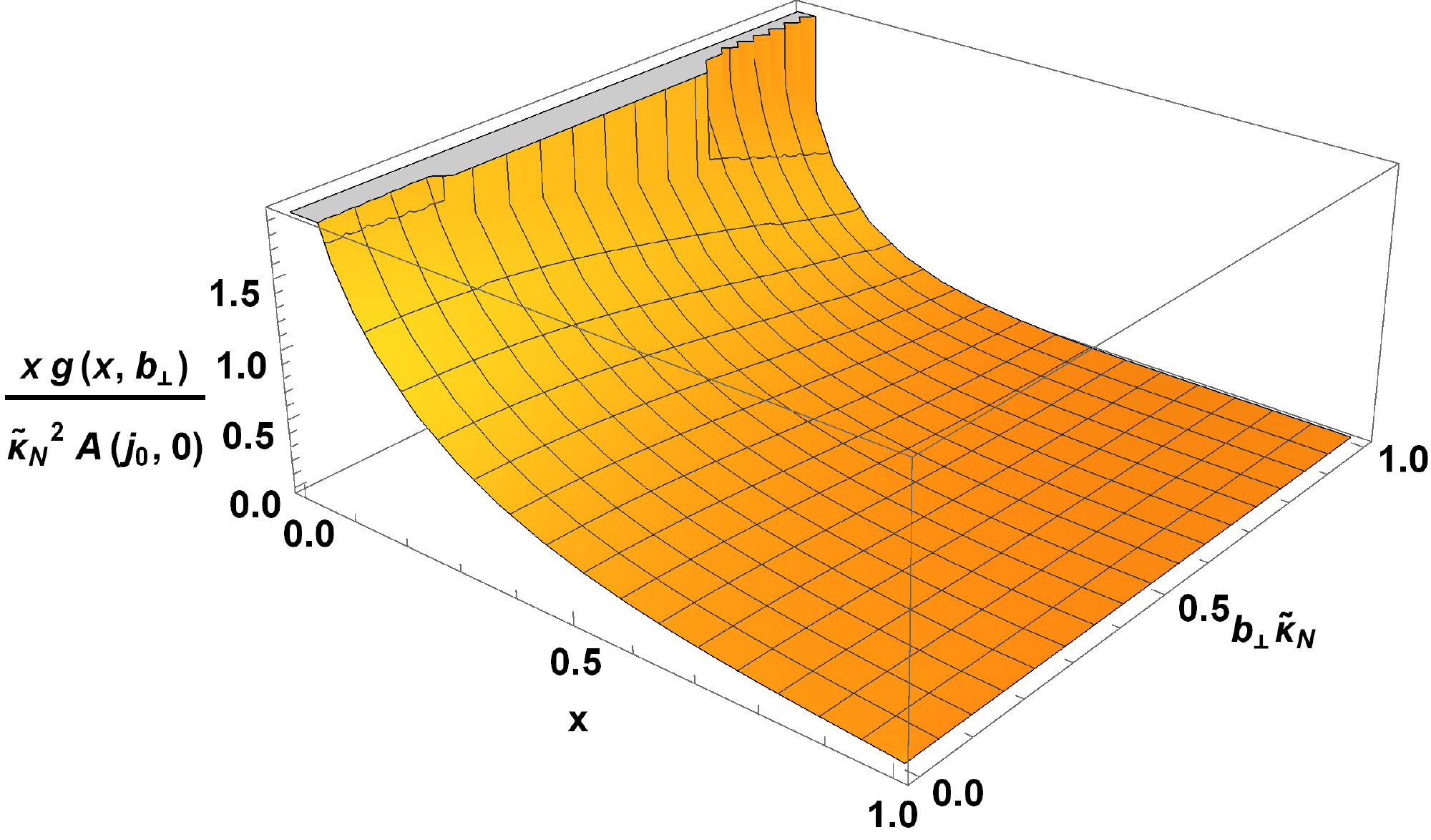}%
}\hfill
\subfloat[The small-x gluon distribution inside the proton (\ref{POM3}) as probed by the Pomeron
with $\lambda=11.243$, $\tau=3$, $x=0.01$ and $b_\perp=({b_x^2+b_y^2})^{\frac 12}$.\label{figxgp6}]{%
  \includegraphics[height=6cm,width=.49\linewidth]{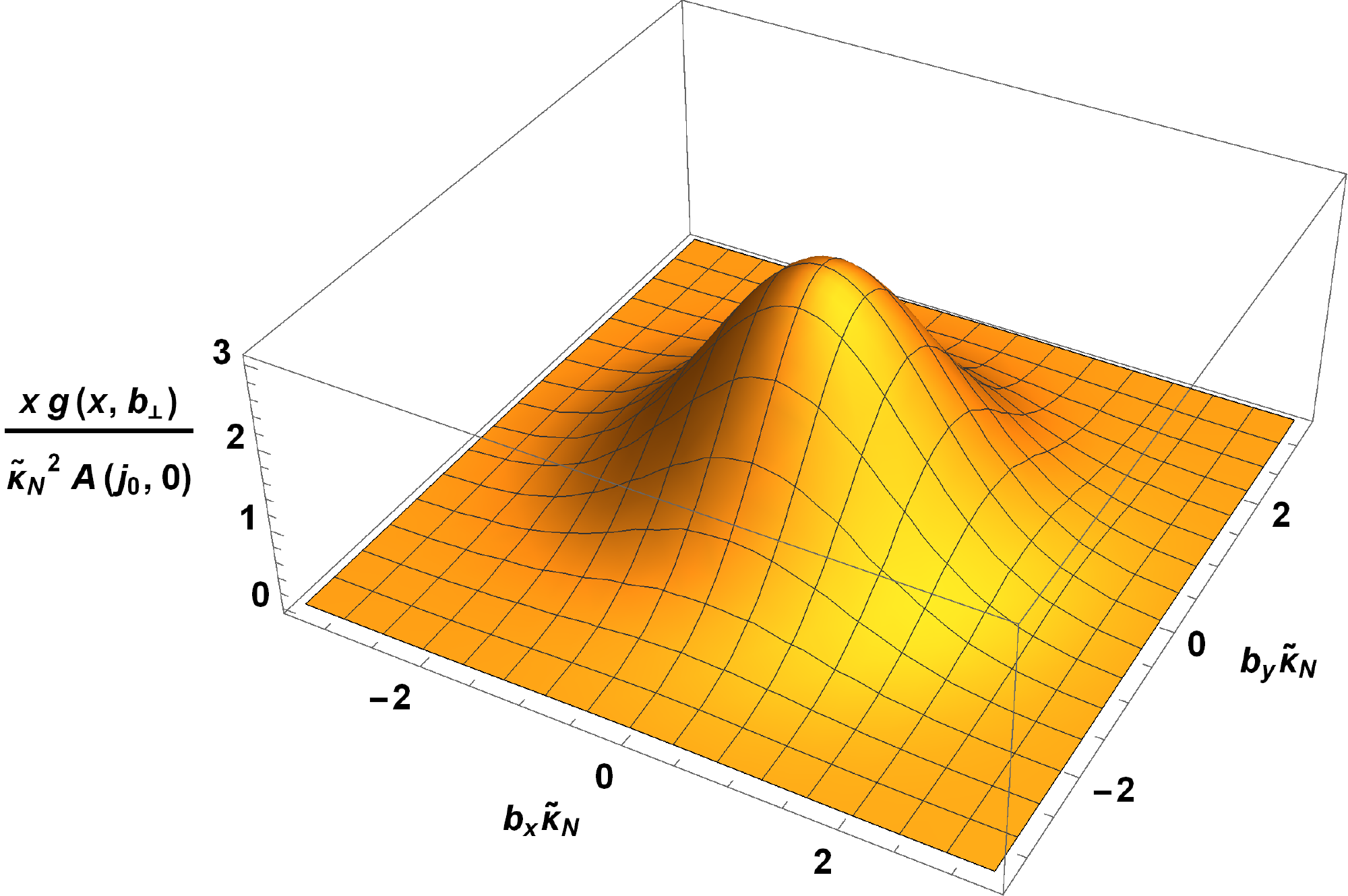}%
}
\caption{Gluon GPD as probed by Pomeron exchange in photoproduction of a heavy meson at high energy.}
\label{fig_pomglue}
\end{figure*}

\subsection{Gluon GPD : $j=j_0$}

Higher spin-j exchanges once resummed yield Pomeron exchange at higher energies. The emerging Pomeron form factor
follows from (\ref{Aj2}) for $j=j_0$ in the form

\begin{widetext}
\bea
\label{POM1}
A_P(K)=&&A(j_0,0)\Gamma(a_K+\Delta(j_0)/2)\times g_5^2\tilde{\kappa}_N^{j_0-2+\Delta(j_0)}A(j_0,K)\nonumber\\
=&&A(j_0,0)\frac{\Gamma(\tau-1/\sqrt{\lambda})}{\Gamma(\tau)}\int_0^1dx \,x^{j_0-1}\,\frac{1}{x^{\alpha_P(t)}}\,
\bigg(\frac{1-x}{1+x}\bigg)^{\tau-\frac 1{\sqrt{\lambda}}}\frac 1{1-x^2}
\bigg((\tau-1)(1+x)+\bigg(\tau-\frac 1{\sqrt{\lambda}}\bigg)(1-x)\bigg)\nonumber\\
\eea
\end{widetext}
with the Pomeron trajectory $\alpha_P(t)=1-2/\sqrt{\lambda}+t/m_0^2$, and 
with $m_0$ fixed by the $2^{++}$ glueball mass in (\ref{SOFTMFG2}). By analogy with the $j=2$ exchange, we suggest that the gluon content of the proton as probed by the Pomeron for small $K$ is concentrated at small-x, and  follows from the dominant Pomeron exchange which is manifest  in
(\ref{POM1}) as

\begin{widetext}
\bea
\label{POM2}
xg(x,K)\sim A(j_0,0)\frac{\Gamma(\tau-1/\sqrt{\lambda})}{\Gamma(\tau)}\,\frac{1}{x^{\alpha_P(t)}}\,
\bigg(\frac{1-x}{1+x}\bigg)^{\tau-\frac 1{\sqrt{\lambda}}}\frac 1{1-x^2}
\bigg((\tau-1)(1+x)+\bigg(\tau-\frac 1{\sqrt{\lambda}}\bigg)(1-x)\bigg)\nonumber\\
\eea
The corresponding transverse gluon density is

\bea
\label{POM3}
b_\perp^2x{g}(x,b_\perp)\sim A(j_0,0)\frac{\Gamma(\tau-1/\sqrt{\lambda})}{\Gamma(\tau)}\,
\bigg(\frac{1-x}{1+x}\bigg)^{\tau-\frac 1{\sqrt{\lambda}}}\frac 1{1-x^2}
\bigg((\tau-1)(1+x)+\bigg(\tau-\frac 1{\sqrt{\lambda}}\bigg)(1-x)\bigg)\,
\frac{2(\tilde \kappa b_\perp)^2}{\pi x^{1-\frac{2}{\sqrt{\lambda}}}}
\frac{e^{-\frac{2\tilde{\kappa}_N^2b_\perp^2}{{\rm ln} \frac 1x}}}{{\rm ln}\frac 1x}\nonumber\\
\eea
\end{widetext}

In Fig.~\ref{fig_pomglue}a we show the behavior of the transverse gluon density probed by the Pomeron in (\ref{POM3}),
for $\lambda=11.243$, $\tau=3$. The same density is shown in  Fig.~\ref{fig_pomglue}b  for fixed $x=0.01$. We note that
the low-x contribution probed by the Pomeron at high energy or equivalently large rapidity $\chi={\rm ln}(s/s_{\rm tr})\gg1$
far from threshold, is substantially larger than the one probed by the graviton close to threshold at small rapidity
$\chi={\rm ln}(s/s_{\rm tr})\sim 1$. Also, we note that at high energy, the transverse gluon density probed by the Pomeron
is diffusive-like throughout.

\subsection{Skewed Gluon GPD}

The gluonic skewed GPD for the energy momentum tensor with finite skewness $\xi=K_z/2\sqrt{m_N^2+K^2/4}$,
 are related to the invariant form factors in (\ref{EMT2})
through (see Eq.~3.127 and Eq.~3.151 in \cite{Belitsky:2005qn}, and references therein)

\bea
&&\int_0^1dx H^g(x, \xi, K)=A(K)+\xi^2 D(K)\rightarrow (1-4\xi^2) A(K)\nonumber\\
&&\int_0^1dx E^g(x, \xi, K)=B(K)-\xi^2 D(K)\rightarrow 4\xi^2 A(K)\nonumber\\
\eea
with the rightmost results following from our holographic results for the invariant form factors,
$B(K)=0$ and $D(K)=4C(K) =-4A(K)$. In terms of (\ref{xgb1}-\ref{XGX111}) (graviton) or (\ref{POM2}) (Pomeron), we have for the skewed gluonic distributions

\bea
H^g(x, \xi, K)=&&(1-4\xi^2)\,xg(x,K)\nonumber\\
E^g(x, \xi, K)=&&4\xi^2\,xg(x,K)
\eea
which amounts to the gluonic contribution to Ji$^\prime$s sum rule~\cite{JI} as

\be
J^{gluon}(0)=\frac 12 \int_0^1 dx (H^g(x,\xi, 0)+E^g(x, \xi, 0))=\frac 12 A(0)\nonumber\\
\ee
As we noted in (\ref{RATIOA0}), the extraction of $A(0)$ from the threshold photoproduction data is model dependent.

\section{Conclusions}

We have analyzed heavy meson photoproduction for all $\sqrt{s}$, using a bottom-up approach holographic construction. We have used the Witten diagrams in AdS$_5$  for diffractive photoproduction of $J/\psi$, shown in Fig.~\ref{wdiagram2}, and explicitly computed the differential cross section for the heavy meson production, first near threshold, 
where it is dominated by the exchange of  massive $2^{++}$ glueballs as spin-2 gravitons in bulk, and second away from threshold,
where the exchange involves a tower of spin-j states that  transmute to the Pomeron.
Our construction is general, and carries readily to heavier meson production such as $\Upsilon$. We have presented
direct predictions for this production near and away from threshold.

Our analysis allowed for the explicit derivation of all three holographic gravitational form factors $A(k), B(k), C(k)$.
In the double limit of a large number of colors and strong coupling, the holographic approach is
dual to quenched QCD, so the form factors are mostly gluonic. Indeed, we have found that the form factors
$A(k)$ and $D(k)=4C(k)$  compare well to the gluonic gravitational form factors from the recent lattice QCD simulations~\cite{MIT}.
The  exception is the form factor $C(k)$ where a strong mixing on the lattice with the
low-lying  scalar-isoscalar was noted.
We have used the $D(k)$ form factor to determined the  distribution of the pressure and shear inside the proton.
The results are comparable to those extracted  recently from the lattice~\cite{Shanahan:2018nnv},
and empirical data  in~\cite{Burkert:2018bqq}.

We have found that the differential cross section for the photoproduction of
heavy meson production,  is solely dependent on the invariant form factor $A(k)$ in our holographic analysis.
The agreement of the differential and total cross sections with the recently
reported GlueX data for $J/\Psi$ production near threshold~\cite{GLUEX}, suggests that the heavy meson production is controlled
by the tensor $2^{++}$ glueball as a graviton exchange in bulk. Indeed, it is the graviton Regge trajectory at low $\sqrt{s}$ that transmutes to the Pomeron Regge trajectory at large $\sqrt{s}$ in holography, thereby providing a unified description of the photoproduction process of heavy mesons at all energies. These results complement those presented originally in~\cite{DJURIC,LEE} away from threshold, and are overall consistent with some of the general observations presented recently 
in~\cite{Hatta:2018ina,Hatta:2019lxo} close to threshold.

From a pertinent integral representation of $A(k)$ in the soft-wall model, we have determined the  GPD of the
gluons in the proton as probed by the spin-2 glueball or graviton near threshold, and the Pomeron way above threshold  in the photoproduction process. The holographic construction clearly shows that the GlueX experiment~\cite{GLUEX} directly probes the tensor gluonic contribution of the energy form factor in the nucleon state as a bulk Dirac fermion.  Conversely, we have used the GlueX data in combination with our holographic cross section result to extract in an almost model independent way the gravitational form factor $A(k)$ modulo $A(0)$, and consequently the gluon GPD of the proton. The value of $A(0)$ as it relates to the gluonic contribution to the proton spin is model dependent, and cannot be reliably extracted from the threshold data in the photoproduction process. Our observations are overall consistent with the original arguments presented in~\cite{BRODSKY} using short distance QCD methods.

The forthcoming high statistics measurements from SoLID~\cite{MEZIANI}  will provide further insights and
checks on the present holographic analysis.

\section{Acknowledgements}

We thank Zein-Eddine Meziani for discussions.
This work was supported by the U.S. Department of Energy under Contract No.
DE-FG-88ER40388.

\section{Appendix: wavefunctions and propagators in holographic QCD}

\subsection{Dirac fermion/proton}

The normalized wavefunctions for the bulk Dirac fermion are~\cite{CARLSON}

\bea
\Psi(p,z)&=&\psi_R(z)\Psi^0_{R}(p)+ \psi_L(z)\Psi^0_{L}(p)\,,\nonumber\\
\bar\Psi(p,z)&=&\psi_R(z)\bar\Psi^0_{R}(p)+ \psi_L(z)\bar\Psi^0_{L}(p)\,,\nonumber\\
\eea
where for the  {\bf hard-wall}
\bea
\psi_R(z)&=&\frac{\sqrt{2} z^{5/2} J_{\tau-2}(m_N z)}{z_0 J_{\tau-1}(m_N z_0)}\,, \nonumber\\
\psi_L(z)&=&\frac{\sqrt{2} z^{5/2} J_{\tau-1}(m_N z)}{z_0 J_{\tau-1}(m_N z_0)}\,,
\eea
with the Bessel functions $J_{\alpha}(m_N z)$, and for the {\bf soft-wall}

\be
&&\psi_R(z)=\frac{\tilde{n}_R}{\tilde{\kappa}_{N}^{\tau-2}} z^{\frac{5}{2}}\xi^{\frac{\tau-2}{2}}L_0^{(\tau-2)}(\xi)\,,\nonumber\\
&&\psi_L(z)=\frac{\tilde{n}_L}{\tilde{\kappa}_{N}^{\tau-1}} z^{\frac{5}{2}}\xi^{\frac{\tau-1}{2}}L_0^{(\tau-1)}(\xi)\,,\nonumber\\
\ee
with the generalized Laguerre polynomials $L_n^{(\alpha)}(\xi)$, $\tilde{n}_R=\tilde{n}_L \tilde{\kappa}_{N}^{-1}\sqrt{\tau-1}$, and $\tilde{n}_L=\tilde{\kappa}_{N}^{\tau}\sqrt{{2}/{\Gamma(\tau)}}$. The bulk wave functions are normalized for the {\bf hard-wall} as
\be
\int_{0}^{z_0} dz\,\sqrt{g}\,e^{\mu}_{a}\,\psi_{R/L}^2(z)=\delta^{\mu}_a\,,\nonumber\\
\ee
and for the {\bf soft-wall} as
\be
\int_{0}^{\infty} dz\,\sqrt{g}\,e^{-\phi}\,e^{\mu}_{a}\,\psi_{R/L}^2(z)=\delta^{\mu}_a\,,\nonumber\\
\ee
with $\phi=\tilde{\kappa}_{N}^2z^2$, and the inverse vielbein $e^{\mu}_{a}=\sqrt{\abs{g^{\mu\mu}}}\delta^{\mu}_a$ (no summation intended in $\mu$).

For both the hard-wall and soft-wall models, we have the twist parameter $\tau=3$, $\Psi^0_{R/L}(p)= P_{\pm}u(p)$, $\bar\Psi^0_{R/L}(p)=\bar u(p)P_{\mp}$, and $P_{\pm}=(1/2)(1\pm \gamma^5)$. We also work with the normalizations of the boundary constant spinors for both the hard-wall and soft-wall models as
\be
&&\bar u(p)u(p)=2m_N\,,\nonumber\\
&&2m_N\times\bar u(p')\gamma^{\mu}u(p)=\bar u(p')(p'+p)^{\mu}u(p)\,.\nonumber\\
\ee

\subsection{Photon/spin-1 mesons}

\subsubsection{Hard wall}

 For time-like momenta ($q^2>0$), the non-normalizable wave function for the virtual photon is generally given by $A_{\mu}=V(q,z)\,n_{\mu}e^{-iq\cdot x}$ where \cite{Grigoryan:2007vg, Abidin:2008ku}

\be
V(q,z)=-g_5\sum_n \frac{F_n\phi_n(z)}{q^2-m_n^2} , \label{vpt1}
\ee
with $V(0,z)=V(q,0)=1$, the decay constant of the vector mesons $F_n=(1/g_5)(-\frac{1}{z'}\partial_{z'}\phi_n(z'))|_{z'=\epsilon}$, and the normalized wave functions of the vector mesons $A_{\mu}=\phi_n(z)\,n_{\mu} e^{-q\cdot x}$
\be
\phi_n(z)=c_{n}z J_1(m_n z)\equiv J_{A}(m_n,z)  \,,
\ee
with $c_{n}=\frac{\sqrt{2}}{z_0 J_1(m_n z_0)}$, which satisfy the normalization condition

\be
\int dz\,\sqrt{g}\,(g^{xx})^2\,\phi_n(z)\phi_m(z)=\delta_{nm}\,.\nonumber\\
\ee

In the hard-wall model, the summation in (\ref{vpt1}) can be carried out analytically and is given by

\bea
V(q,z)
= \frac{\pi}{2}zq\left(\frac{Y_0(qz_0)}{J_0(qz_0)}J_1(qz)-Y_1(qz)\right)\,.\nonumber\\
\label{vpt2}
\ee

For space-like momenta ($q^2=-Q^2$), the non-normalizable wave function for the virtual photon is generally given by $A_{\mu}=\mathcal{V}(Q,z)\,n_{\mu}e^{-q\cdot x}$ where
\be
\mathcal{V}(Q,z)=g_5\sum_n \frac{F_n\phi_n(z)}{Q^2+m_n^2}\,. \label{vps1}
\ee
For example, in the hard-wall model, the summation in (\ref{vps1}) can be carried out analytically and is given by

\bea
\mathcal{V}(Q,z)
= Qz\left(\frac{K_0(Qz_0)}{I_0(Qz_0)}I_1(Qz)+K_1(Qz)\right)\,,\nonumber\\
\label{vps2}
\ee
with the normalization ${\cal V}(0,z)={\cal V}(Q,0)=1$.

The bulk-to-bulk propagator for the massive mesons, for time-like momenta ($q^2>0$), can be written as

\be
G_{\mu\nu}(z,z^\prime)={\cal T}_{\mu\nu}G(z,z^\prime)=\bigg(-\eta_{\mu\nu} + \frac{k_{\mu} k_{\nu}}{m_n^2}\bigg)G(z, z^{\prime})\,,\nonumber\\
\ee
with

\be
G(z,z')=\sum_n \frac{\phi_n(z)\phi_n(z')}{q^2-m_n^2} . \label{vbbt1}
\ee
For space-like momenta  $q^2=-Q^2$ in (\ref{vbbt1}). Also recall that ${\cal V}(q,z)=\frac{1}{z'}\partial_{z'}G(z,z')|_{z'=\epsilon}$\,.
Note that for $z\rightarrow 0$, we can write (\ref{vbbt1}) as
\begin{widetext}
\be
G(z\rightarrow 0,z')&&\approx \frac{\phi_n(z\rightarrow 0)}{-g_5F_n}\sum_n \frac{-g_5F_n\phi_n(z')}{q^2-m_n^2}=\frac{z^2}{2}\sum_n \frac{-g_5F_n\phi_n(z')}{q^2-m_n^2}=\frac{z^2}{2}V(q,z') , \nonumber\\\label{vbbt2}
\ee
\end{widetext}
where we used

\be
F_n=(1/g_5)(-\frac{1}{z'}\partial_{z'}\phi_n(z'))|_{z'=\epsilon}=-\frac{1}{g_5}c_nm_n\,,\nonumber\\
\ee
and $\phi_n(z\rightarrow 0)\approx \frac{1}{2}c_nm_nz^2$ for the hard-wall. Defining the decay constant as $f_n=-\frac{F_n}{m_n}$,  we have
\be
\phi_n(z)=\frac{f_n}{m_n}\times g_5m_{n}z J_1(m_n z)\,,
\ee
as required by vector meson dominance (VMD).
For space-like momenta ($q^2=-Q^2$), we have

\be
G(z\rightarrow 0,z')\approx \frac{z^2}{2}\sum_n \frac{g_5F_n\phi_n(z')}{Q^2+m_n^2}=\frac{z^2}{2}\mathcal{V}(Q,z') . \nonumber\\\label{vbbt2}
\ee

\subsubsection{Soft wall}

 Similar relationships hold for the soft-wall model where the normalized wave function for vector mesons is given by \cite{Grigoryan:2007my}

\be
\phi_n(z)=c_{n}\tilde{\kappa}_V^2z^2 L_n^1( \tilde{\kappa}_V^2z^2)\equiv J_{A}(m_n,z)  \,,\nonumber\\
\ee
with $c_{n}=\sqrt{{2}/{n+1}}$ which is determined from the normalization condition (for the soft-wall model with background dilaton $\phi=\tilde{\kappa}_V^2z^2$)
\be
\int dz\,\sqrt{g}e^{-\phi}\,(g^{xx})^2\,\phi_n(z)\phi_m(z)=\delta_{nm}\,.\nonumber\\
\ee
Therefore, we have

\be
F_n=\frac 1{g_5}\bigg(-e^{-\phi}\frac{1}{z^\prime}\partial_{z^\prime}\phi_n(z^\prime)\bigg)_{z^\prime=\epsilon}=-\frac{2}{g_5}c_n(n+1)\tilde{\kappa}_V^2\,,\nonumber\\
\ee
with $\phi_n(z\rightarrow 0)\approx c_n\tilde{\kappa}_V^2z^2(n+1)$. If we define the decay constant as $f_n=-{F_n}/{m_n}$,  we have

\be
\phi_n(z)=\frac{f_n}{m_n}\times 2g_{5}\tilde{\kappa}_V^2z^2 L_n^1( \tilde{\kappa}_V^2z^2)\,,
\ee
as required by vector meson dominance (VMD).

Note that for $z\rightarrow 0$, we can write the bulk-to-bulk propagator (\ref{vbbt1}) as

\begin{widetext}
\be
G(z\rightarrow 0,z')&&\approx \frac{\phi_n(z\rightarrow 0)}{-g_5F_n}\sum_n \frac{-g_5F_n\phi_n(z')}{q^2-m_n^2}=\frac{z^2}{2}\sum_n \frac{-g_5F_n\phi_n(z')}{q^2-m_n^2}=\frac{z^2}{2}V(q,z') . \nonumber\\\label{vbbt2sw}
\ee
\end{widetext}
For space-like momenta ($q^2=-Q^2$), we have the bulk-to-bulk propagator near the boundary

\be
G(z\rightarrow 0,z')\approx \frac{z^2}{2}\sum_n \frac{g_5F_n\phi_n(z')}{Q^2+m_n^2}=\frac{z^2}{2}\mathcal{V}(Q,z')\,, \nonumber\\\label{vbbt2sw}
\ee
where~\cite{Grigoryan:2007my}

\bea
\mathcal{V}(Q,z)
=\kappa_V^2z^2\int_{0}^{1}\frac{dx}{(1-x)^2}x^a{\rm exp}\Big[-\frac{x}{1-x}\kappa_V^2z^2\Big]\,,\nonumber\\
\label{vps2sw}
\ee
with the normalization ${\cal V}(0,z)={\cal V}(Q,0)=1$.

\subsection{Tansverse-traceless graviton/spin-2 glueballs}

\subsubsection{Hard wall}

 For time-like momenta ($k^2>0$), the non-normalizable wave function for the virtual tansverse-traceless graviton is generally given by $h_{\mu\nu}=h(k,z)\epsilon_{\mu\nu}^{TT}e^{-ik\cdot x}$ where \cite{Hong:2004sa}
\be
h(k,z)=-\sqrt{2}{\kappa}\sum_n \frac{F_n\psi_n(z)}{k^2-m_n^2} , \label{hpt1}
\ee
with the normalization $h(0,z)=h(k,0)=1$, which could be relaxed.
The decay constant of the spin-2 glueball with mass $m_n$  is

\be
F_n=\frac{1}{\sqrt{2}\kappa_V}\bigg(-\frac{1}{z^{\prime 3}}\partial_{z^\prime}\psi_n(z^\prime)\bigg)_{z^\prime=\epsilon}
\ee
 and the normalized wave functions of the spin-2 glueballs $h_{\mu\nu}=\psi_n(z)\,\epsilon_{\mu\nu}^{TT} e^{-ik\cdot x}$

\be \label{jh}
\psi_n(z)=c_{n}z^2 J_2(m_n z)\equiv J_{h}(m_n,z) \,,
\ee
with $c_{n}=\frac{\sqrt{2}}{z_0 J_2(m_n z_0)}$, which satisfy the normalization condition

\be
\int dz\,\sqrt{g}\,\abs{g^{xx}}\,\psi_n(z)\psi_m(z)=\delta_{nm}\,.\nonumber\\
\ee

In the hard-wall model, the summation in (\ref{hpt1}) can be carried out analytically and is given by\cite{CARLSON,Abidin:2008ku,Hong:2004sa}

\bea
h(k,z)
=\frac{\pi}{4}k^2z^2\left(\frac{Y_1(kz_0)}{J_1(kz_0)}J_2(kz)-Y_2(kz)\right)\,.\nonumber\\
\label{hpt2}
\ee

For space-like momenta ($k^2=-K^2$), the non-normalizable wave function for the virtual transverse-traceless graviton is generally given by
$h_{\mu\nu}=\mathcal{H}(K,z)\,\epsilon_{\mu\nu}^{TT}e^{-ik\cdot x}$ where

\be
\mathcal{H}(K,z)=\sqrt{2}\kappa\sum_n \frac{F_n\psi_n(z)}{K^2+m_n^2}\,. \label{hps1}
\ee
In the hard-wall model, the summation in (\ref{hps1}) can be carried out analytically and is given by\cite{CARLSON,Abidin:2008ku,Hong:2004sa}

\bea
\mathcal{H}(K,z)
= \frac{1}{2}K^2z^2\left(\frac{K_1(Kz_0)}{I_1(Kz_0)}I_2(Kz)+K_2(Kz)\right)\,.\nonumber\\
\label{hps2}
\ee

For time-like momenta ($q^2>0$), the bulk-to-bulk propagator for the massive spin-2 glueballs, can be written as \cite{CARLSON,Abidin:2008ku,Hong:2004sa}

\be
G_{\mu\nu\alpha\beta}^{TT}(z,z')={1 \over 2} \left({\cal T}_{\mu\alpha} {\cal T}_{\nu\beta} + {\cal T}_{\mu\beta} {\cal T}_{\nu\alpha} -
\frac 23 {\cal T}_{\mu\nu} {\cal T}_{\alpha\beta}\right)G(z,z^\prime)\,,\nonumber\\
\ee
with ${\cal T}_{\mu\nu} = -\eta_{\mu\nu} + k_{\mu} k_{\nu}/m_n^2$ and
\be
G(z,z')=\sum_n \frac{\psi_n(z)\psi_n(z')}{k^2-m_n^2} . \label{hbbt1}
\ee
For space-like momenta, we simply replace $k^2=-K^2$ in (\ref{hbbt1}). Also remember that

\be
h(k,z)=\frac{1}{z'^3}\partial_{z'}G(z,z')|_{z'=\epsilon}\,.
\ee

Note that for $z\rightarrow 0$, we can write (\ref{hbbt1}) as

\be
G(z\rightarrow 0,z')\approx \frac{z^4}{4}\sum_n \frac{-\sqrt{2}\kappa F_n\psi_n(z')}{k^2-m_n^2}=\frac{z^4}{4}h(k,z') , \nonumber\\\label{hbbt2}
\ee
where we used

\be
F_n=\frac{1}{\sqrt{2}\kappa}\bigg(-\frac{1}{z^{\prime 3}}\partial_{z^\prime}\psi_n(z^\prime)\bigg)_{z^\prime=\epsilon}=-\frac{1}{2\sqrt{2}\kappa}c_nm_n^2\,,\nonumber\\
\ee
and $\psi_n(z\rightarrow 0)\approx \frac{1}{8}c_nm_n^2z^4$ for the hard-wall. Hence, for space-like momenta ($k^2=-K^2$), we have

\be
G(z\rightarrow 0,z')\approx \frac{z^4}{4}\sum_n \frac{\sqrt{2}\kappa_VF_n\phi_n(z')}{K^2+m_n^2}=\frac{z^4}{4}\mathcal{H}(K,z') . \nonumber\\\label{hbbt3}
\ee

\subsubsection{Soft wall}

Similar relationships hold for the soft-wall model where the normalized wave function for spin-2 glueballs is given by \cite{BallonBayona:2007qr} (note that the discussion in \cite{BallonBayona:2007qr} is for general massive bulk scalar fluctuation but can be used for spin-2 glueball which has an effective bulk action similar to massless bulk scalar fluctuation)
\bea
 \label{wfSW}
&&\psi_{n}(z)=c_n\,z^{4}L_{n}^{\Delta(j)-2}(2\xi)\,,\nonumber\\
\eea
with
\be
c_n=\Bigg(\frac{2^{4}\tilde{\kappa}_{N}^{6}\Gamma(n+1)}{\Gamma(n+3)}\Bigg)^{\frac 12}\,,
\ee
which is determined from the normalization condition (for soft-wall model with background dilaton $\phi=\tilde{\kappa}_N^2z^2$)
\be
\int dz\,\sqrt{g}e^{-\phi}\,\abs{g^{xx}}\,\psi_n(z)\psi_m(z)=\delta_{nm}\,.\nonumber\\
\ee
Therefore we have

\be
F_n=\frac{1}{\sqrt{2}\kappa}\bigg(-\frac{1}{z^{\prime 3}}\partial_{z^\prime}\psi_n(z^\prime)\bigg)_{z^\prime=\epsilon}=-\frac{4}{\sqrt{2}\kappa}c_n L_{n}^{2}(0)\,,\nonumber\\
\ee
with $\psi_n(z\rightarrow 0)\approx c_n\,z^{4}L_{n}^{2}(0)$. 
For space-like momenta ($q^2=-Q^2$), we have the bulk-to-bulk propagator near the boundary
\be
G(z\rightarrow 0,z')\approx \frac{z^4}{4}\sum_n \frac{\sqrt{2}\kappa F_n\phi_n(z')}{K^2+m_n^2}=\frac{z^4}{4}\mathcal{H}(K,z') , \nonumber\\\label{hbbt3SW2}
\ee
where, for the soft-wall model,~\cite{CARLSON,Abidin:2008ku,BallonBayona:2007qr}

\begin{widetext}
\bea
\mathcal{H}(K,z)
&&=4z^{4}\Gamma(a_K +2)U\Big(a_K+2,3;2\xi\Big)
=\Gamma(a_K+2)U\Big(a_K,-1;2\xi\Big)\nonumber\\
&&=\frac{\Gamma(a_K+2)}{\Gamma(a_K)}
\int_{0}^{1}dx\,x^{a_K-1}(1-x){\rm exp}\Big(-\frac{x}{1-x}(2\xi)\Big)\,,
\label{BBSWj2}
\eea
\end{widetext}
with $a_K={a}/{2}={K^2}/{8\tilde{\kappa}_N^2}$, and we have used the transformation $U(m,n;y)=y^{1-n}U(1+m-n,2-n,y)$. (\ref{BBSWj2}) satisfies the normalization condition ${\cal H}(0,z)={\cal H}(K,0)=1$.

\subsection{Trace-full graviton/spin-0 glueballs}

\subsubsection{Hard wall}

For time-like momenta ($k^2>0$), the non-normalizable wave function for the virtual trace-full graviton is generally given by
$h_{\mu\nu}=k^2 f(k,z)\epsilon_{\mu\nu}^{T}e^{-ik\cdot x}$ where

\be
f(k,z)=2\sqrt{2}\kappa\sum_n \frac{F_n\psi_n(z)}{k^2-m_n^2} , \label{fpt1}
\ee
with $f(0,z)=f(k,0)=1$, the decay constant of the spin-0 glueballs $F_n=\frac{1}{2\sqrt{2}\kappa}(\frac{1}{z'^3}\partial_{z'}\psi_n(z'))|_{z'=\epsilon}$, and the normalized wave functions of the spin-0 glueballs $h_{\mu\nu}=\psi_n(z)\,\epsilon_{\mu\nu}^{T} e^{-k\cdot x}$ which satisfy the normalization condition
\be
\int dz\,\sqrt{g}\,\abs{g^{xx}}\,\psi_n(z)\psi_m(z)=\delta_{nm}\,,\nonumber\\
\ee
with the normalized wave functions for the spin-0 glueballs
\be \label{jf}
\psi_n(z)=c_{n}z^2 J_2(m_n z)\equiv J_{f}(m_n,z)  \,,
\ee
where $c_{n}=\frac{\sqrt{2}}{z_0 J_2(m_n z_0)}$. In the hard-wall model, the summation in (\ref{fpt1}) can be carried out analytically and is given by

\bea
f(k,z)
= \frac{\pi}{4}k^2z^2\left(\frac{Y_1(kz_0)}{J_1(kz_0)}J_2(kz)-Y_2(kz)\right)\,.\nonumber\\
\label{fpt2}
\ee

For space-like momenta ($k^2=-K^2$), the non-normalizable wave function for the virtual trace-full graviton is generally given by $h_{\mu\nu}=\mathcal{F}(K,z)\,\epsilon_{\mu\nu}^{T}e^{-k\cdot x}$ where

\be
\mathcal{F}(K,z)=-2\sqrt{2}\kappa\sum_n \frac{F_n\psi_n(z)}{K^2+m_n^2}\,. \label{fps1}
\ee
The summation in (\ref{fps1}) can be carried out analytically and is given by

\bea
\mathcal{F}(K,z)
= \frac{1}{2}K^2z^2\left(\frac{K_1(Kz_0)}{I_1(Kz_0)}I_2(Kz)+K_2(Kz)\right)\,.\nonumber\\
\label{fps2}
\ee

For time-like momenta ($q^2>0$), the bulk-to-bulk propagator for the massive spin-0 glueballs, can be written as $G_{\mu\nu\alpha\beta}^{T}(z,z')=\eta_{\mu\nu}\eta_{\alpha\beta}G(z,z')$ where
\be
G(z,z')=\sum_n \frac{\psi_n(z)\psi_n(z')}{k^2-m_n^2} . \label{fbbt1}
\ee
with

\be
f(k,z)=\bigg(\frac{1}{z^{\prime 3}}\partial_{z^\prime}G(z,z^\prime)\bigg)_{z^\prime=\epsilon}\,.
\ee

Note that for $z\rightarrow 0$, we can write (\ref{fbbt1}) as
\be
G(z\rightarrow 0,z')\approx \frac{z^4}{4}\sum_n \frac{2\sqrt{2}\kappa F_n\psi_n(z')}{k^2-m_n^2}=\frac{z^4}{4}f(k,z') , \nonumber\\\label{fbbt2}
\ee
where we used
\be
F_n=\frac{1}{2\sqrt{2}\kappa}\bigg(\frac{1}{z^{\prime 3}}\partial_{z^\prime}G(z,z^\prime)\bigg)_{z^\prime=\epsilon}=\frac{1}{4\sqrt{2}\kappa}c_nm_n^2\,,\nonumber\\
\ee
and $\psi_n(z\rightarrow 0)\approx \frac{1}{8}c_nm_n^2z^4$ for the hard-wall. Hence, for space-like momenta ($k^2=-K^2$), we have
\be
G(z\rightarrow 0,z')\approx \frac{z^4}{4}\sum_n \frac{-2\sqrt{2}\kappa F_n\phi_n(z')}{K^2+m_n^2}=\frac{z^4}{4}\mathcal{F}(K,z') . \nonumber\\\label{fbbt3}
\ee

\subsubsection{Soft wall}

 Note that similar relationships hold for the trace-full graviton/spin-0 glueball in the soft-wall model.
 We do not detail them here as they are similar to the ones  given for the spin-2 glueballs  modilo  normalization constants.

\subsection{Dilaton/spin-0 glueballs}

\subsubsection{Hard wall}

 For time-like momenta ($k^2>0$), the non-normalizable wave function for the virtual dilaton is generally given by
\be
\varphi(k,z)=\sqrt{2}\kappa\sum_n \frac{F_n\psi_n(z)}{k^2-m_n^2} , \label{dpt1}
\ee
with $\varphi(0,z)=\varphi(k,0)=1$, the decay constant of the spin-0 glueballs $F_n=\frac{1}{\sqrt{2}\kappa}(\frac{1}{z'^3}\partial_{z'}\psi_n(z'))|_{z'=\epsilon}$, and the normalized wave functions of the spin-0 glueballs $\psi_n(z)$ which satisfy the normalization condition
\be
\int dz\,\sqrt{g}\,\abs{g^{xx}}\,\psi_n(z)\psi_m(z)=\delta_{nm}\,,\nonumber\\
\ee
with the normalized wave functions for the spin-0 glueballs
\be \label{jpsi}
\psi_n(z)=c_{n}z^2 J_2(m_n z)\equiv J_{\varphi}(m_n,z)  \,,
\ee
where $c_{n}=\frac{\sqrt{2}}{z_0 J_2(m_n z_0)}$.
For example, in the hard-wall model, the summation in (\ref{dpt1}) can be carried out analytically and is given by
\bea
\varphi(k,z)
=\frac{\pi}{4}k^2z^2\left(\frac{Y_1(kz_0)}{J_1(kz_0)}J_2(kz)-Y_2(kz)\right)\,.\nonumber\\
\label{dpt2}
\ee

For space-like momenta ($k^2=-K^2$), the non-normalizable wave function for the virtual dilaton is generally given by
\be
\mathcal{D}(K,z)=-\sqrt{2}{\kappa}\sum_n \frac{F_n\psi_n(z)}{K^2+m_n^2}\,. \label{dps1}
\ee
For example, in the hard-wall model, the summation in (\ref{dps1}) can be carried out analytically and is given by

\bea
\mathcal{D}(K,z)
= \frac{1}{2}K^2z^2\left(\frac{K_1(Kz_0)}{I_1(Kz_0)}I_2(Kz)+K_2(Kz)\right)\,.\nonumber\\
\label{dps2}
\ee

For time-like momenta ($q^2>0$), the bulk-to-bulk propagator for the massive spin-0 glueballs, can be written as
\be
G(z,z')=\sum_n \frac{\psi_n(z)\psi_n(z')}{k^2-m_n^2} . \label{dbbt1}
\ee
We recall  that $\varphi(q,z)=\frac{1}{z'^3}\partial_{z'}G(z,z')|_{z'=\epsilon}$\,.

Note that for $z\rightarrow 0$, we can write (\ref{dbbt1}) as
\be
G(z\rightarrow 0,z')\approx \frac{z^4}{4}\sum_n \frac{\sqrt{2}\kappa F_n\psi_n(z')}{k^2-m_n^2}=\frac{z^4}{4}\varphi(k,z') , \nonumber\\\label{dbbt2}
\ee
where we used
\be
F_n=\frac{1}{2\sqrt{2}\kappa}\bigg(\frac{1}{z^{\prime 3}}\partial_{z^\prime}G(z,z^\prime)\bigg)_{z^\prime=\epsilon}=\frac{1}{4\sqrt{2}\kappa}c_nm_n^2\,,\nonumber\\
\ee
and $\psi_n(z\rightarrow 0)\approx \frac{1}{8}c_nm_n^2z^4$ for the hard-wall. Hence, for space-like momenta ($k^2=-K^2$), we have
\be
G(z\rightarrow 0,z')\approx \frac{z^4}{4}\sum_n \frac{-2\sqrt{2}\kappa F_n\phi_n(z')}{K^2+m_n^2}=\frac{z^4}{4}\mathcal{D}(K,z') . \nonumber\\\label{dbbt3}
\ee

\subsubsection{Soft wall}

Note again, that similar relationships hold for the dilaton/spin-0 glueballs in the soft-wall model, but we do not go into details here as it is very similar to the spin-2 glueballs upto normalization constants.

\section{Appendix: Contributions to holographic photoproduction}

Here most of the results will be given for the soft wall model explicitly.
The results for the hard wall model follow by setting $\phi=0$.

\subsection{Dilaton contribution}


The dilaton contribution  to the holographic photoproduction amplitude can be determined from
Fig.~\ref{wdiagram2} by replacing the spin-2 glueball propagator by spin-0 glueball propagator of dilaton as

\begin{widetext}
\bea
\label{AmpD}
&&i{\cal A}^{\varphi}_{A p\rightarrow  A p} (s,t)=\sum_n i\tilde{{\cal A}}^{\varphi}_{Ap\rightarrow A p} (m_n,s,t)\nonumber\\
&&i\tilde{{\cal A}}^{\varphi}_{ A p\rightarrow Ap} (m_n,s,t)=(-i)V_{\varphi AA}(q_1,q_2,k,m_n)\times \tilde{G}_{\varphi}(m_n,t)\times (-i)V_{\varphi\bar\Psi\Psi}(p_1,p_2,k,m_n)\,,\nonumber\\
\eea
\end{widetext}
with the bulk vertices  ($k=p_2-p_1=q_1-q_2$)

\begin{widetext}
\be
 \label{DAA}
V_{\varphi AA}(q, q', k, m_n)\equiv &&\left(\frac{\delta S^k_{\varphi AA}}{\delta \varphi(k,z)}\right)\,J_{\varphi}(m_n,z)=\sqrt{2\kappa^2}\times\frac{1}{4}\int dz\sqrt{g}\,e^{-\phi}z^4K(q,q',n,n',z)J_{\varphi}(m_n,z) \,,\nonumber\\
V_{\varphi\bar\Psi\Psi}( p_1, p_2, k, m_n)\equiv && \left(\frac{\delta S^k_{\varphi\bar\Psi\Psi}}{\delta (\partial_z\varphi(k,z))}\right)\,\partial_z J_{\varphi}(m_n,z)+\left(\frac{\delta S^k_{\varphi\bar\Psi\Psi}}{\delta (\varphi(k,z))}\right)\,J_{\varphi}(m_n,z)\nonumber\\
=&&\frac{\sqrt{2\kappa^2}}{2}\int dz\,\sqrt{g}\,e^{-\phi}z\,\bar\Psi(p_2,z)\Big(\gamma^5\,\partial_zJ_{\varphi}(m_n,z)+k_{\alpha}\gamma^{\alpha}\,J_{\varphi}(m_n,z)\Big)\Psi(p_1,z)\,,\nonumber\\
\ee
\end{widetext}
and the bulk-to-bulk propagator

\bea
\label{Gphi}
G_{\varphi}(m_n,t,z,z^\prime)=&& J_{\varphi}(m_n,z)\tilde{G}_{\varphi}(m_n,t)J_{\varphi}(m_n,z^\prime)\,,\nonumber\\
\tilde{G}_{\varphi}(m_n,t)=&&\frac{i}{t-m_n^2+i\epsilon}\,,
\eea
For $z'\rightarrow 0$, and $t=-K^2$ in (\ref{Gphi}), we can use (\ref{dbbt3}), which simplifies (\ref{AmpD}) as
\begin{widetext}
\be
\label{nAmpD}
&&i{\cal A}^{\varphi}_{Ap\rightarrow A p} (s,t)\approx
(-i)\mathcal{V}_{\varphi AA}(q_1,q_2,k)\times (i)\times(-i)\mathcal{V}_{\varphi\bar\Psi\Psi}(p_1,p_2,k)\,,	\nonumber\\
&&\mathcal{V}_{\varphi AA}(q_1,q_2,k)=\sqrt{2\kappa^2}\times\sqrt\frac{1}{4}\int dz\sqrt{g}\,e^{-\phi}z^4K(q,q',n,n',z)\frac{z^4}{4}\nonumber\\
&&\mathcal{V}_{\varphi\bar\Psi\Psi}(p_1,p_2,k)=\sqrt{2\kappa^2}\times\frac{1}{2}\int dz\,e^{-\phi}\sqrt{g}\,z\,\bar\Psi(p_2,z)\Big(\gamma^5\,\partial_z \mathcal{D}(K,z)
+k_{\alpha}\gamma^{\alpha}\,\mathcal{D}(K,z)\Big)\Psi(p_1,z)\,.\nonumber\\
\ee
\end{widetext}

\subsection{Graviton contribution}

The graviton contribution in Fig.~\ref{wdiagram2} in the diffractive part of the holographic photoproduction amplitude
was analyzed in~\cite{GAO}  for the Pomeron kinematics in the hard wall model. Here we will
give the results for all kinematics  for both the CFT case in AdS, and the conformally broken
case in walled AdS.

In AdS space, for tansverse-traceless part, Witten$^\prime$s diagrammatic rules give formally

\begin{widetext}
\be
\label{Amph}
&&i{\cal A}^{h}_{Ap\rightarrow  Ap} (s,t)=\sum_n i\tilde{{\cal A}}^{h}_{Ap\rightarrow A p} (m_n,s,t)\nonumber\\
&&i\tilde{{\cal A}}^{h}_{A\rightarrow A p} (m_n,s,t)=(-i)V_{hAA}^{\mu\nu(TT)}(q,q',k,m_n)\times \tilde{G}^{TT}_{\mu\nu\alpha\beta}(m_n,t)\times
 (-i)V_{h\bar\Psi\Psi}^{\alpha\beta(TT)}(p_1,p_2,k,m_n)\,,	\nonumber\\
&&i{\cal A}^{f}_{Ap \rightarrow A p} (s,t)=\sum_n i\tilde{{\cal A}}^{f}_{ Ap\rightarrow A p} (m_n,s,t)\nonumber\\
&&i\tilde{{\cal A}}^{f}_{Ap\rightarrow Ap} (m_n,s,t)=(-i)V_{fAA}^{\mu\nu(T)}(q,q',k,m_n)\times \tilde{G}^{T}_{\mu\nu\alpha\beta}(m_n,t)\times
 (-i)V_{f\bar\Psi\Psi}^{\alpha\beta(T)}(p_1,p_2,k,m_n)\,,\nonumber\\	
\ee
\end{widetext}
with the bulk vertices ($k=p_2-p_1=q-q'$)

\begin{widetext}
\be
&&V_{hAA}^{\mu\nu(TT)}(q,q',k,m_n)\equiv \left(\frac{\delta S_{hAA}^k}{\delta (\epsilon^{TT}_{\mu\nu}h(k,z))}\right)\,J_{h}(m_n,z)=\sqrt{2\kappa^2}\times\frac{1}{2}\int dz\sqrt{g}\,e^{-\phi}z^4K^{\mu\nu}(q,q',n,n',z)J_{h}(m_n,z)\,,\nonumber\\
&&V_{h\bar\Psi\Psi}^{\alpha\beta(TT)}(p_1,p_2,k,m_n)\equiv \left(\frac{\delta S_{h\bar\Psi\Psi}^k}{\delta (\epsilon^{TT}_{\alpha\beta}h(k,z))}\right)\,J_{h}(m_n,z)=-\sqrt{2\kappa^2}\times\frac{1}{2}\int dz\sqrt{g}\,e^{-\phi}z\bar\Psi(p_2,z)\gamma^\alpha p^\beta\Psi(p_1,z)J_{h}(m_n,z)\,,\nonumber\\ \label{vh}\nonumber\\
&&V_{fAA}^{\mu\nu(T)}(q,q',k,m_n)\equiv \left(\frac{\delta S_{fAA}^k}{\delta (\epsilon^{T}_{\mu\nu}f(k,z))}\right)\,J_{f}(m_n,z)=\nonumber\\
&&\sqrt{2\kappa^2}\times\frac{1}{4}\int dz\sqrt{g}\,e^{-\phi}z^4\tilde{k}^2\Big(K^{\mu\nu}(q,q',n,n',z)-\frac{1}{4}\eta^{\mu\nu}K(q,q',n,n',z)\Big)J_{f}(m_n,z)\,,\nonumber\\
&&V_{f\bar\Psi\Psi}^{\alpha\beta(T)}(p_1,p_2,k,m_n)\equiv\left(\frac{\delta S_{f\bar\Psi\Psi}^k}{\delta \partial_z(\epsilon^{T}_{\alpha\beta}f(k,z))}\right)\,\partial_z J_{f}(m_n,z)+ \left(\frac{\delta S_{f\bar\Psi\Psi}^k}{\delta (\epsilon^{T}_{\alpha\beta}f(k,z))}\right)\,J_{f}(m_n,z)=\nonumber\\
&&-\sqrt{2\kappa^2}\times\frac{1}{2}\int dz\sqrt{g}\,e^{-\phi}z\,\tilde{k}^2\bar\Psi(p_2,z)\Big(\eta^{\alpha\beta}\gamma^5\partial_z J_{f}(m_n,z)+\gamma^\alpha p^\beta J_{f}(m_n,z)+\eta^{\alpha\beta}k_{\mu}\gamma^\mu J_{f}(m_n,z)\Big)\Psi(p_1,z)\,,\nonumber\\
\label{vf}
\ee
\end{widetext}
with $p=({p_1+p_2})/{2}$.
The bulk-to-bulk graviton propagator is $G_{\mu\nu\alpha\beta}=G_{\mu\nu\alpha\beta}^{TT}+G_{\mu\nu\alpha\beta}^{T}$.
The transverse and traceless TT-part describes massive $2^{++}$ glueballs~\cite{Raju:2011mp,DHoker:1999bve}

\begin{widetext}
\be
 \label{Gh}
&&G_{\mu\nu\alpha\beta}^{TT}(m_n,t,z,z')=J_{h}(m_n,z)\tilde{G}_{\mu\nu\alpha\beta}^{TT}(m_n,t)J_{h}(m_n,z')\,,\nonumber\\
&&\tilde{G}_{\mu\nu\alpha\beta}^{TT}(m_n,t)=
{1 \over 2} \left({\cal T}_{\mu\alpha} {\cal T}_{\nu\beta} + {\cal T}_{\mu\beta} {\cal T}_{\nu\alpha} -
\frac 23 {\cal T}_{\mu\nu} {\cal T}_{\alpha\beta}\right)\frac{i} {t-m_n^2+i\epsilon}\,,\nonumber
\ee
\end{widetext}
with $\tilde{G}$ the boundary propagator,

\be
{\cal T}_{\mu\nu} = -\eta_{\mu\nu} + k_{\mu} k_{\nu}/m_n^2
\ee
The trace-full T-part $G_{\mu\nu\alpha\beta}^T$ describes massive $0^{++}$ glueballs~\cite{DHoker:1999bve}
\be \label{Gf}
G_{\mu\nu\alpha\beta}^{T}(m_n,t,z,z')=J_{f}(m_n,z)\tilde{G}_{\mu\nu\alpha\beta}^{T}(m_n,t)J_{f}(m_n,z')\,,\nonumber\\
\ee
with the boundary propagator
\be
&&\tilde{G}_{\mu\nu\alpha\beta}^{T}(m_n,t)=\eta_{\mu\nu}\eta_{\alpha\beta}\,\frac{i}{t-m_n^2+i\epsilon}\,.\nonumber
\ee
For $z'\rightarrow 0$, and $t=-K^2$, and $t=-K^2$ in (\ref{Gh}) and (\ref{Gf}). We can use (\ref{hbbt3}) and (\ref{fbbt3}), and simplify  (\ref{Amph}) as

\begin{widetext}
\be
\label{nAmph}
&&i{\cal A}^{h}_{Ap\rightarrow A p} (s,t)\approx(-i)\mathcal{V}^{\mu\nu(TT)}_{hAA}(q_1,q_2,k_z)\times \bigg(\frac{i}{2}\eta_{\mu\alpha}\eta_{\nu\beta}\bigg)\times(-i)\mathcal{V}^{\alpha\beta(TT)}_{h\bar\Psi\Psi}(p_1,p_2,k_z)\,,	\nonumber\\
&&i{\cal A}^{f}_{A p\rightarrow A p} (s,t)\approx(-i)\mathcal{V}^{\mu\nu(T)}_{hAA}(q_1,q_2,k)\times (i\,\eta_{\mu\nu}\eta_{\alpha\beta})\times(-i)\mathcal{V}^{\alpha\beta(T)}_{f\bar\Psi\Psi}(p_1,p_2,k)\,,	\nonumber\\
\ee
\end{widetext}
with

\begin{widetext}
\be
&&\mathcal{V}^{\mu\nu(TT)}_{hAA}(q_1,q_2,k_z)=\sqrt{2\kappa^2}\times\frac{1}{2}\int dz\sqrt{g}\,e^{-\phi}z^4K^{\mu\nu}(q,q',n,n',z)\frac{z^4}{4}\,,\nonumber\\
&&\mathcal{V}^{\alpha\beta(TT)}_{h\bar\Psi\Psi}(p_1,p_2,k_z)=-\sqrt{2\kappa^2}\times\frac{1}{2}\int dz\,\sqrt{g}\,e^{-\phi}z\,\bar\Psi(p_2,z)\gamma^\mu p^\nu\,\Psi(p_1,z)\mathcal{H}(K,z)\,,
\nonumber\\
&&\mathcal{V}^{\mu\nu(T)}_{fAA}(q_1,q_2,k)=\sqrt{2\kappa^2}\times\frac{1}{4}\int dz\sqrt{g}\,e^{-\phi}z^4\tilde{k}^2\Big(K^{\mu\nu}(q,q',n,n',z)-\frac{1}{4}\eta^{\mu\nu}K(q,q',n,n',z)\Big)\frac{z^4}{4}\,,\nonumber\\
&&\mathcal{V}^{\alpha\beta(T)}_{f\bar\Psi\Psi}(p_1,p_2,k)=-\sqrt{2\kappa^2}\times\frac{1}{2}\times\frac{1}{2}\int dz\,\sqrt{g}\,e^{-\phi}z\,\tilde{k}^2\bar\Psi(p_2,z)\Big(\eta^{\alpha\beta}\gamma^5\partial_z\mathcal{F}(K,z)+\gamma^\alpha p^{\beta}\mathcal{F}(K,z)\nonumber\\
&&+\eta^{\alpha\beta}k_{\mu}\gamma^\mu\mathcal{F}(K,z)\Big)\Psi(p_1,z)\,.
\ee
\end{widetext}

\section{Appendix: Elements of the Reggeization}

\subsubsection{Hard wall}

 The reggeization of the graviton exchange is obtained through the substitution~\cite{Polchinski:2001tt}

\be
J_h(m_n(j),z) \rightarrow \tilde{\psi}_n(j,z)=z^{-(j-2)}\psi_n(j,z)\nonumber\\
\ee
followed by the summation over all spin-j exchanges using the Sommerfeld-Watson formula

\be
\label{SPIN}
\frac 12\sum_{j\geq 2}(s^j+(-s)j)\rightarrow  -\frac {\pi} 2\int_{\mathbb C}\frac{dj}{2\pi i}\left(\frac{s^{j-2}+(-s)^{j-2}}{{\rm sin}\,\pi j}\right)\nonumber\\
\ee
for a pertinent choice of the contour ${\mathbb C}$. This requires the analytical continuation of the
exchanged amplitudes to the complex j-plane.
For the hard-wall model, the normalized wave function is given by

\bea
\psi_n(j,z)=&&c_{n}(j)z^2J_{\tilde{\Delta}(j)} (m_n(j)z)\nonumber\\
c_{n}(j)=&&\frac{1}{\sqrt{2}z_0 J_{\tilde{\Delta}(j)}(m_n(j)z_0)}
\eea
for $\partial_z\psi_n(j,z_0)=0$ and

\bea
\tilde{\Delta}(j)\equiv&&\Delta(j)-2\nonumber\\
=&&\Big({4+2\sqrt{\lambda}(j-2)}\Big)^{\frac 12}=\sqrt{2\sqrt{\lambda}(j-j_0)}\nonumber\\
\eea
with $j_0=2-\frac{2}{\sqrt{\lambda}}$ and $j\geq 2$.


{\bf For time-like momenta $k^2>0$}: we can also determine the non-normalizable wave function for the virtual tansverse-traceless spin-j glueball, as
\be
h(j,k,z)=-\sqrt{2}\kappa\sum_n \frac{F_n(j)\psi_n(j,z)}{k^2-m_n^2(j)} ,\nonumber\\\label{hpt1j}
\ee
which satisfies the boundary conditions $\partial_z h(j=2,k,z_0)=0$. We define a decay constant function (not exactly the decay constant) of the spin-j glueballs as

\be
F_n(j)\equiv \frac{C(j,k,\epsilon)}{\sqrt{2}\kappa}(-\sqrt{g}\,e^{-\phi}\,\abs{g^{xx}}\,\partial_{z'}\psi_n(j,z'))|_{z'=\epsilon}\,,\nonumber\\
\ee
The normalized wave functions of the spin-j glueballs $\psi_n(j,z)$ satisfy the normalization condition

\be
\int dz\,\sqrt{g}e^{-\phi}\,\abs{g^{xx}}\,\psi_n(j,z)\psi_m(j,z)=\delta_{nm}\,.\nonumber\\
\ee
For example, in the hard-wall model, the summation over $n$ in (\ref{hpt1j}) can be carried out analytically and is given by

\begin{widetext}
\bea
h(j,k,z)=-\sqrt{2}\kappa\sum_n \frac{F_n(j)\psi_n(j,z)}{k^2-m_n^2(j)}\
=z^2\Bigg(\frac{A(j,k,z_0)}{B(j,k,z_0)}J_{\tilde{\Delta}(j)}(kz)-Y_{\tilde{\Delta}(j)}(kz)\Bigg)\,,
\label{hpt2j}
\ee
\end{widetext}
with

\be
&&A(j,k,z_0)=\partial_{z}\big(z^2Y_{\tilde{\Delta}(j)}(kz)\big)|_{z=z_0}\,,\nonumber\\
&&B(j,k,z_0)=\partial_{z}\big(z^2J_{\tilde{\Delta}(j)}(kz)\big)|_{z=z_0}\,,
\ee
We also define

\be
&&C(j,k,\epsilon)=h(j,k,\epsilon)\approx -\epsilon^2 Y_{\tilde{\Delta}(j)}(k\epsilon)
\,.\nonumber\\
\ee

{\bf For space-like momenta $k^2=-K^2$}: the non-normalizable wave function for the virtual transverse-traceless graviton is generally given by
\be
\mathcal{H}(j,K,z)=\sqrt{2}\kappa\sum_n \frac{\mathcal{F}_n(j)\psi_n(j,z)}{K^2+m_n^2(j)}\,,\nonumber\\ \label{hps1j}
\ee
which satisfies the IR boundary conditions $\partial_z \mathcal{H}(j=2,K,z_0)=0$. We have defined a decay constant function (for space-like momenta) of the spin-j glueballs as
\be
\mathcal{F}_n(j)\equiv-\frac{\mathcal{C}(j,k,\epsilon)}{\sqrt{2}\kappa}(-\sqrt{g}\,e^{-\phi}\,\abs{g^{xx}}\,\partial_{z'}\psi_n(j,z'))|_{z'=\epsilon}\,,\nonumber\\
\ee

In the hard-wall model, the summation in (\ref{hps1j}) reduces to

\begin{widetext}
\bea
\mathcal{H}(j,K,z)=\sqrt{2}\kappa\sum_n \frac{\mathcal{F}_n(j)\psi_n(j,z)}{K^2+m_n^2(j)}
= z^2\Bigg(\frac{\mathcal{A}(j,K,z_0)}{\mathcal{B}(j,K,z_0)}I_{\tilde{\Delta}(j)}(Kz)+K_{\tilde{\Delta}(j)}(Kz)\Bigg)\,,
\label{hps2j}
\eea
\end{widetext}
 with
\be
&&\mathcal{A}(j,K,z_0)=\partial_{z}\big(z^2K_{\tilde{\Delta}(j)}(Kz)\big)|_{z=z_0}\,,\nonumber\\
&&\mathcal{B}(j,K,z_0)=\partial_{z}\big(z^2I_{\tilde{\Delta}(j)}(Kz)\big)|_{z=z_0}\,,
\ee
We also define
\be
&&\mathcal{C}(j,K,\epsilon)=\mathcal{H}(j,K,\epsilon)\approx\epsilon^2 K_{\tilde{\Delta}(j)}(K\epsilon)\,.\nonumber\\
\ee

{\bf For time-like momenta $k^2>0$}: the bulk-to-bulk propagator for the massive spin-j glueballs, can be written as

\begin{widetext}
\bea
\bar{G}_{\mu\nu\alpha\beta}^{TT}(j,z,z^\prime)=&&{1 \over 2} \left({\cal T}_{\mu\alpha}(j) {\cal T}_{\nu\beta}(j) + {\cal T}_{\mu\beta}(j){\cal T}_{\nu\alpha}(j) -
\frac 23 {\cal T}_{\mu\nu}(j) {\cal T}_{\alpha\beta}(j)\right)\bar{G}(j,z,z^\prime)
\nonumber\\
\bar{G}(j,z,z^\prime)=&&z^{-(j-2)}G(j,z,z^\prime)z^{\prime {-(j-2)}}=\sum_n \frac{\tilde{\psi}_n(j,z)\tilde{\psi}_n(j,z^\prime)}{k^2-m_n^2(j)} ,
\label{hbbt1j}
\eea
and ${\cal T}_{\mu\nu}(j) = -\eta_{\mu\nu} + k_{\mu} k_{\nu}/m_n^2(j)$. For space-like momenta, we simply replace $k^2=-K^2$ in (\ref{hbbt1j}). Also remember that

\be
h(j,k,z)=C(j,k,\epsilon)\,\sqrt{g}\,e^{-\phi}\,\abs{g^{xx}}\,\partial_{z'}G(j,z,z')|_{z'=\epsilon}\,.\nonumber\\
\ee
Note that for $z\rightarrow 0$, we can approximately write the bulk-to-bulk propagator $G(j,z,z')=\sum_n \frac{ \psi_n(z)\psi_n(z')}{k^2-m_n^2(j)}$ in (\ref{hbbt1j}) in terms of the unnormalized bulk-to-boundary propagator $h(j,k,z')$ as

\be
&&G(j,z\rightarrow 0,z')\approx \frac{\psi_{n}(z\rightarrow 0)}{(-\sqrt{2}\kappa)F_n(j)}\times
(-\sqrt{2}\kappa)\sum_n \frac{ F_n(j)\psi_n(z')}{k^2-m_n^2(j)}=\frac{2^{-\tilde{\Delta}(j)}\times k^{\tilde{\Delta}(j)}\times z^{\tilde{\Delta}(j)+2}}{\tilde{\Delta}(j)+2}\,h(j,k,z') , \nonumber\\\label{hbbt2j}
\ee
where we used

\be
F_n(j)=\frac{C(j,k,\epsilon)}{\sqrt{2}\kappa}(-\sqrt{g}\,e^{-\phi}\,\abs{g^{xx}}\,\partial_{z'}\psi_n(j,z'))|_{z'=\epsilon}=-\frac{1}{\sqrt{2}\kappa}\frac{2^{\tilde{\Delta}(j)}}{\pi}c_n(j)\,\Big(\frac{m_n(j)}{k}\Big)^{\tilde{\Delta}(j)}\Big(2+\tilde{\Delta}(j)\Big)\frac{\Gamma(\tilde{\Delta}(j))}{\Gamma(1+\tilde{\Delta}(j))}\,,\nonumber\\
\ee
\end{widetext}
with $\psi_n(z\rightarrow 0)\approx \frac{2^{-\tilde{\Delta}(j)}}{\Gamma[1+\tilde{\Delta}(j)]}c_n(j)(m_n)^{\tilde{\Delta}(j)}z^{\tilde{\Delta}(j)+2}$, and $C(j,k,\epsilon)=\frac{1}{k^2}(k\epsilon)^{2-\tilde{\Delta}(j)}\frac{2^{\tilde{\Delta}(j)}}{\pi}\Gamma(\tilde{\Delta}(j))$ for the hard-wall .

{\bf For space-like momenta $k^2>0$}: we also have

\begin{widetext}
\be
\label{hbbt2j}
G(j,z\rightarrow 0,z')\approx \frac{\psi_{n}(z\rightarrow 0)}{(\sqrt{2}\kappa)\mathcal{F}_n(j)}\times(\sqrt{2}\kappa)\sum_n \frac{ \mathcal{F}_n(j)\psi_n(z')}{K^2+m_n^2(j)}=\frac{\frac{2^{1-\tilde{\Delta}(j)}}{\pi}\times K^{\tilde{\Delta}(j)}\times z^{\tilde{\Delta}(j)+2}}{\tilde{\Delta}(j)+2}\,\mathcal{H}(j,K,z') ,
\nonumber\\
\ee
where we used

\be
\mathcal{F}_n(j)=-\frac{\mathcal{C}(j,K,\epsilon)}{\sqrt{2}\kappa}(-\sqrt{g}\,e^{-\phi}\,\abs{g^{xx}}\,\partial_{z'}\psi_n(j,z'))|_{z'=\epsilon}=\frac{1}{\sqrt{2}\kappa}2^{\tilde{\Delta}(j)-1}c_n(j)\,\Big(\frac{m_n(j)}{K}\Big)^{\tilde{\Delta}(j)}\Big(2+\tilde{\Delta}(j)\Big)\frac{\Gamma(\tilde{\Delta}(j))}{\Gamma(1+\tilde{\Delta}(j))}\,,\nonumber\\
\ee
\end{widetext}
with

\bea
\psi_n(z\rightarrow 0)\approx &&\frac{2^{-\tilde{\Delta}(j)}}{\Gamma[1+\tilde{\Delta}(j)]}c_n(j)(m_n)^{\tilde{\Delta}(j)}z^{\tilde{\Delta}(j)+2}\nonumber\\
\mathcal{C}(j,K,\epsilon)=&&\frac{1}{K^2}(K\epsilon)^{2-\tilde{\Delta}(j)}2^{\tilde{\Delta}(j)-1}\Gamma(\tilde{\Delta}(j))\nonumber\\
\eea
for the hard-wall model. Also remember that

\be
\mathcal{H}(j,K,z)=-\mathcal{C}(j,K,\epsilon)\sqrt{g}\,e^{-\phi}\,\abs{g^{xx}}\partial_{z'}G(j,z,z')|_{z'=\epsilon}\,.\nonumber\\
\ee


\subsubsection{Soft wall}

 The same relationships hold for the soft-wall model, where the spin-j glueballs' normalized wavefunctions are given in terms of the generalized Laguerre polynomials as \cite{BallonBayona:2007qr} (note that the discussion in \cite{BallonBayona:2007qr} is for general massive bulk scalar fluctuations but can be used for spin-j glueballs which have an effective bulk action (or bulk equation of motion) similar to massive bulk scalar fluctuations \cite{Polchinski:2001tt})
\be
 \label{wfSW}
&&\psi_{n}(j,z)=c_n(j)\,z^{\Delta}L_{n}^{\Delta(j)-2}(2\xi)\,,
\ee
where $\xi=\tilde{\kappa}_{N}^2z^2$, and the normalization coefficients are
\be
c_n(j)=\Big(\frac{2^{\Delta(j)}\tilde{\kappa}_{N}^{2(\Delta(j)-1)}\Gamma(n+1)}{\Gamma(n+\Delta(j)-1)}\Big)^{\frac 12}\,.
\ee
The non-normalized bulk-to-boundary propagators for spin-j glueballs are given in terms of Kummer's (confluent hypergeometric) function of the second kind and its integral representation as (for space-like momenta $k^2=-K^2$)

\begin{widetext}
\bea
\mathcal{H}(j,K,z)
&&=z^{\Delta}U\Big(a_K+\frac{\Delta(j)}{2},\Delta(j)-1;2\xi\Big)
=z^{\Delta(j)}(2\xi)^{2-\Delta(j)}U\Big(\tilde{a}(j),\tilde{b}(j);2\xi\Big)\nonumber\\
&&=z^{\Delta(j)}(2\xi)^{2-\Delta(j)}\frac{1}{\Gamma(\tilde{a}(j))}
\int_{0}^{1}dx\,x^{\tilde{a}(j)-1}(1-x)^{-\tilde{b}(j)}{\rm exp}\Big(-\frac{x}{1-x}(2\xi)\Big)\,,
\label{BBSWj}
\eea
\end{widetext}
where
\be
&&a_K=\frac{a}{2}=\frac{K^2}{8\tilde{\kappa}_N^2}\nonumber\\
&&\tilde{a}(j)=a_K+2-\frac{\Delta(j)}{2}\nonumber\\
&&\tilde{b}(j)=3-\Delta(j)
\ee
and we have used the transformation $U(m,n;y)=y^{1-n}U(1+m-n,2-n,y)$.
The bulk-to-bulk propagator can also be approximated as (for space-like momenta $k^2=-K^2$)

\begin{widetext}
\be
&&G(j,z\rightarrow 0,z')\approx \frac{\psi_{n}(z\rightarrow 0)}{(\sqrt{2}\kappa)\mathcal{F}_n(j)}\times(\sqrt{2}\kappa)\sum_n \frac{ \mathcal{F}_n(j)\psi_n(z')}{K^2+m_n^2(j)}=\frac{\frac{2^{\Delta(j)-2}\Gamma(a_K+\frac{\Delta(j)}{2})}{\Gamma(\Delta(j)-2)}\times \tilde{\kappa}_{N}^{2\Delta(j)-4}\times z^{\Delta(j)}}{\Delta(j)}\,\mathcal{H}(j,K,z') , \nonumber\\\label{hbbt2jSW}
\ee
\end{widetext}
where we used

\bea
&&\mathcal{F}_n(j)=-\frac{\mathcal{C}(j,K,\epsilon)}{\sqrt{2}\kappa}(-\sqrt{g}\,e^{-\phi}\,\abs{g^{xx}}\,\partial_{z'}\psi_n(j,z'))|_{z'=\epsilon}\,,\nonumber\\
&&\mathcal{C}(j,K,\epsilon)=\mathcal{H}(j,K,\epsilon)
\eea
and the substitution $\psi_n(z\rightarrow 0)\approx c_n(j)\,z^{\Delta}L_{n}^{\Delta(j)-2}(0)$ for the soft-wall model.


 \vfil
\end{document}